\documentclass[aps,letterpaper,twocolumn,prx,nofootinbib,floatfix]{revtex4}
\usepackage{amsmath,hyperref}
\usepackage{amsfonts}
\usepackage{amssymb}
\usepackage[scanall]{psfrag}
\usepackage{psfrag}
\usepackage{graphicx}
\usepackage{color}
\usepackage{tabularx}

\begin{document}
\newcommand{\of}[1]{\left( #1 \right)}
\newcommand{\sqof}[1]{\left[ #1 \right]}
\newcommand{\abs}[1]{\left| #1 \right|}
\newcommand{\avg}[1]{\left< #1 \right>}
\newcommand{\cuof}[1]{\left \{ #1 \right \} }
\newcommand{\bra}[1]{\left < #1 \right | }
\newcommand{\ket}[1]{\left | #1 \right > }
\newcommand{\pil}{\frac{\pi}{L}}
\newcommand{\bx}{\mathbf{x}}
\newcommand{\by}{\mathbf{y}}
\newcommand{\bk}{\mathbf{k}}
\newcommand{\bp}{\mathbf{p}}
\newcommand{\bl}{\mathbf{l}}
\newcommand{\bq}{\mathbf{q}}
\newcommand{\bs}{\mathbf{s}}
\newcommand{\psibar}{\overline{\psi}}
\newcommand{\svec}{\overrightarrow{\sigma}}
\newcommand{\dvec}{\overrightarrow{\partial}}
\newcommand{\bA}{\mathbf{A}}
\newcommand{\bdelta}{\mathbf{\delta}}
\newcommand{\bK}{\mathbf{K}}
\newcommand{\bQ}{\mathbf{Q}}
\newcommand{\bG}{\mathbf{G}}
\newcommand{\bw}{\mathbf{w}}
\newcommand{\bL}{\mathbf{L}}
\newcommand{\ohat}{\widehat{O}}
\newcommand{\up}{\uparrow}
\newcommand{\down}{\downarrow}
\newcommand{\MM}{\mathcal{M}}
\newcommand{\MN}{\mathcal{N}}
\newcommand{\MR}{\mathcal{R}}
\newcommand{\MT}{\mathcal{T}}
\newcommand{\MP}{\mathcal{P}}
\newcommand{\tW}{\tilde{W}}
\newcommand{\tX}{\tilde{X}}
\newcommand{\tY}{\tilde{Y}}
\newcommand{\tZ}{\tilde{Z}}
\newcommand{\tOm}{\tilde{\Omega}}
\newcommand{\barA}{\bar{\alpha}}
\author{Eliot Kapit$^{1,3}$}
\email{ekapit@mines.edu}
\author{Brandon A. Barton$^{2,3}$}
\author{Sean Feeney$^{3}$}
\author{George Grattan$^{3}$}
\author{Pratik Patnaik$^{1,3}$}
\author{Jacob Sagal$^{3}$}
\author{Lincoln D. Carr$^{1,2,3}$}
\author{Vadim Oganesyan$^{4,5,6}$}

\affiliation{$^1$ Department of Physics, Colorado School of Mines, 1523 Illinois St, Golden CO 80401}
\affiliation{$^2$ Department of Applied Mathematics and Statistics, Colorado School of Mines, 1500 Illinois St, Golden CO 80401}
\affiliation{$^3$ Quantum Engineering Program, Colorado School of Mines, 1523 Illinois St, Golden CO 80401}
\address{$^4$ Department of Physics and Astronomy, College of Staten Island, CUNY, Staten Island, NY 10314, USA}
\address{$^5$ Physics program and Initiative for the Theoretical Sciences, The Graduate Center, CUNY, New York, NY 10016, USA}
\address{$^6$ Center for Computational Quantum Physics, Flatiron Institute, 162 5th Avenue, New York, NY 10010, USA}

\title{On the approximability of random-hypergraph MAX-3-XORSAT problems with quantum algorithms}

\begin{abstract}

A canonical feature of the constraint satisfaction problems in NP is approximation hardness, where in the worst case, finding sufficient-quality approximate solutions is exponentially hard for all known methods. Fundamentally, the lack of any guided local minimum escape method ensures both exact and approximate classical approximation hardness, but the equivalent mechanism(s) for quantum algorithms are poorly understood. For algorithms based on Hamiltonian time evolution, we explore this question through the prototypically hard MAX-3-XORSAT problem class. We conclude that the mechanisms for quantum exact and approximation hardness are fundamentally distinct. We review known results from the literature, and identify mechanisms that make conventional quantum methods (such as Adiabatic Quantum Computing) weak approximation algorithms in the worst case. We construct a family of “spectrally filtered” quantum algorithms that escape these issues, and develop analytical theories for their performance. We show that, for random hypergraphs in the approximation-hard regime, if we define the energy to be $E = N_{\mathrm{unsat}}-N_{\mathrm{sat}}$, spectrally filtered quantum optimization will return states with $E \leq q_m E_{\mathrm{GS}}$ (where $E_{\rm GS}$ is the ground state energy) in sub-quadratic time, where conservatively, $q_m \simeq 0.59$. This is in contrast to $q_m \to 0$ for the hardest instances with classical searches. We test all of these claims with extensive numerical simulations. We do not claim that this approximation guarantee holds for all possible hypergraphs, though our algorithm's mechanism can likely generalize widely. These results suggest that quantum computers are more powerful for approximate optimization than had been previously assumed.

\end{abstract}

\maketitle

\tableofcontents

\section{Introduction}

\begin{figure}
     \centering
         \includegraphics[width=0.48\textwidth]{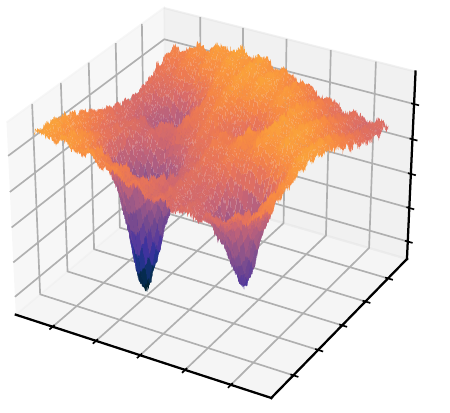}
         \includegraphics[width=0.48\textwidth]{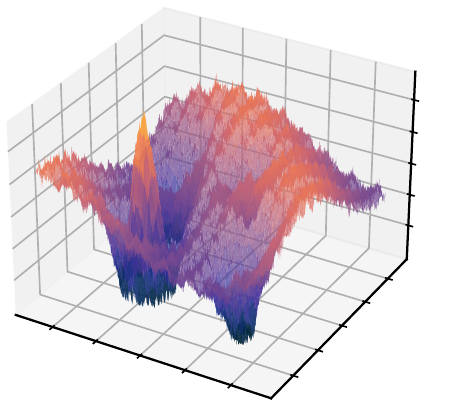}
     \caption{\textit{Illustration of the spectral folding procedure}. (Top) Sketch of the rough energy landscape of an approximation-hard CSP, with a single deep minimum whose basin of attraction is an exponentially small fraction of the configuration space. Directly optimizing this cost function through quantum approaches often misses the deep minimum entirely, due to mechanisms proposed in section~\ref{introsec} and confirmed numerically in section~\ref{qaoasec}. (Bottom) Spectrally folded energy landscape in Eq.~(\ref{nlfold}), where the problem Hamiltonian energies are mirrored around an approximation target $E = A E_{\mathrm{GS}}$. This can be implemented in a gate model algorithm with modest overhead, as shown in the text. Doing so promotes the states near the fold to an exponentially large ground state band while eliminating an interference effect that reduces tunneling into them from trivial initial states; for a wide range of problem instances (and likely, low-order problem classes), this works out to an approximation guarantee. Detailed performance predictions, and numerical benchmarking, are shown in the text.}\label{foldfig}
\end{figure}

Combinatorial optimization of constraint satisfaction problems (CSPs) is an enormously important--and often, enormously difficult--area of modern computer science~\cite{korte2011combinatorial}. Specifically, a huge array of problems in optimization, cybersecurity, machine learning, and more amount to finding low-energy configurations of large collections of $M$ few-body constraints over $N$ binary variables, see Fig.~\ref{fig:hypergraph_incidence_matrix}. Generically, the energy landscape of these cost functions is extremely rough (see Fig.~\ref{foldfig}) with exponentially many local minima, making it very difficult to find the true ground state or even a sufficiently low energy configuration.  

From the point of view of statistical physics these cost functions are often equivalent to the Hamiltonian of a \emph{disordered spin glass}, and the core hardness mechanism comes from the inability of the system to efficiently escape local minima by flipping small numbers of spins at each step~\cite{garey1974some,ullman1975np,grest1986cooling,crescenzi1995compendium,hochba1997approximation,monasson1998optimization,woeginger2003exact,bapst2013quantum,hillar2013most,venturelli2015quantum,bellitti2021entropic,jones2022random}. In the hardest problems--and often, even, in the typical case--the time to find the solution grows exponentially for all known classical and quantum methods. Remarkably, these problems are not only hard to solve, but also hard to approximate, where approximation is defined as finding any configuration  within a defined fraction of the global optimum~\cite{arora2009computational}. And while a host of clever algorithms have been proposed over the years to attack these problems, supplemented by rapid growth in computing power since the invention of integrated circuits, the exponential worst-case difficulty scaling remains, and is believed (in the as yet unproven statement that ${\rm P \neq NP}$) by most computer scientists to be a fundamental and insurmountable fact of our reality, at least for classical computers.

To make further progress, a host of heuristic quantum algorithms have been developed, such as analog quantum annealing~\cite{finnila1994quantum,kadowakinishimori1998,das2008colloquium,johnson2011quantum,boixo2014evidence,hauke2019perspectives,king2023quantum}, and its closed system and digital cousins, adiabatic quantum computing (AQC)~\cite{farhigoldstone2000,albashlidar2017} and quantum approximate optimization algorithms (QAOAs)~\cite{farhi2014quantum}. In all of these methods the spin glass problem Hamiltonian (diagonal in the computational $z$ basis) is combined with a transverse field term (typically, but not exclusively, a uniform field along $x$), which allows the system to escape from local minima through collective quantum fluctuations. In many cases this takes the form of multiqubit tunneling, a collective process where large clusters of qubits all change configuration simultaneously to tunnel from one minimum to another \cite{grattan2023exponential}. In some artificial problem instances these processes have been shown to produce exponential speedups compared to classical simulated annealing, with polynomial speedups observed in experiment for short-ranged graphs~\cite{albash2018demonstration,king2018observation,ebadi2022quantum,king2023quantum,bauza2024scaling}. However, generally applicable beyond-quadratic speedups for NP optimization problems have not been realized, and given the present noisy state of quantum hardware and the projected severe overhead of error correction~\cite{babbush2021focus}, this degree of speedup is insufficient for practical advantage over classical machines.

In this work, we approach the question of approximation hardness with the goal of identifying both core intuitive mechanisms ensuring it, and opportunities for exponential quantum advantage. In particular, for MAX-3-XORSAT (a particularly difficult CSP class, defined in detail in section~\ref{3xordef}), we argue the following:
\begin{itemize}
\item The hardness mechanism(s) for directly finding the ground state of a given problem Hamiltonian $H_P$ with heuristic quantum methods can be readily identified, and likely cannot be circumvented in the worst cases.
\item Unlike classical algorithms based on local updates, the mechanisms which ensure that it is hard to find the ground state do not generalize to ensure approximation hardness for quantum algorithms.
\item However, there are good reasons to believe that traditional quantum approaches such as AQC or QAOA are not effective approximators (e.g. reliably returning low energy states in polynomial time) in the worst case. We present relatively large scale numerical simulations that support this expectation.
\item Understanding why this is the case suggests a novel \emph{spectrally filtered quantum optimization} strategy, that transforms $H_P$ through the application of nonlinear filter functions and then solves the transformed Hamiltonian through more traditional means. For some variations the performance of spectrally filtered quantum optimization can be predicted analytically and promises an efficient approximation guarantee for an extremely large fraction of instances, well into the classically hard regime. For random hypergraphs in this regime, spectrally filtered quantum optimization provides an exponential speedup for returning low energy states.
\end{itemize}

We support these claims with extensive theoretical analysis and numerical tests of all core predictions, to the largest feasible system sizes for simulating quantum algorithms, and to the largest sizes needed to ensure asymptotic scaling has been reached for classical algorithms. In doing so, we define instance construction rules that ensure classical approximation hardness, at least for algorithms based on local updates, and numerically establish that high depth Trotterized AQC (TAQC) does not exhibit meaningful quantum advantage in finding either exact or approximate solutions to these hard instances. We analytically and numerically show that spectrally filtered quantum optimization can efficiently approximate these problems, in practice in a linearly or sub-quadrically growing number of cost function evaluations, depending on formulation. We present all these results as constructive evidence in support of our core conjecture:

\textit{\textbf{Core conjecture: Quantum approximability of random MAX-3-XORSAT instances} -- Let $\MP \of{ N_C, N, \epsilon}$ be the set of all MAX-3-XORSAT instances with $N_C$ unique three-body constraints (with no negations), $N$ binary variables, and where the optimal configurations $G$ satisfy $\of{1 - \epsilon} N_C$ constraints in total. Let $E \of{s} = - N_C \of{ n_{\mathrm{sat}} \of{s} - n_{\mathrm{unsat}} \of{s} }$ be our definition of the classical energy of configuration $s$. Let $\epsilon \ll 1$ be a small constant, let the density of constraints $N_C / N \equiv d_C \gg 1$ be large, and let $q_m > 0$ be a constant independent of $\epsilon$ and $d_C$. Finally, let $H_P$ be a random instance drawn from $\MP \of{ N_C, N, \epsilon}$. Then, with high probability, but not guaranteed for all  instances, a quantum computer with $O \of{N}$ noise-free qubits can return string(s) with $E \leq q_m E \of{G}$ in $O \of{N_C^2 N^2 {\rm polylog} \of{N} }$ or fewer total gates. We further conjecture that $q_m \geq 0.6$, and that for $\epsilon \ll 1$ and $d_C \gg 1$ it is likely close to 1.}

We present two algorithms which would prove this conjecture true if the analytically predicted scaling of either one holds asymptotically as $N \to \infty$. The algorithms are both based on spectral filtering, but their structure, and the methods used to predict performance, are very different; we see the fact that two very different calculations based on different methods return similar results as further suggestive that our conjecture is true. 

The central observation underlying both approaches is that if one is focused on finding approximate solutions using quantum time evolution, the difficulty is controlled mostly by global statistical properties of the problem, many of which are hypergraph-independent. For a ``direct" implementation of the MAX-3-XORSAT cost function, these properties imply an exponential time to solution. However, by optimizing a filtered cost function, $f \of{H_P}$ (where $f \of{E}$ is any real scalar function of the classical energy), one can tune the most important statistical properties and there are straightforward choices of $f \of{E}$ that lead to low-order polynomial scaling, at least for random hypergraphs. We state the above claim as a conjecture rather than a proved statement owing to the fact that both analytical calculations rely on approximation steps to make them tractable, and while these approximations are well-supported theoretically and the final results are matched by numerics up to the largest system sizes we could computationally access, we err on the side of caution nonetheless.

This paper is structured as follows. In section~\ref{introsec}, we first provide an overview of classical and quantum approximation hardness and the MAX-3-XORSAT problem, define the construction rules for the problems studied in this work, and state the principles which guide the parameter choices we make in formulating our algorithms. In section~\ref{specfoldalgdef}, we define spectrally filtered quantum optimization and propose three optimization algorithms based on it. Then, in section~\ref{theorysec}, we develop novel theoretical tools capable of predicting the performance of these algorithms, and establish an approximation guarantee for random hypergraphs. To verify all of these claims, in section~\ref{numresults}, we present extensive numerical tests and simulations for a variety of algorithms and problem parametrizations. A summary of the key simulation results is presented in table~\ref{approxtab}. We finally offer concluding remarks, and include additional technical details in the appendices.

\section{Hardness mechanisms, problem definitions and previous approaches}\label{introsec}

\subsection{Inability of previous methods to find or even approach the ground state at large $N$}

To motivate our novel methods, it is important to first review the qualitative reasons classical and established quantum approaches are unable to efficiently solve or approximate these problems. The classical failure mechanism is straightforward: hard problems display a high density of poor quality local minima, e.g. high energy as compared to the true ground state. Once a local minimum is reached, as no general mechanism for guided local minimum escape exists it is impossible to know in general how close one is to a ground state, either in energy or Hamming distance (number of bit flips separating two states), at least unless the found minimum happens to satisfy a large fraction of constraints. 

We emphasize that we are concerned here with an approximation guarantee, not just a method that works well in practice. For completely random problems, one can often predict the average ground state energy using statistical physics arguments, such as the famous Parisi solution to the Sherrington-Kirkpatrick model~\cite{parisi1979infinite}. If one is able to find configurations close to this energy in a given instance, that is not sufficient to rule out the existence of some other, much deeper minimum far away in configuration space, even if randomly drawing problems with such deep minima is exponentially unlikely. In other words, there is no efficient classical algorithm to know if a given instance is extremal in this way unless P=NP~\cite{arora2009computational} (see section~\ref{3xordef} for more details).

Since there are in many cases exponentially many more poor quality minima than low-lying ones, the basin of attraction of the true ground state is generally an exponentially small fraction of total configuration space and thus very hard to find, a phenomenon that has been referred to as an ``entropic barrier" to problem solving~\cite{bellitti2021entropic}. In particular, the authors of~\cite{bellitti2021entropic} showed that for the 3-XORSAT problem they considered, once a local minimum was found it was more efficient to simply restart the algorithm from a random state instead of attempting to climb out of the found minimum through penalized local operations, as in simulated annealing or parallel tempering~\cite{earl2005parallel}. We expect this may be a generic feature in some of the hardest CSP classes. And interestingly, these arguments apply equally well to approximation hardness, not just finding the optimal solution. While for a given class and system size approximation is nominally an easier problem, given that there are many more valid approximate solutions, both tasks scale exponentially in the worst case, for fundamentally the same reason. We now turn to quantum algorithms, for which the situation is considerably more complex. 

\subsection{Mechanisms for quantum solution hardness: exponentially small gaps and transverse field chaos}

We consider a broad class of heuristic quantum algorithms, which derive from quantum annealing, AQC and QAOA. These algorithms can be supplemented with additional gate model techniques, such as amplitude amplification~\cite{shaydulin2023evidence}, which improve performance but do not circumvent exponential runtimes. In these algorithms the system is initialized in the ground state of a trivial Hamiltonian, which is then slowly interpolated into $H_P$; the system is then measured. Other variations based on energy matching~\cite{baldwin2018quantum,smelyanskiy2018non,kechedzhi2018efficient,smelyanskiy2019intermittency} initialize a known or planted low energy state of $H_P$ and then use collective quantum tunneling to try to find other low energy states. 

These algorithms are fundamentally distinct from classical approaches in two ways. The first is the presumed mechanism for quantum advantage: collective quantum phase transitions (including multiqubit tunneling), where local minima can be efficiently escaped through many-body quantum effects that have no classical analog. Second and more subtly, where classical local update algorithms all start from a random high energy state and attempt to cool to low energy states, the quantum algorithms considered in this work start below the energy of the problem ground state.  They then attempt to transition into the global minimum and into other low energy states as the energy of the initial state crosses the target state from below as the total Hamiltonian changes.

Unfortunately, when compared to other applications such as quantum simulation or factoring large numbers with Shor's algorithm, the realistic performance advantage of these algorithms is generally much more modest. In the worst case--and for many problem classes, the typical case--the macroscopic quantum tunneling rate into the ground state decreases exponentially in $N$. This is a fairly generic expectation, as in many cases, including the MAX-3-XORSAT problem discussed below, the tunneling rate at the crossing point can be computed using $N$th order perturbation theory and the convergence of such a method implies exponential decay. 

But let us imagine that one could somehow ensure that the gap at the paramagnet-to-spin-glass transition decays polynomially in size of the problem. This occurs naturally in the Sherrington-Kirkpatrick problem~\cite{montanari2021optimization,farhi2022quantum} and, likely, for MAXCUT~\cite{boulebnane2021predicting,basso2021quantum}, and in other classes it can sometimes be engineered through clever algorithm formulation~\cite{PhysRevA.98.042326,hauke2019perspectives,PhysRevLett.123.120501,sels2017minimizing,bao2018optimal,yang2017optimizing,counter2023}. Unfortunately, even this is not sufficient to ensure that the solution can be found in polynomial time. This is because of a phenomenon known as transverse field chaos (TFC)~\cite{altshulerkrovi2010,knysh2016}, where energetic corrections from the transverse field can change the energy hierarchy of classical minima in the quantum spin glass phase, potentially pushing local minima below the energy of the true ground state of $H_P$ (when all transverse terms are turned off). Consequently, optimization methods will steer the system toward these false ground states first, with additional phase transitions that occur as the transverse field is further weakened. As these transitions occur at weak field values, from the analysis in section~\ref{theorysec} they are generically exponentially slow, and indeed, engineering this effect intentionally is an elegant way to craft hard benchmark problems for quantum algorithms~\cite{tang2021unconventional}. This effect can be avoided by restricting the algorithm to unstructured driver Hamiltonians~\cite{farhi2010unstructured} or very weak transverse fields, but in either case performance is very poor. The combination of exponentially small gaps and TFC make it extremely unlikely that any quantum algorithm of this type can reliably and directly find the ground state of NP-complete problems.

\subsection{Approximation hardness and conventional AQC/QAOA}\label{AQChard}

Approximation hardness, however, is another story. For many NP-hard problems guaranteeing an approximation better than random guessing by a constant fraction is also NP-hard~\cite{sahni1976p,haastad2001some,arora2009computational}. As TFC involves crossings between states that were close in energy to begin with, it cannot by itself lead to quantum approximation hardness at this level. So for example, if the random state energy of a given class is chosen to be zero and the ground state $-N$, a sufficiently general convention, then any algorithm which could  return states in polynomial time with energy $ \leq - c N$ for constant $c>0$ for all instances would promise a potentially exponential speedup. Thus, a hypothetical algorithm which always returned states with energy $-N/3$ or below in polynomial time assuming spectral continuity would still promise an exponential speedup for the hardest problem classes even if TFC reduced that guarantee to $-N/4$. In other words, while TFC can prove ruinous for finding an exact solution when the problem exhibits a clustering phase~\cite{XU2020122708,mezard2002analytic,mezard2005clustering,hartmann2006phase,krzakala2007landscape,altarelli2008relationship,ibrahimi2012set,jones2022random} and there are exponentially many states very close to the ground state in energy but well separated from it in Hamming distance, it is not going to push zero energy states into competition with $O \of{1}$ fractions of the ground state energy.\footnote{It's important to note that for some problems where the classical approximation hardness threshold is relatively close to 1, such as MAXCUT, TFC might well become a serious obstacle to achieving quantum advantage for approximation. }

Exponentially small gaps are a more serious problem, but those too cannot so easily be assumed to ensure approximation hardness. This is because the empirical ``difficulty exponents" of phase transitions in low-order CSPs are often quite small; for random hypergraph MAX-3-XORSAT instances, the minimum gap at the paramagnet-spin glass transition scales as approximately $ \Omega_0 \of{N} \sim 2^{-c N}$ with $c\simeq 0.125-0.14$ (see section~\ref{sec:multistage}). Quite generally (see the MSCALE conjecture in Ref. \cite{kapit2021noise}) this is an asymptotically accurate proxy for the tunneling matrix element, which we can use to approximate the tunneling rate from the paramagnetic initial state and the dressed excited states of $H_P$. Crucially, while the fraction of states $p_E \of{N}$ below the approximation threshold is typically exponentially small, the total number of such states is exponentially large, $\sim 2^N \times p_E \of{N}$.  Thus an optimistic first approximation would be to look for an energy $E$ where $\Omega_0^2 \of{N} \times 2^N \times p_E \of{N}$ stops decaying exponentially in $N$, which should be a relatively easy task. 

But this naive analysis is inaccurate because, as we explain momentarily, destructive quantum interference and a weakening transverse field suppress matrix elements to excited states as compared to ground state mixing, and the decay exponents grow with excitation energy. The exponential number of target states in approximation problems should still allow an algorithm suffering from both TFC and exponentially decaying matrix elements to guarantee an efficient approximation, provided, at least, that those decay exponents are not too large. 
\footnote{In a technical sense, the success of the novel spectrally folded trial minimum annealing algorithm developed in this work pivots on the mechanism to suppress destructive quantum interference and maintain a strong transverse field at the transition point, thus preserving or even improving small tunneling rate exponent in the estimate above.  It is worth noting that a dimensionless Thouless criterion similar (but distinct) to the rate-based one above has been advocated as an accurate estimate for the many-body localization-delocalization transition\cite{pal2010many,serbyn2017thouless} -- localization requires the matrix element for tunneling to be smaller than the level spacing for which inverse density of states is often an accurate proxy. While we generally expect excited states above a typical local minumum to be (intra-well) delocalized, they are not inter-well delocalized and so the question of efficient approximability we are interested in is likely closely related to whether or not we are able to delocalize the many-body dynamics across different local minima, particularly for the trial minimum annealing formulation in section~\ref{TMAdef}. Note, however, that of the two criteria the rate-based one is the more conservative of the two. We intend to revisit this analogy between approximability and delocalization in the future.}

Given all this, should we expect that AQC or QAOA will efficiently approximate NP-hard problems? Formally, it has not been proven that these algorithms are not efficient approximators when the circuit depth is allowed to grow as a low-order polynomial in system size, though there are good reasons to be doubtful, supported by a number of recent works~\cite{basso2022performance,boulebnane2022solving,anshu2023concentration,benchasattabuse2023lower}. Let us consider, qualitatively, how AQC and related methods solve such a problem, assuming the limit of quasi-continuous time; we present a quantiative analysis below in Sec.~\ref{sec:multistage}. The system is initialized in the paramagnetic ground state of a uniform transverse field Hamiltonian $H_D$, and the system evolves in time interpolating between $H_D$ and $H_P$ by lowering the coefficient of one and raising the coefficient of the other, raising the energy of the paramagnetic state until it crosses the problem ground state from below. We again emphasize the fundamental distinction of crossing from below, rather than cooling from above, in quantum and classical algorithms here. The minimum gap at the transition is expected to decay exponentially in $N$ and is approximately given by the overlap of the dressed problem ground state $\ket{G_D}$ with the paramagnet state $\ket{S}$; we perform this calculation in Sec.~\ref{sec:multistage}. If the dressings are weak enough that $\ket{G_D} = \ket{G}$ (the Grover limit~\cite{rolandcerf2002}), then $\left < S | G_D \right > = 2^{-N/2}$; however for random MAX-3-XORSAT problems the perturbative corrections spread $\ket{G_D}$ over a more significant (if still exponentially small) fraction of Hilbert space and reduce the decay exponent to around a quarter of that in the Grover case.

Assuming that we evolve time too quickly (e.g. not exponentially long) and miss the primary phase transition, we can ask how efficiently $\ket{S}$ will mix with the problem's excited states, as these rates ultimately determine the algorithm's efficacy as an approximator. Calculating them directly is very difficult, but we can qualitatively predict that they should be much smaller for two reasons. First, these crossings occur as the transverse field strength $\kappa$ is reduced toward zero, and since the dressings that reduce the decay exponent all scale with extensive powers of $\kappa$, they will be much reduced by any reduction in $\kappa$ itself. Second, when considering a dressed excited state $\ket{E_D}$, the perturbative dressings that come from mixing with states with lower energy now have opposite sign and destructively interfere with other corrections in the overlap with $\ket{S}$, in contrast to $\ket{G_D}$ where all higher order terms are positive definite. Both of these effects can, and do, considerably worsen the mixing rates with excited states and make AQC a poor approximator in the worst cases, and often in practice. That said, we do note a recent encouraging result demonstrating an empirical scaling advantage over classical heuristics in finding approximate solutions to 2d spin glasses, using real quantum annealing hardware \cite{bauza2024scaling}.

\subsection{The MAX-3-XORSAT problem and approximation-hard instance construction}\label{3xordef}

Given the severe overhead of fault tolerance~\cite{babbush2021focus}, quantum hardware is expected to exhibit enormous prefactor disadvantages as compared to parallel silicon, particularly when the comparison is made to hardware with equivalent financial value (e.g. millions of USD). Quantum algorithms thus have the most promise when the problem is hard or outright impossible for classical machines. NP-hard constraint satisfaction problems are no exception, so when benchmarking a proposed quantum algorithm it is important to ensure that the problem classes we consider are sufficiently hard for classical machines, and in the present NISQ era, exhibit their exponential difficulty at small enough $N$ that numerical simulations of quantum algorithms can demonstrate meaningful improvements. MAX-3-XORSAT~\cite{arora2009computational} problems are ideal for benchmarking quantum algorithms because their exponential difficulty scaling is obvious at small $N$ for both classical and prior quantum approaches, in contrast to other problems where the asymptotic exponential scaling often does not set in until system sizes that are prohibitively large for simulation.

\begin{figure}
    \centering
    \includegraphics{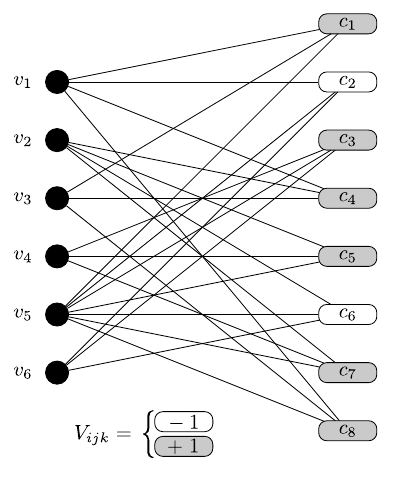}
    \caption{Graphical representation of a random 3-uniform hypergraph $G_{\mathcal{H}} = (\mathcal{V},\mathcal{E})$ used in the MAX-3-XORSAT problem. The set of vertices $\mathcal{V}$ labelled $\{v_1,\dots,v_6\}$ are connected to a hyperedge (i.e., constraint) from the set $\mathcal{E}$ on the right labeled $\{c_1, \dots, c_8\}$ if the vertex is in constraint. Each constraint necessarily contains a random set of three unique vertices. The legend specifies the value of the variable $V_{ijk} = \pm 1$ where the grey and white filled $c_j$ boxes denote a $(+1)$, and $(-1)$ valued constraint respectively. This example has a constraint density of $N_C=\frac{4}{3}N$ for 6 vertices and 8 constraints, although we use other values of $N_C$ throughout this work.}
    \label{fig:hypergraph_incidence_matrix}
\end{figure}

The MAX-3-XORSAT problem consists of a hypergraph of $N_C$ three-body constraint terms, as sketched in figure~\ref{fig:hypergraph_incidence_matrix}:
\begin{eqnarray}
    H_P = -\sum_{ijk}^{N_C} V_{ijk} Z_i Z_j Z_k, \; \; V_{ijk} = \pm 1.
\end{eqnarray}
A constraint is said to be satisfied if, for a given bitstring, $V_{ijk} Z_i Z_j Z_k = 1$, and unsatisfied otherwise. A random state satisfies half the constraints on average and is thus energy zero. This is called a hypergraph because $V_{ijk}$ has three indices rather than the usual two found in graph theory. Thanks to the linearity of the problem, one can use Gaussian elimination to check if a solution exists that satisfies all the constraints in $O \of{N^3}$ time, but if the problem is not fully satisfiable, finding the lowest energy state(s) is NP hard. 

Further, it was shown by H\aa stad~\cite{haastad2001some} that if the true ground state satisfies a fraction $\of{1-\epsilon}$ of the constraints, then finding any configuration that satisfies more than $\of{1/2 + \epsilon}$ of them is also NP-hard (see also \cite{khot2007linear,allen2015refute,d2022ihara}).  The hardest instances are thus those with small but finite $\epsilon$, e.g. almost satisfiable problems, as both finding the true ground state, and even finding an approximate solution, is exponentially difficult. Note that if the problem graph is sparse (e.g. $N_C/N$ is on the order of 1) finding approximate solutions can still be easy, since one can randomly select a fraction of the constraints $< c N$ (for some $O \of{1}$ constant $c$), and solve that new, much easier problem; solutions to this sub-problem will satisfy half the remaining constraints, on average.

To ensure that we are studying problems that are both hard to solve and hard to approximate for all known methods, inspired by Refs. \cite{alekhnovich2003more,khot2007linear,allen2015refute,barak2015beating,d2022ihara} we consider a family of instances we call planted partial solution problems (PPSPs). To construct a PPSP, we choose a small unsatisfied fraction $\epsilon$ and pick a random hypergraph of $N_C \gg N$ unique triplets; we use $\epsilon = 0.1$ in all simulations here. We then pick a random bitstring $G$ and randomly select $\of{1-\epsilon} N_C$ of the constraints to be satisfied in $G$, by picking the sign of $V_{ijk}$ appropriately, with the rest unsatisfied. If $\epsilon$ is small and $N_C /N \gg 1$, $G$ will be the problem ground state with very high probability, as the SAT/UNSAT transition for this problem is at $N_C/N \sim 0.92$~\cite{dubois20023} and at densities much higher than the SAT/UNSAT threshold ground states for random graphs satisfy $N_C/2 + O \of{ \sqrt{N \times N_C}}$ constraints. When we refer to random hypergraph problems throughout this work, we refer to this construction rule: a random, potentially fairly dense, hypergraph where one can optionally randomly chose an anomalously large fraction of constraints to be satisfied by matching the signs to a randomly chosen ground state bitstring. We note that \cite{d2022ihara} recently discovered a hardness threshold at $N_C \propto N^{3/2}$, where at higher densities the problem again becomes amenable to classical optimization.

Our PPSP construction is necessary because truly approximation hard problems--where the practical polynomial time approximation difficulty approaches the random guessing limit of the complexity class separation--are rare in the space of all possible instances. Sufficiently sparse problems are approximation-easy, and for denser random problems one can always find strings that satisfy $N_C/2 + O \of{\sqrt{N \times N_C}}$ constraints in polynomial time~\cite{haastad2002advantage,barak2015beating}, with a smaller prefactor in front of the $\sqrt{N \times N_C}$ than the prefactor in the average satisfied in the ground state. We formulated our PPSP construction to ensure our algorithm was being benchmarked on instances with a plausible claim to true classical approximation hardness. We note that commonly studied 3-regular problems~\cite{bellitti2021entropic,kowalsky20213} do not display strong approximation hardness, as they are sparse, and can be solved efficiently if satisfiable. And intriguingly, we show that, by some metrics, the performance of our novel algorithm progressively improves with increasing $N_C/N$ in this regime, at constant $\epsilon$.

Finally it is important to note an important phenomenological difference between truly random problems and random approximation-hard instances drawn from $\MP \of{ N_C, N, \epsilon}$, for which our PPSPs are merely an efficient construction method. As already mentioned, for truly random problems with large $d_C$, the true ground state $G$ satisfies $\frac{N_C}{2} + O \of{\sqrt{N_C N}}$ constraints, and the system exhibits clustering physics in the sense that there are exponentially many well-separated local minima with similar scaling, e.g. close to the energy of $G$. Finding any one cluster is not difficult, so these instances are not approximation-hard in practice, though finding the optimal solution (which may be highly degenerate) is generally hard. In contrast, for random instances with sufficiently large $d_C$ and small $\epsilon$, it is overwhelmingly likely that $G$ is unique, and indeed it may not even be possible for a proper clustering phase to emerge near the ground state energy. However, if we look at energies well above $E_G$ we find that these problems do exhibit clustering physics and tend to have exponentially many well-separated local minima satisfying the same $\frac{N_C}{2} + O \of{\sqrt{N_C N}}$ as in random problems. We refer to this as the clustering energy throughout this work; the core phenomenological difference between these instances and random problems is thus that one minimum is extensively lower than all the others, and finding it, or configurations close to it in energy, is the goal of our algorithms. That the clustering energy in PPSPs matches that of random problems is consistent with the fact that deciding if a given instance is a truly random problem or a PPSP (in our language) is itself classically hard \cite{alekhnovich2003more,allen2015refute,d2022ihara}.

\subsection{Parameter choices for algorithm formulation and benchmarking}\label{nofinetune}

Before presenting our main theoretical and numerical results, we want to discuss the principles that guide how we choose parameters when formulating and testing our novel algorithms. Our choices of the runtime scaling, and other parameters such normalizing the ground state to $-N$ in the classical problem, and the timestep $dt$, are motivated by the desire to carefully separate \emph{fundamental} limitations on algorithm performance from \emph{incidental ones}. Here, we consider fundamental limitations to be the core physics of the problem, principally exponentially small transition rates; these are the limitations that we wrote this paper to explore and which require new theoretical and algorithmic insight to address. Conversely, incidental limitations are more prosaic effects that degrade performance, such as Trotter error from larger timesteps, or the proliferation of local excitations by varying parameters too quickly. These limitations are more generic, and have easier solutions such as smaller timesteps and slower parameter variation, though determining optimal parameter choices (particularly when noise must be taken into account) can still be challenging. However, it is often not straightforward to determine, in practice, why a given algorithm is not performing as expected, particularly when the only useful information available at the end is a set of probabilities of finding various target states. 

The parameter choices throughout this work are guided by the goal of having the most easily interpretable results, to check the basic veracity our theories. We did not try to determine the optimal use of quantum resources for a given problem. By far the most important consideration in this vein is the runtime per shot. Obviously, the choice of optimal runtime scaling in any algorithm is often subtle, and our core conjecture--that random hypergraph MAX-3-XORSAT instances can with high probability be efficiently approximated in quantum polynomial time--is somewhat loose on the degree of the polynomial scaling. Our choice of generally linear scaling is motivated by the fundamental physics of the problems under consideration.

Specifically, all of the algorithms in this work consist of parameter sweeps, where states in a target manifold $\MT$ are found through collective phase transitions from a trivial initial state as it energetically crosses the states in $\MT$. In the case of multi-stage filtered optimization (MSFO), $\MT$ is simply the classical problem ground state, where for the folded optimization algorithms $\MT$ includes all states near the approximation target $A E_{GS}$ (both algorithms are detailed in the next section). Following the arguments in~\cite{kapit2021noise}, the total success probability for runtime $t_F \of{N}$ is expected to scale as
\begin{eqnarray}\label{PTF}
P \of{t_F} \propto \sum_{j \in \MT} \frac{\abs{\Omega_{0j}^2}}{W_j \of{N}} t_F \of{N},
\end{eqnarray}
where $\Omega_{0j}$ is the matrix element to transition to state $j$ in $\MT$. If the algorithm proceeds by varying dimensionless control parameter $s$, then the energy scale $W_j \of{N}$ is proportional to the slope of $\partial \of{E_0 - E_j}/\partial s$ in the vicinity of the transition
\begin{eqnarray}\label{defW}
W_j \propto \abs{ \bra{0} \frac{\partial H}{\partial s} \ket{0} - \bra{j} \frac{\partial H}{\partial s} \ket{j} }.
\end{eqnarray}
Here, $\ket{0}$ is our initial state and $\ket{j}$ is the target state, with Eq.~\ref{defW} measured in the vicinity of the avoided crossing but not precisely at it, where that expression formally vanishes. Near a single transition at $s = s_c$, the energy gap $\Delta \of{s}$ is typically expected to scale as 
\begin{eqnarray}
\Delta \of{s} \propto 2 \sqrt{ \Omega_{0j}^2 + W_j^2 \of{s-s_c}^2 },
\end{eqnarray}
and can be extracted by fitting to this form. For the problems and energy scales chosen in this work, all $W_j$ are asymptotically $O \of{N}$ in all studied cases.

This scaling, in our estimation, immediately sets a floor of $t_F \of{N} \propto N$. Even in a very ``easy" problem where $\Omega_{0j}$ is constant with $N$, for algorithms of this type for any runtime $t_F \of{N}$ that scales sublinearly $P \of{t_F}$ will subsequently decay with system size. Further, for sufficiently fast ramps, one needs to consider local diabatic heating that arises from varying Hamiltonian parameters too quickly, as a process where the system successfully transitions into a target state in $\MT$ but is then excited out of it by creating local excitations similarly reduces $P \of{t_F}$, causing it to decay more quickly than Eq.~\ref{PTF} predicts. Fortunately, for a system which is locally gapped on either side of the phase transition (as a PPSP is, though not necessarily with spectral filtering added), linear runtime is generally enough to make this issue negligible. 

Runtimes longer than linear, in contrast, present their own interpretational challenges. For the MSFO defined in the next section, the prediction of constant $P \of{t_F}$ with $t_F \of{N} \propto N^{1.46}$ comes from a calculation capable of predicting that precise scaling, but for folded optimization with an approximation target $A < 1$ (see below for definitions) our predictions are not able to predict the degree of polynomial scaling directly. For instance, the choice of $A \geq 0.6$ in the spectral folding TMA formulation comes from a calculation with many approximations expected to underestimate performance, and the empirical decay exponents for finding approximate solutions below the achievable approximation ratio $q_a$ are often quite small. Given that these algorithms are not amenable to more efficient classical simulation techniques such as tensor network methods, owing to their relatively high circuit depths and nonlocal structure, and the need to average over many instances to get good statistics, we were limited to system sizes in the high twenties at most in numerical simulations. This can make it tricky to resolve small exponents from transient polynomial scaling when trying to determine $q_a$. Our choice of linear scaling in this respect is further motivated by the fact that in the folded regime, both $\avg{\abs{\Omega_{0j}}^2}_j$ and the number of states in $\MT$ tend to exhibit simple exponential decay and growth, respectively, and consequently polynomial prefactors in $P \of{t_F}$ are expected to be minimized for $t_F \of{N} \propto N$ since $W_j \of{N}$ scales linearly as well. 

Further, though this is not a rigorous comparison, the very simplest classical improvement on random guessing--greedy or quasi-greedy steepest descent--usually takes $O \of{N}$ steps to halt, and if restricted to a constant number of cost function evaluations per shot, will asymptotically return an approximation ratio $q_a = 0$ regardless of $d_C$ and $\epsilon$ as $N \to \infty$. This is true even if the problem is convex, and more complex methods generally involve longer per-shot runtimes. And as remarked earlier, sophisticated mathematical analyses have provided similar worst-case performance bounds for QAOA and TAQC at constant depth for $k$-XORSAT problems \cite{basso2022performance,boulebnane2022solving,anshu2023concentration,benchasattabuse2023lower}, though these restrictions do not apply if we allow the circuit depth to grow with system size. For all these reasons, we thus used exclusively linear runtimes for all numerical experiments directly probing approximation hardness, and subquadratic runtimes for MSFO as guided by a rigorous theoretical prediction.

Finally, we want to address the guiding principles we used for tuning the schedules and other hyperparameters of our algorithms. It has been known since the adiabatic formulation of Grover's algorithm~\cite{rolandcerf2002} that schedule fine tuning can produce a quadratic speedup over a uniform sweep, and more general variational approaches such as ADAPT-QAOA~\cite{zhu2022adaptive} (which use large fine-tuned angles and draw operators at each step from a much larger set) can potentially offer more significant speedups, albeit with the challenge of a much larger search space for optimizing control parameters. And for many problems, these algorithms exhibit \emph{concentration}, where the set of angles that is optimal for one randomly generated instance is close to optimal for another with high probability~\cite{farhi2022quantum,anshu2023concentration}. 

We do not employ such fine tuning in this work for four reasons. First, as argued in~\cite{kapit2021noise}, quadratic speedups from schedule fine tuning are fragile (at least, for gaps that decay exponentially) and unlikely to be viable at large $N$ except in very narrow circumstances. Second, as the truly approximation hard instances of MAX-3-XORSAT (and we suspect, many other CSPs) are extremal in the space of random problems, it is less obvious that concentration arguments would apply to the cases we consider. So even if, for example, these instances exhibit concentration and some set of angles is near-optimal for a specific $\epsilon$ and $d_C = N_C/N$ scaling, we do not think it can be easily assumed that those angles would generalize to other extremal parametrizations. Third, we are interested in results we can explain, and extrapolate to larger $N$, through analytical arguments, and beyond-quadratic performance improvements from angle fine tuning are very difficult to understand in this respect. And finally, the overhead of fine tuning for specific problem instances or classes can be both severe and difficult to calculate a priori. If such fine tuning is necessary for good algorithmic performance its overhead needs to be carefully accounted for when estimating the degree of quantum speedup. We are not aware of any results showing significant quantum speedups for MAX-3-XORSAT using QAOA and angle fine tuning.

We thus restrict all of our simulations in this work to employ schedules governed by smooth, simple functions to control the relative magnitudes of $H_D$ and $H_P$. Unsurprisingly, the best scaling and prefactor choices differ somewhat for Trotterized AQC (TAQC) and variations of spectrally filtered optimization, and the results here are not optimal for any specific algorithm variation or PPSP subclass, but rather represent a decent choice, found by intuition and trial and error, for a broad set of parameters.

\section{Spectrally filtered quantum optimization algorithms}\label{specfoldalgdef}

\subsection{Key preliminaries -- filter functions}
The core idea of spectrally filtered quantum optimization is to modify how the Hamiltonian is applied to the quantum state through the introduction of a filter function, somewhat analogous to the methods used in \cite{hastings2018short,dalzell2023mind}, though our goals and filtering choices are very different. Specifically, the algorithms we consider solve problems through simulating the time evolution of a quantum state, as $\ket{\psi} \to e^{-i H \of{t} dt} \ket{\psi}$, with the exponentiated Hamiltonian discretized as a series of layers e.g. $e^{i a H_D} e^{i b H_P}$. The driver Hamiltonian, and any other additional Hamiltonian terms, are not changed by spectral filtering, so we will ignore them for now and focus on the problem Hamiltonian itself. Specifically, we write $\ket{\psi}$ in the computational basis as $\ket{\psi} = \sum_{m=0}^{2^N-1} c_m \ket{m}$, where $m$ is the decimal integer representation of a given bitstring. Then, for (arbitrary) control angle $\gamma$, the wavefunction expressed in the computational $Z$ basis evolves as
\begin{align}
&e^{i \gamma H_P} \ket{\psi} = \sum_{m=0}^{2^N-1} e^{i \gamma E \of{m}} c_m \ket{m}, \\ 
&E\of{m} = \bra{m} H_P \ket{m} = - \sum_{ijk}^{N_C} \bra{m} V_{ijk} Z_i Z_j Z_k \ket{m}.
\end{align}
In other words, the phase of each component state advances proportionally to its energy under the problem Hamiltonian, and that energy is computed at each step by applying a sequence of gates to implement each constraint. 

\begin{figure*}
    \centering
    \includegraphics[width=2\columnwidth]{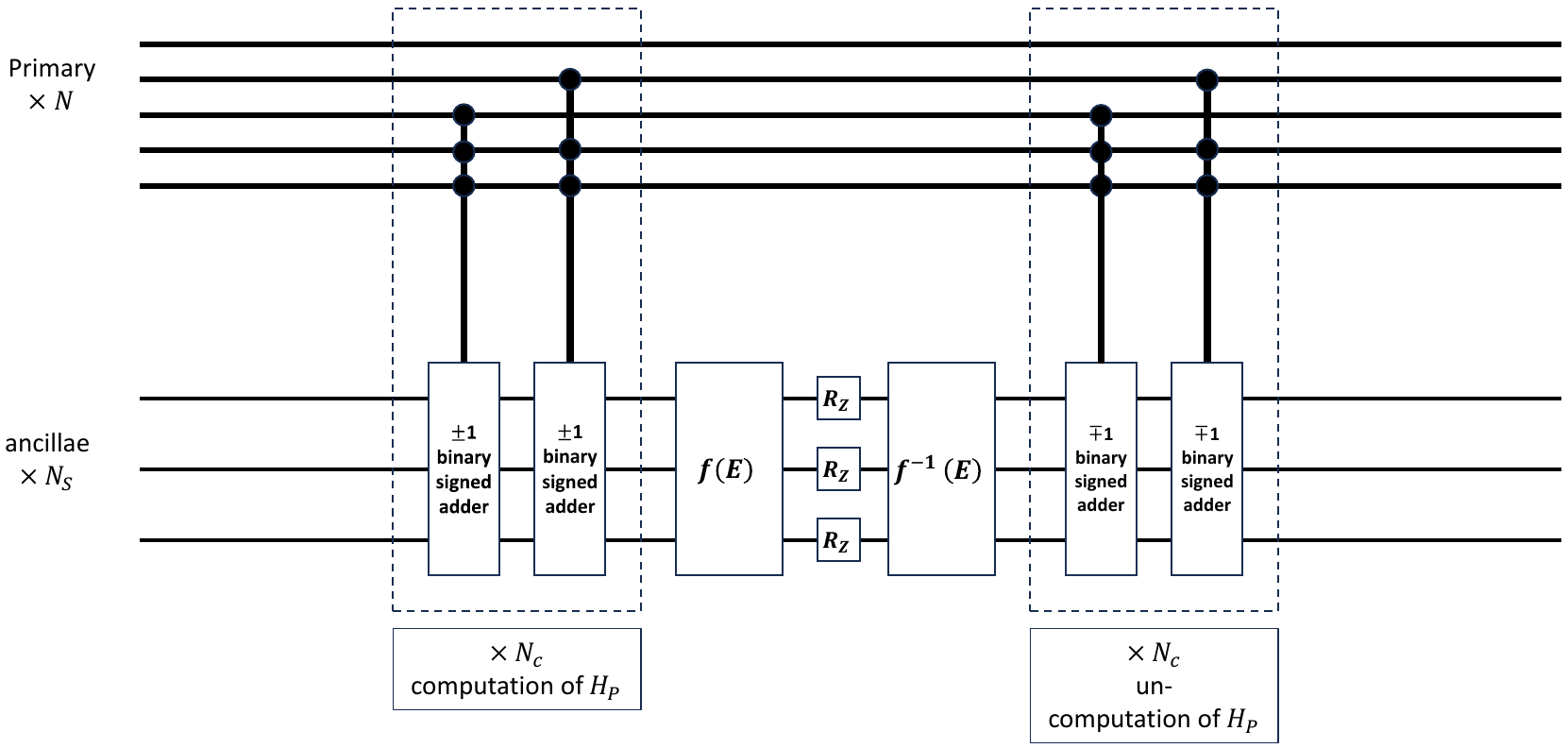}
    \caption{A schematic for implementing spectrally deformed time evolution $U= \exp(i f(H_P) dt)$. A set of gates maps the value of each of the $N_C$ constraints to adding or subtracting 1 to a register of ancilla qubits, using binary signed adder circuits controlled by the value of $V_{ijk} Z_i Z_j Z_k$. A sequence of additional gates computes $f \of{E}$ from $E$ (likely using more ancilla qubits), and then a set of local $Z$ rotations is applied to the register storing $f \of{E}$ to advance the phase of each component of $\ket{\psi}$ proportionally. The computation of $f \of{E}$ and entangling with ``constraint-controlled" gates are then uncomputed, returning the ancillas to their initial state and disentangling them from the $N$ primary qubits over which the problem is defined. The net result of this entire process is to enact the operation in Eq.~\ref{actFE}, evolving time under an arbitrary function of the diagonal Hamiltonian (transverse field layers and other operations on the primary qubits are not shown). For relatively simple functions, the net overhead of this entire process (compared to enacting $\exp \of{i H_P dt}$ directly) is polylogarithmic in $N$.}
    \label{fig:folder}
\end{figure*}

In spectrally filtered optimization, the phase of each component instead advances proportional to an arbitrary function $f$ of the diagonal $H_P$,
\begin{eqnarray}\label{actFE}
\ket{\psi} \to e^{i \gamma f \of{H_P}} \ket{\psi} = \sum_{m=0}^{2^N-1} e^{i \gamma f\of{E \of{m}} } c_m \ket{m}.
\end{eqnarray}
This can be accomplished by introducing a register of auxiliary qubits, applying a gate sequence that maps the sum of the constraint terms to a fraction of that register to store $E$, using a second fraction of that register to compute $f \of{E}$, applying a sequence of controlled-phase gates to advance the phase by $f \of{E}$, and then uncomputing the previous steps to return the register to its initial state. The entire process is sketched in Fig.~\ref{fig:folder}. Provided $f$ is a relatively simple function, this adds a multiplicative overhead which is polylogarithmic in $N$, since $E \of{m}$ is bounded by a polynomial in $N$ and each arithmetic operation takes $O \of{\log{N}}$ steps.\footnote{We note that for any choice of $f$ more complex than multiplying $E$ by a constant (something that does not require auxiliary qubits to begin with), any spatial locality the graph might have is lost in this step, since $E \of{m}$ is a global quantity which we are deforming with $f$.}

The algorithms we consider in this work use two classes of filter function, which we call \emph{nonlinear folding} and \emph{warping} transformations, both of which assume that the classical problem has been normalized \footnote{Formally, this choice assumes that we know the fraction of the $N_C$ constraints which are satisfied in the ground state, something that we cannot know in advance of running our quantum optimization algorithm! However, we can simply repeatedly run the algorithm with different normalization choices to guess its value, a prefactor overhead of at most $O \of{N_C}$.} so that $-N \leq E \leq N$. We define the nonlinear folding transformation as:
\begin{eqnarray}\label{nlfold}
F_{f} \of{E, N, A, x} \equiv N \, {\rm sign} \of{E} \, \of{1 - \abs{\abs{\frac{E}{N}} - A}^x }.
\end{eqnarray}
Here, $A \leq 1$ sets the approximation target of this algorithm, and $x$ sets the (continuous) degree of the fold; the choice $F_f \of{E,N,1,1}$ applied to the normalized $H_P$ has no physical or algorithmic consequences. Critically, it is defined using the conventions that random states have energy zero, so returning a state with energy $A E_{\mathrm{GS}}$ approximates, by a factor of $A$, the degree to which the true ground state itself improves on random states. These normalization choices ensure that the energy difference between the new ground states of the folded problem, and random states, is $N$ as in the original renormalized problem. Making this choice simplifies the analysis significantly. We similarly define the warping transformation as:
\begin{eqnarray}\label{nlwarp}
F_{w} \of{E,N,w} \equiv N^{1-w} \times {\rm sign} \of{E} \times \abs{E}^w.
\end{eqnarray}

A good choice of $A$ is important for spectrally folded quantum optimization to succeed; if (absent warping) $A$ is chosen to be too close to 1, then we risk failing to well-approximate $H_P$ due to the interference effects mentioned in section~\ref{AQChard}. A choice of $A$ which is too small will return a suboptimal approximation ratio, and if $A$ is too close to zero, cause instabilities from having a poorly defined problem to solve. Fortunately, for random hypergraph MAX-3-XORSAT instances drawn from $\MP \of{N_C, N, \epsilon}$ we can predict the threshold $A$ for which we expect a polynomial depth TMA circuit to return states with $E \simeq A E_{\mathrm{GS}}$ from first principles. The ideal value of $A$ depends both on the problem class and on the variation of spectral folding employed; for MAX-3-XORSAT, $A \geq 0.6$ is achievable as derived below in section~\ref{TMAPTAS}. This is a significant leap over the best known classical approximation algorithm for this problem~\cite{barak2015beating}, which offers a weaker guarantee with much more restricted viability, in the worst case, equivalent to $A \to 0$ in our notation.  It is also a significant leap over recent quantum approaches to this problem~\cite{anshu2021improved,marwaha2022bounds}. We note that minimizing $\of{H-E}^2$ is not itself a novel idea and has been used in classical and quantum algorithms for finding states close to specific energies in chemical and many-body systems~\cite{wang1994electronic,wang1994solving,mcclean2016theory,santagati2018witnessing,zhang2021adaptive,tilly2022variational,tazi2023folded}. To our knowledge, however, the use of spectral folding for approximate optimization of CSPs is novel, both in concept and in the analysis we present below to choose $A$ and understand at a deeper level why it presents significant advantages over optimizing the problem $H_P$ directly. We similarly are not aware of any previous work employing non-quadratic folding, or where a warping transformation as in Eq.~\ref{nlwarp} is applied to optimize a classical CSP.

From hereon, we let $H_{\mathrm{fold}}$ be the spectrally folded problem Hamiltonian. Having defined it, there are a number of ways we can attempt to find its ground states. We present three -- first, for completeness, the direct conventional approach, inspired by AQC, and then two new protocols.
\subsection{Direct method}
The simplest choice, and one that performs well empirically, is AQC-inspired state preparation, where we interpolate in Trotterized evolution between the transverse field driver and the folded problem over a time $t_F$:
\begin{align}\label{foldAQC}
&    H \of{t} = f \of{t} H_D + g \of{t} H_{\mathrm{fold}}, 
H_D = - \sum_j X_j, \\
&f \of{0} = g \of{t_F} = 1, \; \;  f \of{t_F} = g \of{0} = 0.
\end{align}
To simulate this prescription classically, it is easiest to simply construct the phase oracle as a diagonal operator applied to the $N$ qubit state at each timestep. As we shall show later, with $x=2$, $A = 0.75$, and $t_F$ growing linearly with $N$, this algorithm performs well in practice with numerical simulation, though we cannot predict its performance analytically. To achieve a better approximation of the ground state--and, crucially, a formulation where performance can be predicted from first principles--we need to apply a warping filter as well, or consider tunneling between a spin glass minimum and the folded ground state band. 

\subsection{Multi-stage filtered optimization}\label{MSFOdef}

\begin{figure}
\includegraphics[width=\columnwidth]{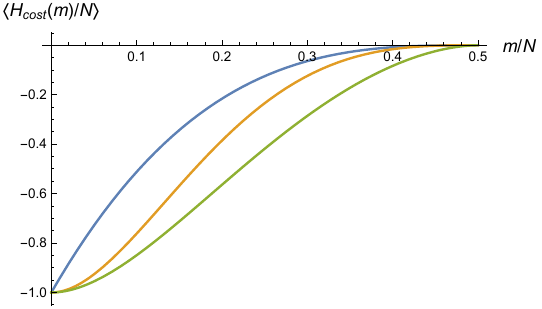}
\caption{Visualization of the composed spectral filters in Eq.~\ref{composedfilters}, as a function of $m$ random flips away from the ground state $G$, normalized so that the ground state energy is $H_{\mathrm{cost}}(m=0)/N=-1$. Blue curve: unfiltered, hypergraph-independent mean energy Eq.~\ref{avgE}. Gold curve: same energy with a nonlinear fold (Eq.~\ref{nlfold}) applied, with $A=1$ and $x=2$. Green curve: applies a spectral warping filter (Eq.~\ref{nlwarp}) after the nonlinear fold, with $w=0.6$; it is this cost function which is optimized in the first stage of our novel multi-stage optimization algorithm.}\label{filterfig}
\end{figure}

We now present a more complex, and more powerful, formulation that we call multi-stage filtered optimzation (MSFO). We will first state the steps of the algorithm, then in section~\ref{sec:multistage} provide a detailed explanation and analysis of its stages, and expected performance. As mentioned earlier, we assume that we know the fraction $\epsilon$ of clauses which are unsatisfied in the ground state, so we can properly set the normalization of $H_P$ so that the ground state energy is $-N$ and random states have energy zero. In the worst case, there is an overhead of $O \of{N_C}$ for guessing this value; the analysis below proceeds for the trials where it was guessed correctly. MSFO follows the following sequence, per-shot:
\begin{itemize}
\item First, we initialize the uniform superposition $\ket{S}$, the ground state of $H_D = - \kappa \sum_j X_j$; here $\kappa$ is the transverse field strength. We use $\kappa = 1$ for simulations later in this work but other $O \of{1}$ choices can of course be used. For the first stage, we optimize the following cost function, comprised of two composed spectral filters defined in Eqs.~(\ref{nlfold},\ref{nlwarp}) applied successively to $H_P$ (see FIG.~\ref{filterfig} for an illustration), with parameters $x \geq 2$ and $w < 1$:
\begin{eqnarray}\label{composedfilters}
H_{\mathrm{cost}}^{x,w} &=& \sum_{n=0}^{2^N - 1} F_w \of{ E_f \of{n}, N, w} \ket{n} \bra{n}, \\
E_f \of{n} & \equiv & F_f \of{ E \of{n}, N, 1, x}. \nonumber
\end{eqnarray}
\item We then choose a time $T_1$ and timestep $dt$ for the first stage, as well as a function $s \of{t}$ which ramps from 0 to 1 over the time $T_1$. In this work, we choose the simplest possible choice $s \of{t} = t/T_1$, but undoubtedly benefits can be gained by further tuning. In this stage, we evolve time in Trotterized evolution from $t=0$ to $t=T_1$ by iterating
\begin{eqnarray}\label{timeevolvestage1}
\ket{\psi \of{t + dt} } = e^{-2 \pi i dt \kappa H_D} e^{-2 \pi i dt s \of{t} H_{\mathrm{cost}}^{x,w}} \ket{\psi \of{t}}.
\end{eqnarray}
\item We have now concluded stage 1. For stage 2, we pick a new duration $T_2$, and two new functions $s_x \of{t}$ and $s_w \of{t}$. These functions are used to ramp off the nonlinear folding and warping transformations, so that at the end of stage 2, the classical cost function is physically equivalent to the un-filtered $H_P$. Both of these functions are defined such that $ s_x \of{0} = 1$, $ s_x \of{T_2} = 1/x$ and $ s_w \of{0} = 1$, $ s_w \of{T_2} = 1/w$. We evolve time from $t=0$ to $t = T_2$ by iterating:
\begin{eqnarray}\label{timeevolvestage2}
\ket{\psi \of{t + dt} } = e^{-2 \pi i dt \kappa H_D} e^{-2 \pi i dt H_{\mathrm{cost}}^{x s_x \of{t},w s_w \of{t}}} \ket{\psi \of{t}}.
\end{eqnarray}
\item In practice, linear interpolation is completely sufficient for the second stage. The nonlinear folding and warping transformations have now been smoothly turned off. For our final stage, we pick a third evolution time $T_3$, and evolve time as in the previous stages, this time ramping down the transverse field strength $\kappa$ to zero. We then measure the system in the computational basis, and this shot is concluded.
\end{itemize}
The simple linear schedules mentioned above are not optimal, but they do not need to be. In the next section we show that, for $H_P$ as defined in our core conjecture, $T_1$ increasing subquadratically with $N$, and provided that we can ignore transverse field chaos, e.g. corrections pushing local minima of $H_{\mathrm{cost}}^{x,w}$ below $G$ and consequently steering the system toward those states first, evolution in the first stage will find the quantum ground state of $H_D + H_{\mathrm{cost}}^{x,w}$ with $O \of{1}$ probability. The runtime of this stage is set by the scaling of a single transition that occurs when $E_{0,D} \simeq - \kappa N$ crosses $E_{G,D}$, the energy of the dressed ground state $\ket{G_D}$ of $H_{\mathrm{cost}}^{x,w}$, and the minimum gap scales proportionally to $\left < S_D | G_D \right >$, where $\ket{S_D}$ is ground state of $H_D$ as dressed by corrections from $H_{\mathrm{cost}}^{x,w}$. As discussed in section~\ref{sec:multistage}, the first stage forms the bottleneck of the algorithm and subsequent stages are much easier, both to understand conceptually and in their minimum gate count. The asymptotic gate count of this algorithm depends on the choice of $x$ and $w$ and is considered in section~\ref{sec:multistage}.

\subsection{Spectrally folded trial minimum annealing}\label{TMAdef}
For our third and final algorithm, we consider trial minimum annealing (TMA), originally proposed in~\cite{kapit2021systems2}. We explore the TMA formulation in depth because we can predict its scaling analytically. In this scheme, a simple classical algorithm is used to find an initial local minimum of $H_P$; the quality of the minimum does not matter and for approximation-hard instances we assume it is far above the true ground state energy, in the worst case asymptotically approaching random guessing. Let this classical minimum state be $\ket{L}$. We will use the linear folding prescription ($x = 1$, no warping transformation applied) in Eq.~\ref{nlfold} for $H_P$ itself, and add to it a new, diagonal \emph{lowering Hamiltonian} $H_L$ which has $\ket{L}$ as its ground state, and assign to it a time-dependent coefficient $C \of{t}$. Finally, we choose an approximation target $A \leq 1$ that specifies the energy $-A N$ (where $E_{GS} = -N$ is negative in our conventions) that our algorithm aims to find states near. Our total cost function Hamiltonian is
\begin{eqnarray}\label{TMAhamdef}
H_{\mathrm{cost}} \of{t} = \frac{\abs{H_P + A N}}{A} + C \of{t} H_L.
\end{eqnarray}
This folding form is a simplification of the generalized nonlinear folding function Eq.~\ref{nlfold}, with $x=1$; the symmetrization there is unnecessary in practice for the values of $A \geq 0.6$ considered in this work.

To go further, we need to specify a form for $H_L$. For this analysis will choose a new random hypergraph of $N_C$ triples which, critically, has no correlation to the hypergraph of $H_P$; we choose the same $N_C$ as the problem for convenience here but any $O \of{N}$ quantity should be fine.  We choose the signs of the new constraints so that $\ket{L}$ satisfies all of them. $H_L$ is not included in the folding procedure so applied separately in time evolution. We then choose $C \of{t = 0}$ such that the initial energy of $\ket{L}$ (defined by $H_{\mathrm{fold}} + C \of{t} H_L$) is well below $-N$ but remains $O \of{N}$. Our algorithm simulates appropriately discretized time evolution in the following sequence:
\begin{itemize}
\item Initialize $\ket{L}$, with $H_{\mathrm{cost}}$ always on, and evolve time smoothly ramping up the transverse field from 0 to $\kappa$ in time $t_r$. We assume $t_r$ increases linearly with $N$ and choose $\kappa \leq \kappa_c$; $\kappa_c \simeq 1.29$ for MAX-3-XORSAT but can vary for other problems, and we expect weak variation from one instance to the next. We want to choose $\kappa$ at or just below this value, so we remain in the dressed problem phase (DPP, defined in section~\ref{theorysec}) at all times. Leaving and then re-entering the DPP does not mean the algorithm will fail but makes predictions harder. This smoothly evolves the state to $\ket{L_D}$, the dressed version of $\ket{L}$.
\item Evolve time for a total time $T$, likely also $O \of{N}$, where $C \of{t}$ is smoothly ramped down to zero, ensuring that $\ket{L}$ crosses the hyperspherical shell of ground states of $H_{\mathrm{fold}} = \abs{H_P + A N}/A$. Note that this crossing occurs when the ground state energy of $C \of{t} H_L$ is $O \of{-N}$, and if we assume the initial minimum was uncorrelated with the true ground state $\ket{G}$, the mean Hamming distance between $\ket{L}$ and any of the ground states of the folded Hamiltonian is $N/2$ flips. Consequently, $H_L$ adds an $O \of{\sqrt{N}}$ energy uncertainty to these states that has no meaningful impact on the approximation ratio. The algorithm succeeds if, at this stage, the many-body state transitions into any state near the fold energy through a collective tunneling process, and we predict the dynamics and scaling of this stage in the next section.
\item Finally, ramp the transverse down to zero smoothly over $t_r$ and measure the system in the $z$ basis. For an appropriate $O \of{1}$ choice of $A$ and a random problem hypergraph, this algorithm will return states with energies close to $A E_{\mathrm{GS}}$ with constant probability. We can optionally repeat the algorithm many times, starting from different choices of $\ket{L}$, to ensure a fairer sampling of states in that energy range.
\end{itemize}
\subsection{Resource estimates for spectrally folded trial minimum annealing}
The total gate count of this algorithm is as follows. We have a factor of $O \of{N + N_C \; {\rm polylog} \of{N}}$ per timestep for the layers of transverse field, $H_{\mathrm{fold}}$ and $H_L$ terms, which we simplify to $N_C \; {\rm polylog \of{N}}$. We obtain, in the worst case, a factor of $O \of{N_C}$ for the number of guesses one needs to make to correctly set the normalization for a chosen $A$. We assume, on empirical grounds, that the total quantum evolution time is $O \of{N}$, for the reasons discussed in section~\ref{nofinetune}. Finally, in the worst case we expect $dt$ may need to decrease polynomially at large $N$, but $dt$ constant or increasing logarithmically is empirically and intuitively fine in the typical case (see also \cite{granet2024benchmarking}). Taken together, and we emphasize assuming that the algorithm is capable of returning states with $E < A E_{\mathrm{GS}}$ in constant probability, we estimate a total runtime between $O \of{N_C^2 N^2 \; {\rm polylog} \of{N}}$ in the worst case and $O \of{N_C N \; {\rm polylog} \of{N}}$ in more typical cases. Assuming constant per-shot success probability, the asymptotic runtimes of MSFO and TMA are thus similar. Justifying that assumption for both algorithms is in some sense the core task of this paper, which we now begin.

\section{Analytical performance predictions}\label{theorysec}

We are now ready to present the main results of this paper: analytical performance predictions for MSFO and spectrally folded TMA. For all the reasons that we specified in the previous sections, we consider approximate optimization to be a much more promising objective for achieving exponential speedups with quantum algorithms than attempts to directly find the exact ground state of hard CSPs. But this too would seem to be an enormously difficult task, given the complexity of performing a rigorous calculation for any individual disordered problem hypergraph. We make this challenge tractable through the following methods. We first narrow our focus to random instances drawn from the space of those which are formally approximation-hard, which allows us to take advantage of certain statistical properties of these random hypergraphs. Second, we present two algorithmic approaches which, by design, are constructed to minimize the influence of specific details of any given instance on final results. 

The two approaches are very different. MSFO uses a sequence of composed filtering transformations to overcome the first order transition that makes more traditional attempts to find the ground state with quantum time evolution take exponential time, at the cost of potentially worsening TFC. For this algorithm we can predict the time to solution scaling through analogy to mean field models, a claim we subsequently verify in numerical simulation. In contrast, TMA applies a folding transformation to promote an exponentially large band of excited states to ground states of the transformed cost function, and then finds them through collective tunneling, at the cost of being unable to find any lower energy states directly. To predict the scaling and parameter choices of this algorithm, we have developed a somewhat novel resummed extensive order perturbation theory based on previous \emph{forward approximation} results~\cite{bravyi2011schrieffer,pietracaprina2016forward,baldwinlaumann2016,baldwinlaumann2017,scardicchio2017perturbation,baldwin2018quantum,grattan2023exponential}.

\subsection{Cost per flip curve}\label{sec:costflip}

Underpinning both the formulation of our novel algorithms, and our analysis of their performance, is an asymptotically exact, hypergraph-independent \cite{hastings2018short} prediction of the average energy $E_{\mathrm{avg}} \of{x}$ for $x$ random classical bit flips away from the ground state. Specifically, for MAX-3-XORSAT, our problem is defined as a hypergraph of $N_C$ $p$-body constraints (e.g. $ V_{ijk} Z_i Z_j Z_k$) over $N$ variables, where $p = 3$ here, and each constraint returns $\pm 1$ and flips to the opposite value when any one of the spins flips. Let us say the system is in some classical configuration $s$; the energy is then given by $E \of{s} = N_C \of{n_{\mathrm{unsat}}-n_{\mathrm{sat}}}$, where a sat constraint returns 1 in this notation, and $n$ implies a density.

Now we flip one spin at random. Each spin participates in, on average, $p N_C/N$ constraints, and consequently, the average energy change for a single spin flip is
\begin{eqnarray}\label{avgsingleflip}
\Delta E_{\mathrm{avg}} = + 2 p \frac{N_C}{N} \of{N_{\mathrm{sat}} - N_{\mathrm{unsat}}} = - 2 p E.
\end{eqnarray}
This is an exact statement, though again it applies only to averages. Now imagine we have flipped $x$ spins from our initial configuration. If we flip one more spin at random, once again $\Delta E_{\mathrm{avg}} \of{x} = -2 p E \of{x} \Delta x $. However, we have already flipped $x$ spins, so when we flip one more at random, with probability $x/N$ we have flipped a spin back and are computing the energy change associated with reducing $x$ by 1. Consequently
\begin{eqnarray}\label{perflip}
\of{1 - \frac{2x}{N}} \frac{\Delta E_{\mathrm{avg}} \of{x}}{\Delta x} = - 2 p E_{\mathrm{avg}} \of{x}.
\end{eqnarray}
If we combine this with $E_{\mathrm{avg}} \of{0} = E \of{G}$ and $E_{\mathrm{avg}} \of{N/2} = 0$ (as $N/2$ flips reaches fully random and uncorrelated configurations), we can interpret this as a differential equation discretized over $N/2$ steps. This differential equation has a straightforward solution: starting from the ground state $G$, for $x$ unique random flips away the average energy is
\begin{eqnarray}\label{avgE}
E_{\mathrm{avg}} \of{x} = E \of{G} \of{1- \frac{2x}{N}}^p.
\end{eqnarray}
Note that this statement is hypergraph independent, and is only an average; individual trajectories will of course display substantial variations. It is a rederivation of a familiar result for dense hypergraphs with Gaussian distributed constraint energies~\cite{derrida1980random,baldwin2018quantum}, but is applicable in much broader contexts and is easy to confirm numerically. 

We do caution however that Eq.~\ref{avgE} is only asymptotically exact, owning to the discretization into single flips in Eq.~\ref{perflip}. At small $N$, deviations from it can be meaningful, particularly for larger $p$; see appendix~\ref{appendix:costflip} for more details. Throughout this work, we rescale all problems by a multiplicative constant so that the ground state energy is $-N$. This rescaling assumes that we can correctly guess the fraction $\epsilon$ of unsatisfied constraints in $G$, as mentioned earlier.

\subsection{Multi-stage filtered optimization performance analysis}\label{sec:multistage}

\subsubsection{First stage scaling}

With this result in hand, we can now predict the scaling of MSFO, beginning with the first stage. To demonstrate that we find the ground state of $H_D + H_{\mathrm{cost}}^{x,w}$, we must predict the scaling of the quantum phase transition that occurs during the ramp up of the cost function Hamiltonian. Such transitions are governed by energetics of flip sequences in the neighborhood of $G$. The cost per flip a few flips away from the ground state determines the location of the transition (as a function of $\kappa$ and $t/T_1$). For example, in the unfiltered problem,
\begin{eqnarray}
E_{G,D} \simeq -N \of{ 1 + \frac{\kappa^2}{2p}+\frac{\kappa^4}{8p^3} + ...}\,.
\end{eqnarray}
This crosses the energy of the paramagnetic state, which is $-N \kappa$ and minimally perturbed by the problem Hamiltonian, at $\kappa_c \simeq 1.29$ for our family of $p=3$ PPSPs. In contrast, the high energy structure of states far from $G$ can significantly influence the scaling of the minimum gap, as it is the overlap of the two competing states that determines the width of the resulting avoided crossing.

We first consider the low energy properties of $H_{\mathrm{cost}}^{x,w}$ in Eq.~\ref{composedfilters}, as the precise location of the transition for any given $H_P$ will determine the magnitudes of the high-energy wavefunction corrections that ultimately set whether or not the system can successfully make the transition into the ground state of $H_D + H_{\mathrm{cost}}^{x,w}$ in time $T_1$. As stated in our core conjecture, we are concerned with approximating random instances from the space of problems where classical approximation is hard both formally and, we presume, in practice: almost-satisfiable instances ($\epsilon \ll 1$) with high densities of constraints $d_C \gg 1$. We can therefore assume a random hypergraph, and if we draw constraints from the set of $\binom{N}{3}$ unique triples (for a total of $d_C N$ constraints), then the distribution of the number of constraints each variable participates in is well approximated as Poissonian, with mean $3 d_C$ constraints per variable. Since the hardest instances to approximate are those where $d_C$ is large but most constraints are satisfied in $G$, we can therefore infer that the local distribution of the cost per flip away from $G$ is also (approximately) Poissonian,\footnote{Formally the distribution is not exactly Poissonian, owing to the fact that no triples are repeated in the hypergraph. At large $d_C$ we expect this will further suppress the tail of the Poisson distribution, as if we construct the hypergraph by drawing constraints randomly one by one, the number of unique additional constraints to draw decreases with local degree. If anything this will further reduce the local degree variance, and it is the suppression of local variance with $d_C$ which is important to our claims.} with mean 6 (reflecting the normalization of ground state energy $-N$) and variance $2/\of{d_C \of{1 - 2 \epsilon}}$. Large (e.g. extensive in $N$) fractional fluctuations in the cost per flip, are, in other words, highly suppressed for random approximation-hard instances. Obvious potential exceptions to this are instances with either small $d_C$ or larger $\epsilon$, but both of those conditions push us away from the realm of approximation hard instances, at least in practice.

That said, our goal is to confidently predict the structure of the phase transition, and even small (percentage-wise) variations in the ratio of the coefficients of $H_D$ and $H_{\mathrm{cost}}^{x,w}$ can meaningfully impact the scaling of the transition. The structure of the composed spectral filters Eq.~\ref{composedfilters} is chosen to further wash out the influence of the details of $H_P$ on the transition into the ground state of $H_D + H_{\mathrm{cost}}^{x,w}$. Specifically, the when $x=2$ the mean classical cost per flip cost drops to $O \of{1/N}$ about $G$, and $O \of{1}$ energies are not reached until $O \of{\sqrt{N}}$ classical flips, at which point superpolynomially many states are already contributing to the many-body wavefunction. One can also, for instance, choose $x=3$ in the filter function Eq.~\ref{nlfold}, at which point the mean classical excitation energy near $G$ is $O \of{1/N^2}$ and $O \of{1}$ energies are not reached until $O \of{N^{2/3}}$ flips; the performance of these two formulations is qualitatively similar at reasonably large $d_C$. Given a filtered local excitation energy which decays as $O \of{1/N}$ or faster, and the exponential suppression of extensive variations in the local excitation energy for random $H_P$ drawn from $\MP \of{ N_C, N, \epsilon}$, we can assert the following:

\textit{Claim 1 (regularity of phase transitions under spectral filtering): for a given choice of $x \geq 2$ and $w < 1$ in Eqs.~\ref{nlfold}-\ref{composedfilters}, and a unit coefficient in front of $H_{\mathrm{cost}}^{x,w}$ so that the energy difference between $G$ and random configurations is $N$, let $\kappa_c$ be the average transverse field strength where the phase transition from the paramagnet $\ket{S_D}$ into the dressed problem ground state $\ket{G_D}$ occurs, as determined by the position of the minimum gap. Then, with very high probability, but not guaranteed for all instances, for a given random instance $H_P$ drawn from $\MP \of{ N_C, N, \epsilon}$, the transverse field strength $\kappa$ where the phase transition occurs will be equal to $\kappa_c$ up to corrections that scale as $O \of{1/N^d}$ for some $d$ that depends on $x$ and $w$.}

To further motivate this claim, we make the following observation. The critical transverse field strength $\kappa_c$ is defined to be the value of $\kappa$, for unit prefactor $H_{\mathrm{cost}}^{x,w}$, where $E_{S,D} = E_{G,D}$. Since $\bra{S} H_{\mathrm{cost}} \ket{S} = 0$ and $\bra{G} H_D \ket{G} = 0$, the shifts to these two energies that set $\kappa_c$ come entirely from the higher order corrections to each state induced by the other Hamiltonian. And if we use second order perturbation theory as an intuitive guide (but not, we emphasize, as a rigorous calculational tool here), we can see that $E_{S,D}$ is very modestly perturbed by $H_{\mathrm{cost}}$, whereas the shifts to $E_{G,D}$ from $H_D$ are significant. For simplicity, let $w=1$, and first let $x=1$ as well, so that our filtering has no physical consequence. Respecting our normalization choice that the energy difference between $E_{G}$ and random configurations is $-N$, the value of each $V_{ijk}$ in $H_P$ is $\pm \of{ d_C \of{1-2 \epsilon} }^{-1}$, which is $\ll 1$ for the approximation-hard problems we are concerned with. With this fact in hand, we can readily compute the shift to $E_{S,D}$ using second order perturbation theory, using $H_D$ as the base Hamiltonian and $H_{\mathrm{cost}}$ as the perturbation:
\begin{eqnarray}
E_{S,D} \simeq - N \of{ \kappa + \frac{1}{6 \kappa d_C \of{1 - 2 \epsilon}^2 } }.
\end{eqnarray}
We observe that the correction here is hypergraph independent to this order (for given $\epsilon$ and $d_C$) and vanishes as $d_C$ becomes large. If we now let $x=2$ (but keep $w=1$), $H_{\mathrm{cost}}$ can be expressed in terms of powers of $H_P$, and now contains $O \of{d_C^2 N^2}$ terms with mean coefficient $N^{-1} \of{ d_C \of{1-2 \epsilon} }^{-2}$, and the resulting corrections to $E_{S,D}$ scale as $d_C^{-2}$ and thus vanish even faster for large $d_C$. For non-integer $x$ or $w \neq 1$ the analysis becomes more technical, but the same qualitative result holds: corrections to the energy of the paramagnetic state are very weak in the approximation hard regime.\footnote{At small $d_C$ or larger $\epsilon$ this may not be the case, of course. For instance, for satisfiable ($\epsilon = 0$) three-regular instances, where every spin participates in exactly three constraints, it is easy to show from this analysis that $\kappa_c = 1$ as the magnitudes of the shifts to the two states are identical.}

In contrast, the corrections to $E_{G,D}$ from the transverse field $H_D$ are more significant. For $x=w=1$ we find
\begin{eqnarray}
E_{G,D} \simeq -N - \kappa^2 \sum_{j=0}^{N-1} \frac{1}{\Delta E_{1,j}},
\end{eqnarray}
where $\Delta E_{1,j}$ is the cost to flip spin $j$ starting from $G$ and has mean value $+6$ in our normalization. These corrections \emph{do not} vanish at large $d_C$ and can be quite sensitive to problem structure, at least before any filtering is applied. It is this observation that motivates our filtering choice of $x \geq 2$. While a simple second order perturbative analysis no longer applies since $\Delta E_{1,j} \propto 1/N$ in that limit, making the local cost per flip vanish strongly suppresses the per-instance variation of the classical energy landscape around $G$, and therefore variations in $\kappa_c$. Combined with the statistical suppression of variations in the local density of constraints for random instances, we can therefore conclude that per-instance variations in $\kappa_c$ will be negligible with high probability, again for random problems drawn from $\MP \of{ N_C, N, \epsilon}$ with $d_C$ large and $\epsilon$ very small.

We now turn to the high energy structure. As mentioned above, the minimum gap at the phase transition is $ \Omega_0 \propto \left < S_D | G_D \right >$, where $\ket{G_D}$ is the dressed ground state of the filtered $H_{\mathrm{cost}}$, and $\ket{S}$ is the uniform superposition ground state\footnote{For simplicity, we ignore the perturbative dressings of $\ket{S}$ by $H_{\mathrm{cost}}^{x,w}$, which may be difficult to calculate in practice, particularly when $x >1$ and $w<1$; these expressions are intended to be illustrative and not exact. The actual prediction of $\Omega_0$ for MSFO later in this work relies on a series of well-justified mean field assumptions and does not involve evaluation Eq.~\ref{Om0basic} directly.} of $H_D$. Since $H_D$ is stoquastic all components of $\ket{G_D}$ will have the same sign, and add constructively in computing $\Omega_0$:
\begin{eqnarray}\label{Om0basic}
\Omega_0 \propto \frac{1}{2^{N/2}} \sum_{n = 0}^{2^N - 1} \left < n | G_D \right >.
\end{eqnarray}
As observed in \cite{atia2019high}, while the location of the transition is more sensitive to the low energy structure of $H_{\mathrm{cost}}$, its scaling with $N$ is more sensitive to the high energy structure, since most components of $\ket{S}$ are near $N/2$ flips away from $G$ and thus at energies that are $O \of{N}$ above $G$. And while the individual amplitude of any such state in $\ket{G_D}$ will be exponentially small, there are exponentially many of them and they all add constructively in Eq.~\ref{Om0basic} since they are positive definite for a stoquastic $H_D$ \cite{bravyi2006complexity}. This becomes most obvious if we rewrite Eq.~\ref{Om0basic} by grouping the components of $\ket{G_D}$ by their Hamming distance from $G$ itself:
\begin{eqnarray}\label{Om0rewrite}
\Omega_0 \propto 2^{-N/2} \sum_{m=0}^{N} \binom{N}{m} \tilde{u} \of{\kappa_c,x,w,m},
\end{eqnarray}
where $\tilde{u} \of{\kappa_c,x,w,m}$ is the average of $\left < n | G_D \right >$ over all states $\ket{n}$ which are $m$ flips away from $G$. For the unfiltered case $x=w=1$ we can do this calculation (at least on average), by computing the perturbative corrections to $\ket{G_D}$, to high orders. Again using Eq.~(\ref{avgE}), if we let $\ket{G^{(i,j,k...)}} \equiv X_i X_j X_k ... \ket{G}$, and $\tilde{E}_{\mathrm{avg}} \of{k} \equiv E_{\mathrm{avg}} \of{k} - E_G$, at the transition point we have
\begin{eqnarray}\label{GD}
&\ket{G_D} \simeq \ket{G} + 
\frac{\kappa_c}{\tilde{E}_{\mathrm{avg}} \of{1}} \sum_j \ket{G^{(j)}}  
\nonumber \\&
+2! \frac{\kappa_c^2}{\tilde{E}_{\mathrm{avg}} \of{1} \tilde{E}_{\mathrm{avg}} \of{2}} \sum_{i \neq j} \ket{G^{(i,j)}} \nonumber \\
&+ 3! \kappa_c^3 \prod_{m=1}^3 \frac{1}{\tilde{E}_{\mathrm{avg}} \of{m}} \sum_{i \neq j \neq k} \ket{G^{(i,j,k)}} + ...
\end{eqnarray}
The factorials come from the combinatorics of ordering the $m$ spin flips to reach each term. Now, since all states are present in $\ket{S}$ with equal amplitude $2^{-N/2}$ and all terms in $\ket{G_D}$ are positive definite, we can immediately conclude
\begin{eqnarray}\label{Om0AQC}
    \Omega_0\! \of{N}\! \simeq 2^{-N/2} \!\of{1 + \sum_{m=1}^N \kappa_c^m \binom{N}{m} m! \prod_{n=1}^m \frac{1}{\tilde{E}_{\mathrm{avg}} \of{n}}  }\!.
\end{eqnarray}
For $\kappa_c = 1.29$, this function is well fit by $\Omega_0 = a \sqrt{N} 2^{-b N}$, where $b \simeq 0.14$, in decent agreement with the result for a mean-field $p=3$-spin ferromagnet derived in~\cite{jorg2010energy}. And despite being a very simple calculation, it is a quantitatively good predictor of the scaling for TAQC in finding ground states of PPSPs, as shown numerically later on in section~\ref{qaoasec}.

We emphasize that this prediction is only an approximate average, of course. And for a filtering choice of $x \geq 2$, a more sophisticated method is needed to compute $ \tilde{u} \of{\kappa_c,x,w,m}$, as we can no longer use a simple perturbation expansion about $G$ owing to the vanishing local gap. But it illustrates something important, and non-perturbative: the scaling of $\Omega_0$ is set by the amplitudes in $\ket{G_D}$ of exponentially many configurations extensive distances away from $G$. 

These amplitudes are functions of both $\kappa_c$--which we have already argued should be stable across most appropriately normalized instances--and the mean energy $\tilde{E}_{\mathrm{avg}} \of{m}$. And this mean energy is a graph-independent quantity, given by Eq.~\ref{avgE} (as modified by our filtering process, of course). Since MAX-3-XORSAT is an odd cost function, and our filtering choices are sign preserving, the density of states will be approximately Gaussian distributed around a peak at $E=0$ and $m=N/2$ (with width $O \of{\sqrt{N}}$), and one can argue from the law of large numbers that significant fluctuations about this average are \emph{proportionally} rarer (and thus less relevant) with increasing $m$. Further, for random $H_P$ drawn from $\MP \of{ N_C, N, \epsilon}$, the largest possible string-to-string deviations from the average Eq.~\ref{avgE}, uncorrelated with $G$, will scale as $O \of{N/\sqrt{d_C}}$ and are thus progressively more suppressed the more approximation-hard our random instances become. Distant configurations at each $m$ whose energy exhibits extensive deviations from Eq.~\ref{avgE} are thus a superpolynomially small fraction of the total, with global magnitude suppressed by a factor proportional to $1/\sqrt{d_C}$. Coupled with the regularization of the low energy landscape (which can influence the relative weight of configurations that have particularly high or low intermediate energies that must be crossed to reach them, starting from $G$), we therefore assert:

\textit{Claim 2 (accuracy of mean field calculations): for a random $H_P$ drawn from $\MP \of{ N_C, N, \epsilon}$, with high probability, but not guaranteed for all instances, the minimum gap $\Omega_0$ in the first stage of filtered optimization with $x \geq 2$ and $w \leq 1$ is well approximated by $\Omega_{\mathrm{mf}} \of{N}$, the result computed using a mean field cost function (Eq.~\ref{avgE}) for $H_P$.}

\begin{figure}
\includegraphics[width=\columnwidth]{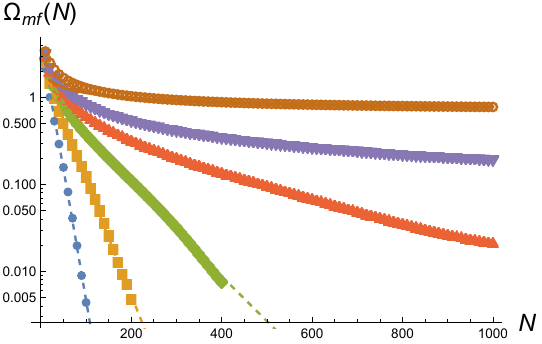}
\caption{Mean field minimum gap $\Omega_{\mathrm{mf}} \of{N}$ in the first stage of MSFO, for nonlinear fold parameter $x=2$, transverse field strength $\kappa = 1$, and warp parameter $w = \cuof{1,0.83,0.75,0.7,2/3,0.6}$ (blue through brown, and monotonically increasing). For $w > 2/3$, $\Omega_{\mathrm{mf}} \of{N}$ displays clearly exponential decay with $N$, with an exponent that steadily decreases with decreasing $w$. Conversely, for $w = 0.6$ the best fit is to simple polynomial decay, $\Omega_{\mathrm{mf}} \of{N} \propto N^{-0.23}$, showing that the spectral warping transformation is capable of softening the exponential decay of the paramagnet-spin glass transition in this problem to low-order polynomial scaling.}\label{mfOmega}
\end{figure}

The ability to predict $\Omega_0$ from mean field analysis is a powerful tool, since we can massively pare down the Hilbert space to just $N+1$ states by employing permutation symmetry~\cite{jorg2010energy}, and predict $\Omega_{\mathrm{mf}} \of{N}$ using exact diagonalization for thousands of spins without much trouble. This allows us to reliably access the real asymptotic behavior of this system, at least in the regime for which we claim these approximations are valid. And we note that, given the arguments above, these approximations become progressively better the further we are into the approximation-hard regime. The resulting functions, plotted in FIG.~\ref{mfOmega}, are well behaved, and it is in their structure that the need for the warping transformation becomes obvious. Namely, for $w=1$ and any value of $x \geq 1$, the transition into the quantum spin glass ground state is first order and $\Omega_{\mathrm{mf}} \of{N}$ decays exponentially in $N$. For $x=2$ and $w=1$ the derived scaling exponent is smaller than for the unfiltered problem with $x=w=1$, but the decay of $\Omega_{\mathrm{mf}} \of{N}$ is still obviously exponential.

The crossover to polynomial scaling occurs when we choose a warping exponent $w < 1$, or more specifically, $w < 2/p$ for a $p$-body spin glass cost function. This is inspired by the observation that, for the sign preserving fractional power implemented in Eq.~\ref{nlwarp}, the first order transition becomes second order when $w < 2/p$ (at exactly $2/p$ the scaling may be stretched exponential for our sign-symmetric cost function), effectively replacing $p$ with $wp$ in Eq.~\ref{avgE}. The intuitive mechanism for the removal of exponential decay is that higher energy states are pushed closer to $G$ in energy, in a nontrivial way that does not alter the overall normalization of a difference of $-N$ between $G$ and random states. This in turn increases their weight in $\ket{G_D}$; we think it notable that the largest \emph{fractional} changes to the magnitudes of the eigenvalues of $H_{\mathrm{cost}}^{x,w}$ are to energies near zero, where of course the density of states is exponentially largest. Note that this sequence of transformations (with $A=1$ in Eq.~\ref{nlfold}) does not change the energetic hierarchy of any classical states, only their scales; local and global minima of $H_P$ are still local and global minima of $H_{\mathrm{cost}}^{x,w}$.

Of course, if we choose $w$ to be too small, we expect performance to worsen, as when $w \to 0$ we lose the ability to discriminate $G$ from states with even tiny negative (but nonzero) energies. In real disordered problems this could manifest, among other issues, as worsening TFC, as local minima are pushed closer to $G$ and it is therefore easier (in principle) for transverse field corrections to push them below the ground state. As remarked earlier, and made increasingly clear over the course of this section, we see TFC as a much more fundamentally difficult problem than first order transitions, but not one that a priori limits quantum advantage for approximation given worst case classical thresholds that approach random guessing. 

Choosing $w$ to be too small can also increase incidental performance degradations, as defined in section~\ref{nofinetune}, such as local heating, since the local excitation energies about $G$ continuously decrease with decreasing $w$. We therefore want to choose $w$ to be small enough to clearly be out of the exponential decay regime for $\Omega_{\mathrm{mf}} \of{N}$, but not smaller; empirically, values in the range of 0.5-0.65 work well for this purpose, depending on protocol choice. For the simple mean-field $p$-spin ferromagnet considered in~\cite{jorg2010energy}, the lower limit to $w$, below which further reductions are counterproductive, can be established as $\simeq 1/p$. The \emph{local} gap will steadily decrease at this point (leading to local heating unless runtime increases) without any corresponding improvement in the global minimum gap. For the disordered problems we consider here, we expect it to be somewhat higher than this, with an optimal value likely controlled by the clustering energy, though we do not presently have the mathematical tools to set an approximate bound. For concreteness, we choose $w=0.6$ in this work, but other values below $2/3$ are effective as well. For $w=0.6$, as seen in FIG.~\ref{mfOmega}, $\Omega_{\mathrm{mf}} \of{N} \propto N^{-0.23}$ in numerical fitting. The entropic barrier is permeable.

This concludes the first stage of our algorithm, the stage which required the most conceptual work and analysis and which introduced a novel speedup mechanism, at least in our estimation. Having found the ground state of $H_D + H_{\mathrm{cost}}^{x,w}$, we now want to convert that into a good approximation, or even solution, to $H_P$, our underlying problem Hamiltonian, and to do so, we turn to stages 2 and 3.

\subsubsection{Second and third stages}

To find solutions to $H_P$, we need to ramp down our transverse field $H_D$ to zero. The two filtering transformations we nested to minimize instance-dependent properties and eliminate the exponentially scaling first order transition both have the effect of reducing the local gap around $G$, and if we want to avoid local heating as we ramp the field down, we need to ramp the field down over a time $T_2 \of{N}$ growing faster than $O \of{N^2}$ as the average local gap scales as $1/N$, with its minimum value across the lattice potentially decaying more quickly given variations in the density of constraints. While this doesn't of course erase an exponential speedup, if getting into the basin of attraction of $G$, or finding any states sufficiently close to it in energy, takes exponential time classically, it's suboptimal and can easily be improved upon. The room for improvement lies in the fact that despite $H_{\mathrm{cost}}^{x,w}$ being classically gapless, $H_{\mathrm{cost}}^{x,w} + \kappa H_D$ has an $O \of{1}$ gap on either side of the phase transition, as does the unfiltered $H_P + \kappa H_D$.

To achieve a faster time to solution, we therefore add two additional stages to the algorithm, though in principle they may be able to be combined into a single extra stage. The key observation motivating these stages is that $G$ is a gapped, presumably unique ground state of $H_P$, and the ground state of $H_D + H_{\mathrm{cost}}^{x,w}$ is spread across a much larger range of configurations, centered around $G$ and with the largest amplitudes in its basin of attraction. If however we are well below $\kappa_c$, which we expect to be for $\kappa = 1$ and $p=3$, then if we smoothly reduce $x$ to 1 and increase $w$ to 1, turning off the filtering transformations, then there is no a priori reason that a phase transition should be crossed, and indeed, the gap does not close at the mean field level when this is done. Assuming this is done sufficiently slowly--and here, we use the same scaling with $N$ as the first stage, though this is likely more than necessary as the gap doesn't close--we are now in the quantum spin glass ground state of $H_P + \kappa H_D$. Had we not implemented any of the filtering transformations, and instead attempted to find the ground state of $H_P$ directly, this step would have taken exponential time, as predicted in Eq.~\ref{Om0AQC} and confirmed later in our TAQC simulations.

Finally, in stage 3, we ramp $\kappa$ down to zero and measure the system. For large $d_C$, the classical problem is locally gapped with high probability, since the variance of the local gap decreases with increasing $d_C$, and so this step too can be done quickly. Assuming we remained in the ground state in stages 1 and 2, we will reach the ground state in this stage and can halt and measure the system. That said, while at large $d_C$ and small $\epsilon$ the classical $G$ is locally gapped with high probability, this is obviously not the case for all possible instances, and instance-dependent variations can impede performance in these steps. We also expect that, in practice, stages 2 and 3 can be combined into a single step (where $\kappa \to 0$, $w \to 1$ and $x \to 1$ all occur simultaneously), but focus on the full three stage version in this work.

\subsubsection{Time to solution scaling}

Assuming claims 1 and 2 hold, it is straightforward to predict the time to solution scaling of MSFO. Specifically, assuming that the first stage is the most computationally challenging step, $P \of{t_F}$ is constant if we choose
\begin{eqnarray}
t_F \of{N} \propto W \of{N} \Omega_{\mathrm{mf}} \of{N}^{-2}.
\end{eqnarray}
For algorithm parameters $x = 2$ and $w=0.6$, the asymptotic scaling of $\Omega_{\mathrm{mf}} \of{N}$ is $\Omega_{\mathrm{mf}} \of{N} \propto N^{-0.23}$, and assuming $W \of{N} \propto N$ this leads to an asymptotic runtime scaling of $O \of{N^{1.46}}$, again for the normalization that the ground state of $H_{\mathrm{cost}}$ is $-N$ and random states have mean energy zero. Given the polynomial scaling and transition stability assumption stated above, this could likely be further improved through schedule fine tuning, or through choosing a slightly smaller value of $w$. We note that this is an asymptotic estimate, and for the relatively small $N$ values accessible to numerical simulation, the scaling remains polynomial with roughly the same degree, but there are additional complexities to consider to obtain accurate predictions. We take up this issue in section~\ref{MSFOsims}.

To get the total gate count of the algorithm, we need to incorporate a few additional factors. First is a factor of $O \of{N_C}$ that comes from the need to guess the satisfied fraction $\epsilon$ to set the energy scale for the two nonlinear folding and warping transformations; there are of course only $O \of{N_C}$ possible values that this can take so at worst case we can just rerun the entire algorithm $O \of{N_C}$ times with different choices and take the best result. There is likewise a factor of $O \of{N_C \times {\rm polylog} \of{N} }$ that comes from computing $H_{\mathrm{cost}} = F_w \of{ F_f \of{ H_P, N, 1, x}, N, w}$ at each timestep using ancillary qubits, as sketched in FIG.~\ref{fig:folder}. Finally, to avoid heating from Trotter error, we may asymptotically need to reduce the timestep $dt$ with increasing $N$, or make the circuit more complex at each timestep. In either case we expect a worst case factor sublinear in $N$. Putting all these considerations together, we arrive at the time to solution scaling of $O \of{N^2 N_C^2 \times {\rm polylog} \of{N}}$ in our core conjecture.

\subsection{Spectrally folded trial minimum annealing performance analysis}

Our multi-stage algorithm employed a sequence of composed filters, with the goal of reaching a single ground state (or at least, states near it in energy and/or hamming distance). If we instead focus only on finding states below a given approximation threshold $A$, the warping transformation is not necessary and a simple linear folding procedure is sufficient to find approximate solutions quickly. To further support our core conjecture, we now present a second algorithm, which employs only a linear folding transformation and, starting from a poor quality local minimum found through simple classical methods, can find good approximate solutions to $H_P$ through collective quantum tunneling processes. This is the spectrally folded trial minimum annealing algorithm (TMA for the remainder of this section) presented in section~\ref{TMAdef}. The performance of these two algorithms is comparable, but they arrive at solutions through completely different mechanisms and the calculations predicting scaling are substantially different.

As outlined in section~\ref{TMAdef}, the TMA calculation starts by employing a simple classical algorithm to find a local minimum state of $H_P$. In this work, we use the quasi-greedy algorithm in section~\ref{greedysec} for this purpose, but essentially any simple method can be used; all that matters is that it's a local minimum and the initial energy of this state does not matter. We assume that our problem $H_P$ is approximation-hard in practice (as if it isn't, there's no reason to use any of these new methods), and the energy of our initial state is much higher than that our approximation target $-A N$, for control parameter $A \leq 1$. Once again we assume we have normalized $H_P$ so that its ground state energy is $-N$. We assume that it is uncorrelated with $G$ and thus $M \sim N/2$ flips away from it. We let this initial state be $\ket{L}$, and apply a classical lowering Hamiltonian $H_L$ to lower it into competition with the folded ground state band. We let the set of all states in or near this band be $\MT$. The transverse field is then turned on, $H_L$ is slowly ramped down to zero, and in the process, states in the folded ground state band are found through collective tunneling processes.

Our performance prediction for TMA is set up as a stability calculation: specifically, we ask if the dressed state $\ket{L_D}$ ($\ket{L}$ dressed by transverse field corrections) is asymptotically stable as it crosses the band center of $\MT$. If \emph{is} unstable and decays, as the only states in energetic competition are those in $\MT$, that means the algorithm has found an approximate solution to $H_P$. The stability calculation itself uses a standard Fermi's Golden Rule analysis and predicts the decay rate by multiplying the average squared per-state decay matrix element $\avg{ \Omega_{0,Lj}^2}_j$ by the total number of states in $\MT$; if the exponential decay of $\avg{ \Omega_{0,Lj}^2}_j$ is balanced or exceeded by the exponential growth of targets in $\MT$, $\ket{L_D}$ is unstable and solutions will be found quickly. Given that Eq.~\ref{avgE} is asymptotically exact, predicting the density of states in $\MT$ is easy,\footnote{Throughout this calculation, we assume a single ground state $G$. The way the calculation is structured, additional deep minima simply add additional targets and make the problem easier; our predictions here are thus in some sense a worst case estimate, at least for random $H_P$ drawn from $\MP \of{N, N_C, \epsilon}$ as specified in our core conjecture.} but predicting the average per-state tunneling rate required the development of a new high-order perturbation theory framework. We begin that calculation now.

\subsubsection{Calculation of the per-state tunneling rate}\label{TMAtunnel}

As in the MSFO calculation, the key ingredients of the per-state tunneling rate are the average cost per flip away from either minimum and the transverse field strength $\kappa$. Given a ground state energy $-N$, we assume that $\kappa < \kappa_c$, the critical value for which $\ket{L_D}$ would transition to a paramagnetic ground state, estimated to be $\kappa_c \simeq 1.29$ earlier. We need to choose $\kappa$ to be less than this value to ensure we remain in the dressed problem phase (DPP) and the calculations performed here are valid.

The average cost per flip curve takes a little more work to determine, but is ultimately also straightforward. As our goal in this section is to compute the average multiqubit tunneling matrix element between $\ket{L_D}$ and a randomly chosen state $\ket{j_D}$ near the fold in $\MT$, starting from one of the two states we need to differentiate between flips that move closer to the other state, and flips that move further away. We call these processes flips of \emph{primary} and \emph{secondary} spins, respectively; a primary spin is one whose value changes between the classical strings $\ket{L}$ and $\ket{j}$, and a secondary spin is one that does not. We note that the typical state in $\MT$ is $M \simeq N/2$ flips away from $\ket{L}$ (we address the issue of varying Hamming distance later on), and assume that $\ket{L_D}$ crosses the band center of $\MT$, where the density of states is exponentially largest, when the classical energy of both states is $\simeq -N$. We revisit this assumption later on.

The mean cost per flip curve away from $\ket{L}$ is trivially given by Eq.~\ref{avgE}, as $H_L$ is a random 3-XORSAT problem Hamiltonian. For flips away from $\ket{T}$, we note that the average state at or near the fold is $x_A$ flips away from $G$. To find $x_A$ we can invert Eq.~\ref{avgE} to find the mean number of flips $x_A N$ for which $\avg{E \of{x_A N}} = - A N$. To be specific,
\begin{eqnarray}\label{defxA}
x_A = \frac{1 - A^{1/3}}{2}.
\end{eqnarray}
We can therefore assume that the typical state in $\mathcal{T}$ is $x_A N$ flips away from $\ket{G}$. If we consider the sequences of primary spin flips connecting $\ket{j_D}$ and $\ket{L_D}$, the typical flip sequence starts $x_A N$ flips away from $\ket{G}$, and notice that with probability $1-x_A$ an additional random flip towards $\ket{L}$ will also move closer to $\ket{G}$. Taking all these effects and the division by $A$ into account, so that the bare unperturbed energy of both $\ket{j}$ and $\ket{L}$ when they cross are both $\sim -N$, a bit of algebra shows that for $y$ flips away from a ground state of the folded Hamiltonian, not only is the total average cost $\Delta E \of{y}$ $A$-independent, it is precisely equal to the cost given by Eq.~(\ref{avgE}). We note that all of the arguments about the statistical regularity of random-hypergraph problems from the previous section apply here as well.

We can are now ready to define the bare classical energy, with no corrections from transverse fields. We start from one of the two states, and consider a state which is $m+n$ random flips away from it. We let $m$ of these flips be ones which move toward the other minimum (e.g. reduce the Hamming distance to it), and the $n$ flips be flips that move away from it. Then the bare average energy $E_{m,n}^{(0)}$ is given by:
\begin{eqnarray}\label{Emnbare}
E_{m,n}^{(0)} = -N \sqof{ \of{ 1 - 2 \frac{m+n}{N}}^p + \of{2 \frac{m-n}{N} }^p }. 
\end{eqnarray}
We assume that the transverse field strength $\kappa$ is below $\kappa_c$, the $p$-dependent critical point where a transition to the paramagnetic state occurs. The ground states are thus the symmetric and antisymmetric combinations of the two dressed classical minima, with splitting $2 \Omega_0$, where $\Omega_0$ decays exponentially in $N$ and our goal in this section is to predict its decay rate.

Computing $\Omega_0$ proceeds through the following steps:
\begin{itemize}
\item We compute the renormalized cost per flip away from either minimum, incorporating transverse field corrections, which we will then use in the energy denominators of our $M$th order perturbation theory. This step is analogous to commonly used resummation schemes in diagrammatic quantum field theory, where self-energy corrections are incorporated into the propagators used to compute higher order processes.
\item We divide the system between primary spins, which flip between the classical minima, and secondary spins, which do not. We then compute the dressed states $\ket{L_D}$ and $\ket{j_D}$ that comprise all the primary flip sequences up to order $M/2$ away from each minimum. It is at this order that the two states have nonzero overlap.
\item These dressed states are then normalized; incorporating this normalization, their overlap gives the primary spin contribution to the tunneling rate, $\Omega_0^{(p)}$.
\item We then compute the secondary spin contributions to tunneling, which take two forms: an increase of the tunneling rate from the constructive contribution of many additional tunneling sequences in which secondary spins participate, and a decrease from normalization corrections and the spread of the classical minima away from the core classical configurations from which the tunneling calculation begins. 
\item Incorporating both sets of secondary spin contributions gives us a closed form expression for $\Omega_0$ which can then be evaluated numerically and compared to exact diagonalization.
\end{itemize}

We first want to compute the energy shifts, in second order perturbation theory, to these states. These corrections arise from a single spin being flipped and flipped back, and are opposite in sign to the cost of the local flip. Let $u_{m,n}$ be the difference between energies of a state $m,n$ and a ground state, incorporating these corrections. Then
\begin{align}\label{umn}
&u_{m,n} = E_{m,n}^{(0)} + N\of{1 + \frac{\kappa^2}{2p}} - \of{ \frac{N}{2} - 2m} \frac{\kappa^2}{\partial_m E_{m,n}^{(0)}} \nonumber \\
&- \of{ \frac{N}{2} - 2n} \frac{\kappa^2}{\partial_n E_{m,n}^{(0)}} + O \of{\kappa^4}.
\end{align}
Note that $\partial_{m} E_{m,n} |_{m,n=0} = \partial_{n} E_{m,n} |_{m,n=0} = 2p$, so $u_{0,0} = 0$. One can observe that if $p=2$, $\partial_{m} E_{m,0} = 4 \of{ 1 - \frac{4m}{N}}$, and the transverse field corrections to state energies are $m$-independent, so that the energy barrier between the two competing ground states is not renormalized by the transverse field. But for $p=3$ and higher these corrections are nontrivial and act to reduce the effective energy barrier between the states, increasing the tunneling matrix element. This process is effectively a resummation of higher order corrections and is necessary to obtain quantitatively accurate results.

We start by computing the dressed states, summing over primary spin corrections only. They take the form
\begin{align}\label{dressG}
&\ket{L_D} \equiv \ket{L} + \sum_{j} \frac{\kappa}{u_{1,0}} X_j \ket{L} + \sum_{j,k \of{j \neq k}} \frac{2 \kappa^2}{u_{1,0} u_{2,0}} X_j X_k \ket{L} 
\nonumber \\
&+ \sum_{j,k,l \of{j \neq k \neq l}} \frac{3! \kappa^3}{u_{1,0} u_{2,0} u_{3,0}} X_j X_k X_l \ket{L} + ... 
\end{align}
If we ignore the local band structure, the expression for $\ket{j_D}$ is functionally identical. Neglecting this structure is expected to underestimate the total collective tunneling rate; we revisit this issue in section~\ref{bandstructure}. We stop our expansion at order $M/2$, which is the lowest nontrivial order needed to connect the states. For simplicity we assume $M$ is even though the argument is easy to generalize to odd $M$ as well. Note that this state is not normalized, and in fact the norm of the state written above is exponentially large, so we will need to incorporate normalization corrections into the definition of the states. Thanks to the dressing of the states, we obtain a primary-spin energy splitting
\begin{align}
&\frac{1}{2} \of{ \bra{L_D} + \bra{j_D} } H_P \of{ \ket{L_D} + \ket{j_D} } = 2 \Omega_0^{(p)}, \nonumber \\
&\frac{1}{2} \of{ \bra{L_D} - \bra{j_D} } H_P \of{ \ket{L_D} - \ket{j_D} } = 0.
\end{align}
Evaluating these expressions, the degeneracy splitting from only considering primary spins is:
\begin{align}\label{Om0primary}
&\Omega_{0}^{(p)} &=& \kappa^{M} \binom{M}{M/2} \frac{1}{u_{M/2,0}} \of{ \frac{M}{2} ! \prod_{k=1}^{M/2 -1} \frac{1}{u_{k,0}} }^2 \times \MN^{(p)}, \nonumber \\
&\MN^{(p)} &=& \of{ 1 + \sum_{k=1}^{M/2} \binom{M}{k} \of{ \kappa^{k} k! \prod_{l=1}^{k} \frac{1}{u_{l,0}} }^2 }^{-1}. 
\end{align} 
Here, $\mathcal{N}_p$ is the normalization correction. This covers the primary spin portion of the macroscopic quantum tunneling rate. 

We now turn to the secondary spins. To introduce secondary spin corrections, consider a single secondary spin $j$ out of the $\of{N-M} \sim N/2$ total, whose bit value is the same in both classical minima. Since the same transverse field is acting on it as all other spins, when we consider the sum of all processes that connect the two minima, we can now divide them between those where $M/2$ primary spins flip from each minima to meet in the middle, with secondary spin $j$ unchanged, and a new set of processes where spin $j$ flips starting from each minimum and the two wavefunctions overlap at the set of states where $M/2$ primary spins have flipped along with $j$. The first set of processes is what was considered in Eq.~\ref{Om0primary}; the second is new, and we want to calculate its matrix element. It is most useful to express these matrix elements as a ratio of the new term to the original, primary-spin-only process, since both decay exponentially in $M$. Let the primary spin perturbative matrix element to reach $M/2$ flips be $\xi_{M/2}$, so that
\begin{eqnarray}
\xi_{M/2} = \kappa^{M/2} \of{ \frac{M}{2} ! \prod_{k=1}^{M/2} \frac{1}{u_{k,0}} }.
\end{eqnarray} 
To define the analogous process where $j$ flips, we need to sum over all the points during the perturbative sequence when that can happen. We thus have:
\begin{eqnarray}\label{defxi}
\xi_{M/2}^{s} = \kappa^{M/2 + 1} \frac{M}{2}! \sum_{n=1}^{M/2} \prod_{k=1}^{n} \frac{1}{u_{k,0}} \prod_{k=n}^{M/2} \frac{1}{u_{k,1}}.
\end{eqnarray}
And noting that we have to make this insertion in the matrix elements from both minima, the total tunneling term is increased by
\begin{align}\label{defGT}
&\Omega_{0}^{(p)} \to \of{ 1 + \of{ \frac{\xi_{M/2}^{s} }{\xi_{M/2}}}^2 } \Omega_{0}^{(p)}, \\
&\of{ 1 + \of{ \frac{\xi_{M/2}^{s} }{\xi_{M/2}}}^2 } \equiv \gamma_T.
\end{align} 
We now need to consider the rest of the secondary spins. Formally of course, the additional energy cost of each secondary spin flip changes as more secondary spins flip, but since these corrections are fairly weak (though they are appreciable and necessary for an accurate prediction of the scaling exponent) the total tunneling rate is going to be dominated by the set of processes where a comparatively small fraction of secondary spins have flipped, and we can thus approximate them as independent contributions. In this limit, since there are $N-M$ secondary spins,
\begin{eqnarray}\label{defGR}
\Omega_{0}^{(p)} \to \Omega_{0}^{(p)} \gamma_T^{N-M}.
\end{eqnarray}

Alongside this, the secondary spin corrections also spread the competing ground state wavefunctions out over Hilbert space, which exponentially reduces the weight of the core classical configurations from which the tunneling calculation begins. To be consistent with the independence approximation made above, we simply compute all the corrections to the ground state from each secondary spin independently and multiply them. Noting that we must apply this calculation to both competing ground states, this reduces the tunneling rate by
\begin{eqnarray}
\Omega_{0}^{(p)} \to \Omega_{0}^{(p)} \of{\frac{\gamma_T}{\gamma_R}}^{N-M}, \; \; \gamma_R \equiv 1 + \of{\frac{\kappa}{u_{0,1}}}^2.
\end{eqnarray}
Note that, if we set the cost per flip of a given secondary spin to some constant $U$, independent of the configuration of the other spins, that would imply it is disconnected from the primary spins as there are no couplings to shift the energy. In this limit a direct evaluation of the two functions shows that $\gamma_T = \gamma_R$ (for any choice of $U$) and this now disconnected spin plays no role in tunneling at all. This factorization of disconnected spins is reassuring, and lends support to the correctness of this approach. Taking into account all these effects, our total tunneling rate is
\begin{align}\label{Om0full}
\Omega_0 = \kappa^{M} \binom{M}{M/2} &\frac{1}{u_{M/2,0}} \of{ \frac{M}{2} ! \prod_{k=1}^{M/2 -1} \frac{1}{u_{k,0}} }^2 \nonumber \\
&\times \MN^{(p)} \times \of{\frac{\gamma_T}{\gamma_R}}^{N-M}.
\end{align}
Taking all of these effects into account yields a highly accurate prediction of the minimum gap scaling for a wide range of values for $p$ and $\kappa$, with only an $O \of{1}$ discrepancy in the prefactor and few percent discrepancies in the scaling exponent (empirically, Eq.~\ref{Om0full} tends to slightly overestimate the decay compared to the exponent extracted from numerical diagonalization). We refer to the Appendix A for more details.

\subsubsection{Achievable approximation ratio with spectrally folded trial minimum annealing}\label{TMAPTAS}

With this result in hand, we will now predict the macroscopic quantum tunneling rate--and thus, achievable approximation ratio--for the spectrally folded Hamiltonian. From this, we can calculate our target value of $A$, assuming that there is a single deep minimum far below the energy of any local minima. Relaxing this assumption should improve the performance of the algorithm by virtue of there being many more target states. We consider the protocol in section~\ref{TMAdef}, with an initial state $\ket{L}$. We assume that the process of ramping the transverse field up and down is itself at least roughly adiabatic, i.e., we can assume approximate spectral continuity with respect to the folding and lowering Hamiltonians, noting that the lowering Hamiltonian will itself create $O \of{\sqrt{N}}$ shifts to the energies of states near the fold.  It follows from our assumption that the ramping process itself does not meaningfully heat the system. We then consider the set $\mathcal{T}$ of all states within $O \of{1}$ shifts of $-A E_{\mathrm{GS}}$ in $H_P$, the states closest to the fold, and compute, as a function of all our various algorithm parameters, the total probability of tunneling into any one of them. 

Since the tunneling rate into any individual state is exponentially small, and the time over which we slowly turn off the lowering Hamiltonian is $T \propto O \of{N}$, we can assume that tunneling will be diabatic with respect to any individual state. A Fermi's Golden rule analysis as in~\cite{kapit2021noise} suggests that the total success probability will be given by Eq.~\ref{PTF}, where again $W \sim O \of{N}$ is the energy range swept over by reducing $C \of{t}$ to 0, and $\Omega_{0,Lj}$ is the tunnel splitting at degeneracy between $\ket{L_D}$ and the target state $\ket{j_D}$, which we assume are an average of $\sim N/2$ flips apart. To go further, we need to compute the average value of $\Omega_{0,Lj}^2$, noting that while of course there will be substantial state-to-state variations, given that there are exponentially many states in $\mathcal{T}$ the average value calculated in Eq.~\ref{Om0full} is going to dominate Eq.~(\ref{PTF}). 

This is again only an average, but noting that the classical cost per flip for any given sequence appears the denominators of equations like (\ref{Om0full}), variations about it are more likely to increase the tunneling rate than decrease it. And as remarked earlier, path-to-path variations about the average are suppressed by $O \of{1/\sqrt{d_C}}$ and thus naturally small in the approximation hard regime. And likewise, since $H_L$ is a random 3-XORSAT problem itself the mean cost per flip away from $\ket{L}$ is going to be given by Eq.~\ref{avgE} as well, so Eqns.~\ref{defxA} through \ref{Om0full} can faithfully predict the average tunneling rate between $\ket{L}$ and a randomly chosen ground state of the folded Hamiltonian.\footnote{We expect that using the average cost per flip in Eq.~\ref{Emnbare} will if anything underestimate the per-state tunneling rate in real disordered problems. This is because all of these energy costs appear in denominators, which leads to likely small asymmetries in how much the deviations from the average in any individual flip sequence contribute to the total matrix element, giving lower energy sequences proportionally higher weight. We do not really expect this effect to be significant but rather highlight it as another point where our prediction is conservative by design.} For more discussion of this approximation, see the appendix.

Of course, this rate decays exponentially; assuming the two states are $M \sim N/2$ flips away for $\kappa = 1.29$, $\Omega_{0} \propto 2^{-b N}$ where $b \simeq 0.2$. But this is balanced by the fact that there are on the order of $\binom{N}{x_A N}$ target states. We can further note that out of these states, while the mean distance to $\ket{L}$ is $M \sim N/2$, ones which are $k$ flips closer have tunneling rates which are larger by a factor of $2^{2 b k}$ on average, and though those states are proportionally rare their increased weight is enough to meaningfully impact our choice of $A$. Since our total runtime is linear, simple diabatic scaling predicts that the probability of tunneling into the typical state $M-k$ flips away is proportional to $\Omega_0^2$, e.g. $2^{-4 b \of{M-k}}$.

We now take this result and plug it into Eq.~(\ref{PTF}), so that we can determine the choice of $A$ where the returned $P_{\mathrm{tot}}$ provides an approximation guarantee. If we use Stirling's approximation to write the binomial coefficients as exponentials, and ignore slowly varying polynomial factors, the total number of states in $\mathcal{T}$ scales as:
\begin{eqnarray}\label{NT}
N_{\mathcal{T}} \propto \exp \left(- \left[ x_A \ln x_A + \of{1-x_A} \ln \of{1-x_A} \right] N\right).
\end{eqnarray}
Likewise, if the average probability of tunneling into a target state $k$ primary spin flips closer to $\ket{L}$ is increased by a factor of at least $2^{4 b k}$, the weighted per-state average of the diabatic tunneling rate into states in $\mathcal{T}$ can be approximated as
\begin{eqnarray}\label{Om0avg}
\frac{\log\avg{\Omega_0^2}}{-N} \approx 2 b \ln 2 - x_A \of{\ln \of{1+ 2^{4 b}} - \ln 2 -2 b \ln 2} .
\end{eqnarray}
Note that this average comes from considering only $x_A N$ flips away from $\ket{G}$ but varying Hamming distance from $\ket{L}$ and thus neglects the influence of comparatively rarer states larger distances from $\ket{G}$. Taking all these terms into account, the probability of returning a state with $E \simeq A E_{\mathrm{GS}}$ as measured relative to the original $H_P$ becomes constant, or at least stops decaying exponentially, when
\begin{align}\label{PTAScondition}
&-2 b \ln 2 - \left[ x_A \ln x_A + \of{1-x_A} \ln \of{1-x_A} \right] + \nonumber \\
&
x_A \of{- \ln 2 -2 b \ln 2 + \ln \of{1+ 2^{4 b}} } = 0.
\end{align}
The achievable approximation ratio is thus determined by the per-state decay exponent $b$, computed in section~\ref{TMAtunnel} as a function of $N$ and $\kappa$ by fitting Eq.~(\ref{Om0full}) to $\Omega_0 \of{N} \propto \sqrt{N} 2^{-b N}$, and then choosing $x_A$ using Eq.~(\ref{defxA}) to solve Eq.~(\ref{PTAScondition}). This analysis only counts states within $O \of{1}$ shifts of $A E_{\mathrm{GS}}$ (recall that $E_{\mathrm{GS}} = -N$ in our normalization) and ignores low-order polynomial prefactors; for $b=0.2$, which again depends on $\kappa = 1.29$ in this calculation, this is solved when $x_A \simeq 0.08$, or $A \simeq 0.59$.

This means that if the true ground state satisfies a constraint fraction $F$ beyond random guessing---e.g. a total fraction $1/2 + F$, so $F$ is at most 1/2 here---our algorithm will return states which satisfy a fraction $1/2 + A F$ with high probability. If we choose $A$ to be too large compared to the target value established by expressions like Eq.~(\ref{PTAScondition}), we risk failing to well-approximate the problem; conversely, choosing $A$ below it will reduce the returned approximation ratio to $A$ and thus perform suboptimally. We do caution that unlike our predictions for MSFO, this analysis cannot predict the degree of polynomial scaling necessary (and does not formally rule out other sub-exponential but super-polynomial scaling forms). That said, based on the reasoning in section~\ref{nofinetune}, and numerical testing out to the largest accessible system sizes, it is likely runtimes that scale linearly with $N$ are sufficient. It strikes us as unlikely that superpolynomial decay will survive once Eq.~\ref{PTAScondition} predicts a positive exponent, in analogy to how for exactly $w=2/3$ the MSFO minimum gap may scale as a stretched exponential, but is firmly in the polynomial scaling regime for $w < 2/3$. And, we emphasize again, this prediction assumes a random, potentially dense hypergraph but is fundamentally independent of the fraction satisfied in $E_{\mathrm{GS}}$ itself and so applies to the planted partial solution instances we use for numerical benchmarking below.

\subsubsection{Further Comments and Caveats}
\label{ssec:caveats}

We expect that this analysis underestimates the choice of $A$ that will return states with $E \leq A E_{\mathrm{GS}}$ with constant probability. This is because our counting here only counts states very close to the fold, when in reality the probability of tunneling into states a small extensive fraction larger than $A E_{\mathrm{GS}}$ is still going to be appreciable due to the continued exponential growth of the number of targets, even if the per-state tunneling rate does tend to decrease with increasing $E$ due to the interference effect mentioned earlier, in which perturbative corrections that mix with states of lower energy have opposite sign. In addition to this consideration, because the target ground states and low lying excitations of the folded Hamiltonian in $\mathcal{T}$ very roughly form a hyperspherical shell, any individual target state will have other states in $\mathcal{T}$ that are relatively close to it in Hamming distance. Consequently we expect these states to have a band dispersion, centered around the mean energy given by the corrections in Eq.~(\ref{umn}). Since we are already assuming off-resonant tunneling, i.e. a per-state tunneling rate $\propto \Omega_0^2$, and so summing over squared matrix elements, if we consider states near the band center where the density is highest this will alter the average tunneling rate by at most a prefactor. However, there are good reasons to suspect that tunneling into extremal states near the bottom of the band can be substantially enhanced, enough to increase the optimal value of $A$. This calculation is difficult to do quantitatively so we do not attempt it here; we instead see the approximation of considering only states near the band center of $\mathcal{T}$ as another choice that likely underestimates the achievable approximation ratio. We discuss this point in more detail in Appendix~\ref{bandstructure}.

We also want to emphasize that this variation is not necessarily the optimal spectrally folded optimization algorithm, but instead merely the one where we were able to analytically compute the threshold $A$. For example, one can perform trial minimum annealing with a simple local $Z$ bias lowering Hamiltonian (e.g. $H_L = \sum_j h_j Z_j$), or standard AQC interpolation using the quadratic folding procedure in Eq.~(\ref{nlfold}) as the cost function. The linear lowering Hamiltonian is expected to have equal or better tunneling rates to a 3-XORSAT-based minimum as the overall cost-per-flip curve is shallower, though the local energy shifts to the ground states of $H_{\mathrm{fold}}$ from $H_L$ are expected to be larger. The total gate count at each time step is lower. Empirical performance in testing up through $N=25$ showed fairly similar performance to 3XOR-based $H_L$ for all other parameters equal, but with more significant non-monotonic behaviors that made fitting difficult; see Sec.~\ref{numresults} for details. That the two schemes could asymptotically converge to the same achievable approximation ratios seems plausible to us but we cannot simulate large enough systems to be sure.

For quadratic folding AQC as in Eq.~(\ref{foldAQC}), if we choose $A=1$ the gap is efficiently computable using the methods in~\cite{jorg2010energy} and decays as $\Omega_0 \sim 2^{-0.104 N}$. Given that, like linear folding, the cost per flip curve is $A$-independent, if we assume that the tunneling rate per state for $A<1$ is basically equal to this, then the total decay exponent vanishes if $x_A \simeq 0.033$ and $A \simeq 0.8$ using the arguments of the previous few paragraphs. We do not think that can be simply assumed as easily as with tunneling between semiclassical minima and a linearly folded problem Hamiltonian, in the DPP, and more theoretical work is needed here to analytically determine the optimal choice of $A$. Interestingly however, our simulation data in Sec.~\ref{specfoldsims} roughly supports this conclusion, with a worst case polynomial time approximation ratio of 0.7 found in our simulations. These simulations show that this method performs similarly to, or slightly worse than, the 3XORSAT-TMA algorithm, which is better able to outperform the approximation guarantee of $\sim 0.6$ derived here.\footnote{For smaller systems we also tested linear folding AQC and quadratic folding TMA. In very preliminary studies we found that quadratic AQC modestly outperformed linear AQC, and linear TMA more significantly outperformed quadratic TMA. So we chose not to pursue those methods for larger simulations and do not present those results here, but they may be viable or even superior for other problem classes.} 

Finally, all of the results we obtain here could be further improved by employing a spectral warping transformation to both $H_P$ and $H_L$, which would reduce the effective degree $p$ toward or even below $2$ (where there is no first order paramagnet-spin glass transition), albeit with the possibility of worsening TFC. In this limit assuming TFC can be ignored the achievable $A$ could become arbitrarily close to 1, as in MSFO. We did not explore this addition in this work, as our goal here was in part to show that folding alone is sufficient for an exponential separation in approximating random instances drawn from $\MP \of{N_C,N, \epsilon}$. Coupled with the fact that our simplifying approximations are all likely to underestimate rather than overestimate performance (particularly the band structure, discussed in Appendix~\ref{bandstructure}), that a large constant $A$ is achievable without reaching for additional algorithmic tools like warping strikes us as further supporting evidence toward the truth of our core conjecture.\footnote{We also expect that the average tunneling rate--and thus, achievable approximation ratio--can likely be further increased by using other, potentially many-frequency, AC methods such as RFQA~\cite{kapit2021systems,kapit2021noise,tang2021unconventional,mossi2023embedding,grattan2023exponential}. For simplicity, we do not incorporate these methods in this work, but they could be a novel way to further improve the performance of this algorithm and are worth exploring in future research.}

In summary, through a relatively novel resummed extensive order perturbation theory, we have shown that random hypergraph MAX-3-XORSAT instances, including extremal ones with planted partial solutions, are efficiently approximable to a fairly large constant fraction through spectrally filtered quantum optimization algorithms. We do not expect this to be the case for TAQC (or QAOA), for the reasons discussed earlier in section~\ref{introsec} and supported by the numerical evidence we present below. We similarly do not expect such guarantees to be possible for directly finding global optima (particularly as $d_C$ becomes small or as $\epsilon$ approaches 1/2), for the reasons set forth in the introduction, though in practice MSFO can often find $G$ in polynomial time when $d_C$ is large and $\epsilon$ is small. Evidently, one cannot so easily summit a mountain in hyperspace, but one can reach the rim of a crater. We now present a series of numerical simulations to further support these claims.

\section{Numerical tests of approximation hardness for traditional methods}\label{numresults}

\subsection{Setup and summary of results}

To confirm our predictions--or at least, verify that any serious issues with our calculations and interpretation of the problem are subtle and not apparent at system sizes within reach of present or near-future classical simulations--we performed a series of numerical simulations of various classical and quantum algorithms applied to our PPSPs. For all quantum simulation tasks we used the Qulacs package~\cite{suzuki2021qulacs}. For smaller systems and algorithm prototyping we ran our simulations on local workstations and cluster nodes; this includes all the spectral folding TMA simulations. For all TAQC and spectral folding AQC simulations presented, we used the Fujitsu Quantum Simulator, a classical HPC system. This allowed us to probe larger system sizes while still averaging over enough instances to have reliable statistics for the problem class. Unless otherwise stated, each datapoint represents the average over 960 or 1000 (for the TAQC and spectral folding AQC results) random problem instances constructed with the prescriptions outlined in section~\ref{3xordef}. In all cases in this work we used an unsatisfied fraction $\epsilon=0.1$ for our partial planted solutions; we expect similar phenomenological behavior for other constant $\epsilon$ values, though the thresholds we measure would vary with $\epsilon$ for classical optimization algorithms and TAQC (but not folded optimization, below its predicted performance floor).

The results of our simulations are summarized in Table~\ref{approxtab}, which lists the best per-shot polynomial time approximation ratio achievable through each algorithm studied, both classical and quantum. To estimate these values, we measured the average per-shot probability $P_q \of{N} = P \of{E \leq q E_{\mathrm{GS}}}$ of returning states with energies at or below $q E_{\mathrm{GS}}$, where $q \leq 1$ is the approximation ratio and $E_{\mathrm{GS}}$ is negative in our conventions. These probabilities were computed by choosing bins of size 0.05 $E_{\rm GS}$.

To determined the polynomial time hardness threshold, we applied a very simple rule where we fitted $P_q \of{N}$ to a simple exponential function, assuming any observed decay corresponded asymptotically to exponential decay, and any positive exponents, e.g. exponential growth, represent small-$N$ growth toward some constant saturation value. For the greedy algorithm with $N_C/N = 1.5 \sqrt{N}$ we fitted decay to an exponential in $\sqrt{N}$ as discussed below. The polynomial time approximation hardness threshold for a given algorithm, unsatisfied fraction $\epsilon$ and $N_C / N$ scaling choice is defined to be $q_a$, the largest values of $q$ for which we do not observe exponential decay. For $N_C/N = 3 \sqrt{N}/2$ this threshold is decaying with system size, and the asterisk next to the result for TAQC is to highlight the fact that we only ran these simulations out to $N=28$ so are likely not capturing the asymptotic threshold. 

This notation highlights a difference from $q_m$ in our core conjecture; $q_a$ corresponds to the best achievable approximation ratio for a given algorithm and problem parametrization, where $q_m$ is the global minimum value of $q_a$ across all spectrally filtered optimization variations and problem parametrizations (again restricted to random instances drawn from $\MP \of{N_C,N,\epsilon}$). We also want to emphasize that $q_a$ is a measure of the performance of an algorithm, whereas the approximation target $A$ in folded optimization is a control parameter, and $q_a$ may well fall short of it. As discussed elsewhere, the number of cost function calls is $O \of{N}$ for all approaches studied , except MSFO. Cases for which a range is quoted are where we felt there was some ambiguity to the fitting, and all values are the result of extrapolating fits to numerical simulations and are naturally somewhat approximate.

\begin{table*}
\begin{tabular}{| c || c | c || c | c | c | c | c || c |}
\hline 
$ N_C / N $ & Classical & TAQC & AQC/0.75 &  AQC/0.85 &  TMA-3/0.75 &  TMA-3/0.85 & TMA-L/0.75 & MSFO \\
\hline
2 & 0.75-0.8 & 0.75 & 0.75 & 0.75 & 0.75 & 0.75 & 0.7 & 1* \\ 
4 & 0.55 & 0.55 & 0.7 & 0.7 & 0.75 & 0.8 & 0.75 & 1* \\
6 & 0.45 & 0.45 & 0.75 & 0.7 & 0.8 & 0.8 & 0.75 & 1* \\
3 $\sqrt{N}/2$ & decaying & 0.25/decaying* & 0.75 & 0.75 & 0.8 & 0.8 & 0.75-0.8 & 1* \\ 
\hline
\end{tabular}
\caption{Approximation hardness thresholds for the classical greedy search, high-depth TAQC and spectral folding variations. This table lists  $q_a$, the largest value of the approximation ratio $q$ before exponential decay is reported, drawn from the numerical experiments in figures~\ref{greedyapproxfig} through~\ref{specfoldtmafig} and~\ref{specfoldaqcfig-85}. Spectral folding results are labeled as protocol/$A$ (where $A$ is the approximation target); AQC is quadratic spectral folding in the AQC formulation, TMA-3 is trial minimum annealing with a linear folded Hamiltonian and 3-XORSAT lowering Hamiltonian, and TMA-L is the same with local $Z$ biases for the lowering Hamiltonian. MSFO is the multi-stage filtered optimization algorithm of section~\ref{MSFOdef}; using a runtime that scales approximately as $N^{3/2}$ (in contrast to the linear runtimes of all other methods) it is capable of finding the ground state directly with constant probability, at least for random instances in the parameter ranges simulated. These results--where the threshold decays for traditional methods but not folded optimization--support the predictions in section~\ref{theorysec} that random hypergraph problems are efficiently approximable through spectrally filtered quantum optimization.}\label{approxtab}
\end{table*}

\subsection{Performance of quasi-greedy classical algorithms}\label{greedysec}
To explore the classical difficulty of our PPSPs, we applied a greedy local search algorithm adapted from~\cite{bellitti2021entropic}. This algorithm is straightforward; we start with a random bitstring. Then, beginning at each step, we calculate $k = N_{\rm{sat},j} - N_{\rm{unsat},j}$ associated with each bit $j$. We then calculate the fraction of bits, $f_k$, belonging to each $k$ value. Note that we only care when $k > 0$ because these are the cases where flipping a particular bit will lower the energy. Using some weight function, $w(k)$, we select a $k$ value with normalized probabilities $\propto w(k)*f_k$ and flip a bit with that $k$ value. If there are multiple bits belonging to the $k$ value chosen, we choose a bit in this set with uniform probability to flip. This is repeated until the configuration finds itself in a minimum, and the algorithm halts. 

 We found the algorithm performed best with a weight profile quadratic with $k$, however this can be experimented with for different results. Notably,~\cite{bellitti2021entropic} found that when applying a highly optimized version of this search to 3-regular 3-XORSAT problems it performed well even when compared to more sophisticated algorithms such as simulated annealing and parallel tempering. The intuitive reason for this can be inferred from the typical energy landscapes of these problems, which are rough and contain exponentially many high energy local minima. Once one is found, it is more efficient to simply restart the algorithm from a new random configuration instead of attempting to ``climb out" using penalized operations in simulated annealing or parallel tempering. As the locations of these minima are uncorrelated with the true ground state, finding one provides no useful information in a ground state search.

\begin{figure}
\includegraphics[width=\columnwidth]{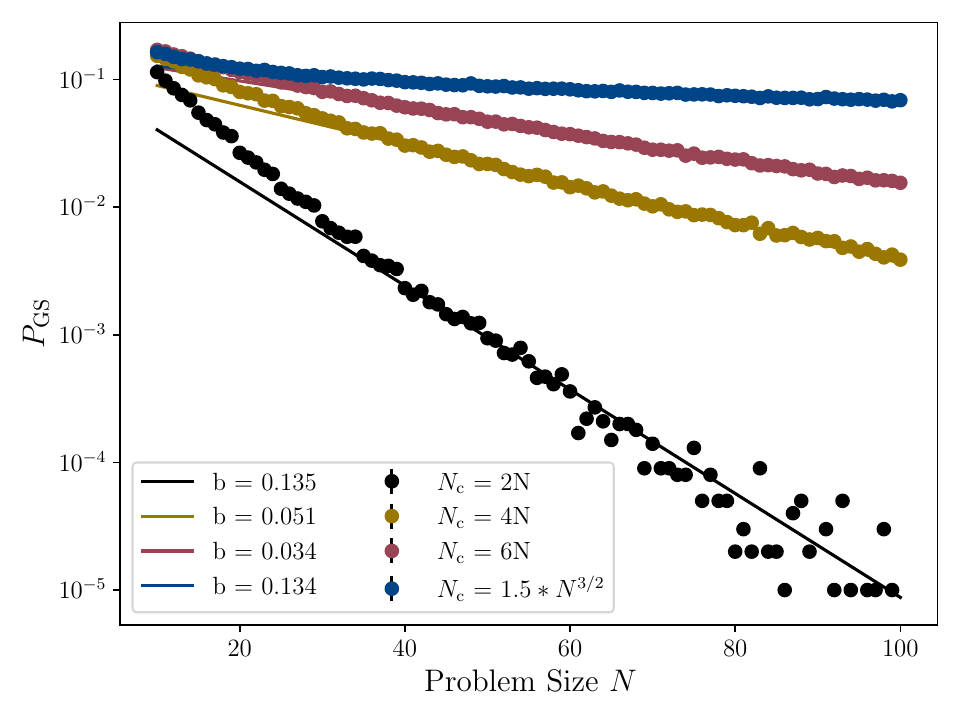}
\caption{Per-shot probability of finding the true ground state with the quasi-greedy classical algorithm defined in section~\ref{greedysec} as adapted from~\cite{bellitti2021entropic}, for the four constraint densities studied in this work and planted solution unsatisfied fraction $\epsilon=0.1$. The probability decays superpolynomially with $N$; fits are to $a 2^{-b N}$ for $N_C = \cuof{2,4,6}N$ and $a 2^{-b \sqrt{N}}$ for $N_C = 1.5 N^{3/2}$; the scaling matches the empirically observed Eq.~(\ref{greedyPGS}) fairly well. Each datapoint is the average of $10^{5}$ shots. As the constraint density increases, the basin of attraction widens, and the problem becomes easier--though still exponentially scaling--for local update classical routines. Each shot consists of $O \of{N}$ local updates so the time to solution scales essentially as the inverse of this probability.}\label{greedypgsfig}
\end{figure}

\begin{figure*}
\includegraphics[width=0.49\textwidth]{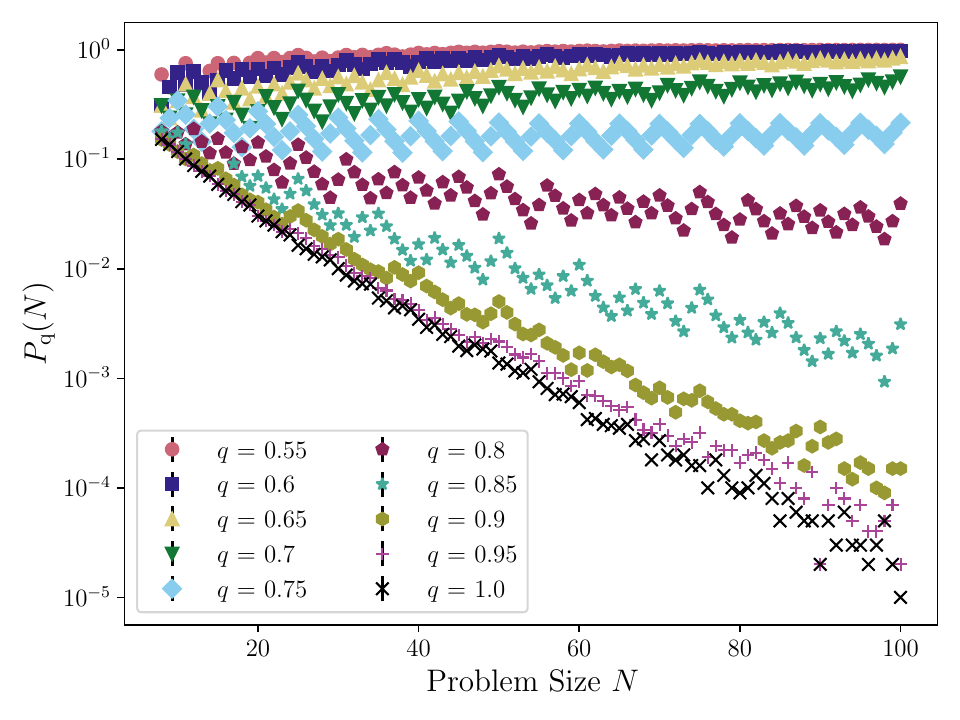}
\includegraphics[width=0.49\textwidth]{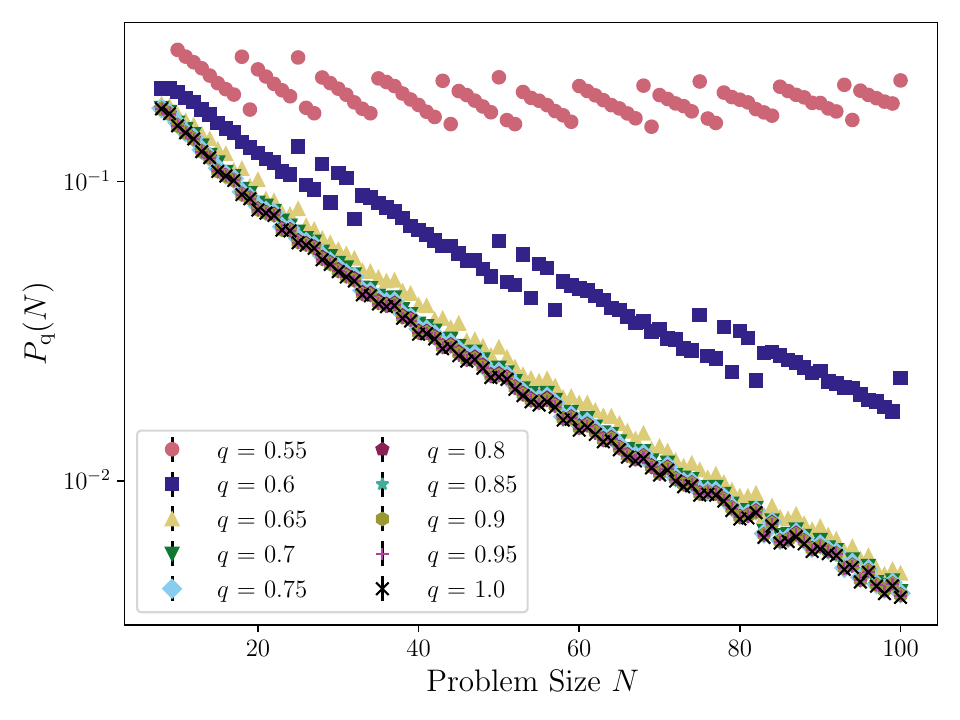}

\includegraphics[width=0.49\textwidth]{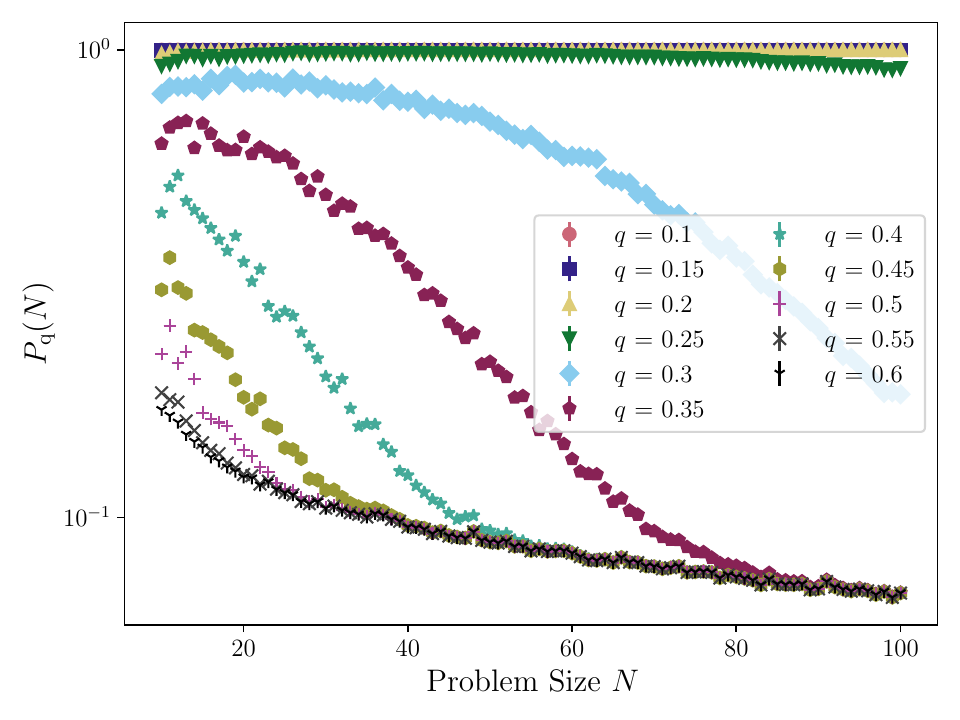}
\includegraphics[width=0.49\textwidth]{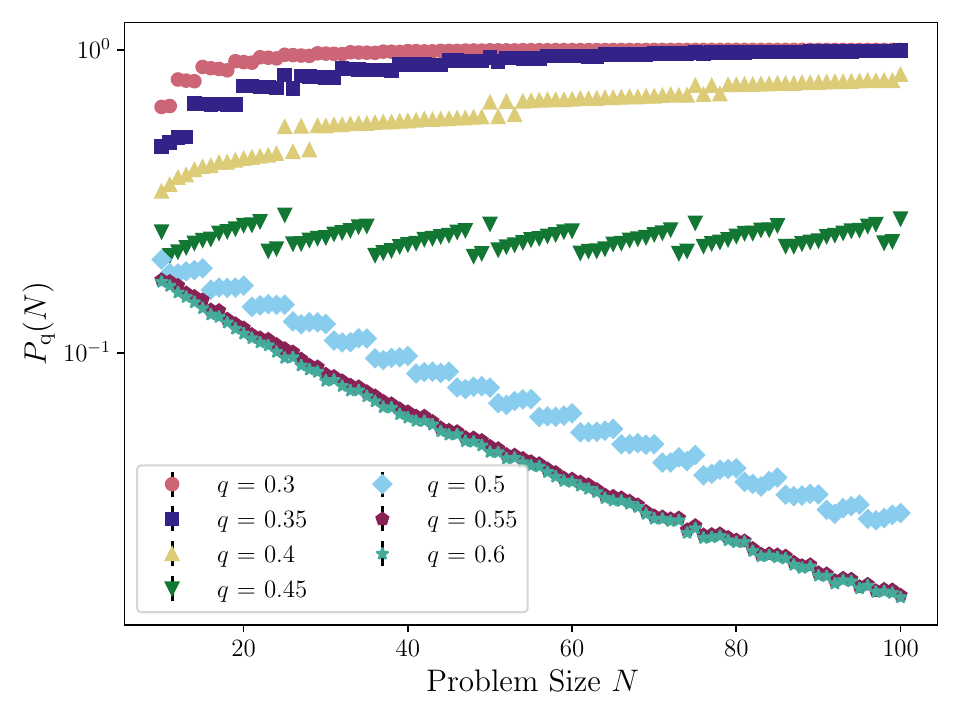}
\caption{Classical approximation hardness of PPSPs using the quasi-greedy classical algorithm, for constraint densities (clockwise from top left) $N_C = \cuof{2N,4N,6N,1.5 N^{3/2}}$. In each figure we plot the probability that a given shot returns an energy below $q E_{\mathrm{GS}}$ for various choices of $q$ (note varying scales on probability $y$-axis). For all the fixed $N_C/N$ problem classes we find an empirical approximation threshold $q = A_g$ below which finding states becomes superpolynomially hard, and that this value decreases as $N_C/N$ increases. For the case $N_C = 1.5 N^{3/2}$ (bottom left), this value steadily drops, as discussed in the text.}\label{greedyapproxfig}
\end{figure*}

This expected inefficiency of simulated annealing/parallel tempering for this problem can easily be inferred from the results plotted in Figs.~\ref{greedypgsfig} and \ref{greedyapproxfig}. Namely, in all cases the algorithm will find a single relatively deep minimum with high probability at each shot, leading to a super-polynomial cost to escape from it in algorithms simulating a thermal bath. Interestingly, as $N_C/N$ increases for fixed unsatisfied ground state fraction $\epsilon$, we find that the decay exponent of the per-shot probability of finding the ground state, $P_{\mathrm{GS}} \of{N}$, monotonically decreases, suggesting that the basin of attraction of the true ground state is widening as the problem becomes more extremal. In fact, for $N_C \geq 2 N$, we empirically observe that the per-shot probability of finding the planted ground state has the approximate scaling
\begin{eqnarray}\label{greedyPGS}
\log \of{ P_{\mathrm{GS}} \of{N} } \simeq - c_g \frac{N^2}{N_C}\,.  \; \; \;  ({\rm PPSPs,} \; N_C \geq 2 N)
\end{eqnarray}
However, the approximation ratio $q_a$--defined as the minimum energy for which the probability of finding states at or below it stays constant as $N$ increase--steadily worsens. We attribute this to there being a high density of local minimia with energies $\geq  q_a E_{\mathrm{GS}}$ (recall $E_{\mathrm{GS}}$ is negative in our conventions), but below that threshold the number of minima quickly decreases and the probability of finding one decreases exponentially. This results in the scaling collapse seen in Fig.~\ref{greedyapproxfig}--for sufficiently low energy there are no minima aside from the ground state, so the approximation probability scales nearly identically to $P_{\mathrm{GS}} \of{N}$. This \emph{high energy clustering phase} is a feature of our PPSP construction, and is responsible for its classical approximation hardness. We again contrast this to problems near the statistical SAT/UNSAT threshold such as three-regular instances, where the \emph{clustering energy}, the lowest energy where there are still exponentially many local minima and they are thus easy to find, is close to $E_{\mathrm{GS}}$ and they are not approximation-hard in practice as a result. We conjecture that high energy clustering behavior is a generic feature of low-degree constraint problem classes that are approximation-hard for local update algorithms.

As shown in Fig.~\ref{greedyapproxfig}, these construction rules yield in a set of instances which are hard to approximate in practice. If we let $N_C /N $ grow slowly with $N$, e.g. as $\ln \of{N}$ or $\sqrt{N}$, then as $N \to \infty$, the probability of finding any states with energies any $O \of{1}$ fraction better than random guessing, decays superpolynomially\footnote{We note again the recent result of \cite{d2022ihara}, who showed that for $N_C / N > O \of{N^{1/2}}$ the problem can be efficiently approximated classically; our PPSPs with $N_C/N = 1.5 N^{1/2}$ are close to this threshold but do not cross it.} in $N$. And since the unsatisfied fraction $\epsilon$ in the ground state is small but nonzero, Gaussian elimination cannot be used to efficiently find the solution, forcing classical computers to rely on local update algorithms stymied by entropic barriers. It is of course possible that some clever algorithm could be written to exploit our PPSP structure to efficiently solve or approximate these instances classically; we merely claim hardness for generic methods based on local updates. Our PPSP construction rules can easily be generalized to other CSPs, and we suggest that they could prove to be a useful tool for exploring practical approximation hardness in other contexts.

\subsection{Performance of high-depth TAQC for this problem}\label{qaoasec}

\begin{figure*}
    \includegraphics[width=0.49\textwidth]{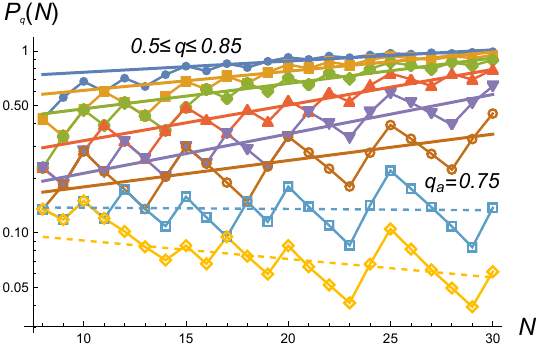}
    \includegraphics[width=0.49\textwidth]{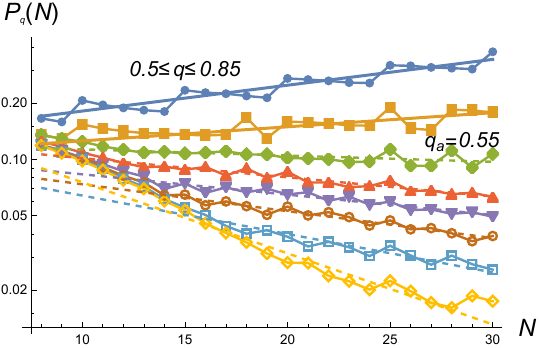}

    \includegraphics[width=0.49\textwidth]{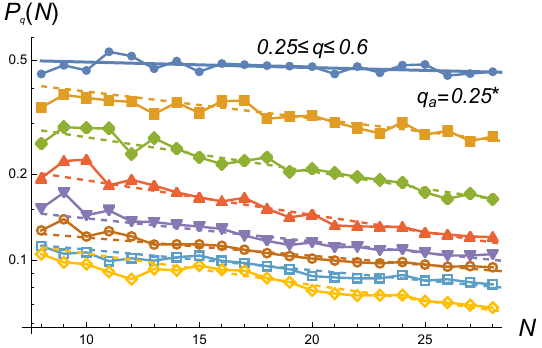}
    \includegraphics[width=0.49\textwidth]{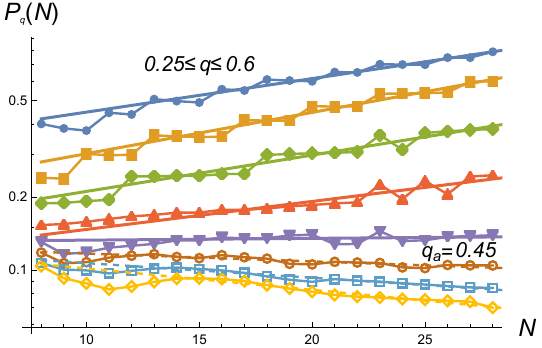}
    \caption{Performance of TAQC as an approximator for $N_C = \cuof{2,4,6,3 \sqrt{N}/2}N$ (clockwise from top left), with problem and algorithm parameters described in section~\ref{qaoasec}. Plotted are the probabilities $P_q \of{N} = P \of{E \leq q E_{\mathrm{GS}}}$ for $q$ running from 0.5 to 0.85 in steps of 0.05 (top panels) and 0.25 to 0.6 (bottom panels). The results for the larger constraint densities thus plot a weaker approximation range. Thick straight lines correspond to simple exponential fits where $P_q \of{N}$ is not decaying; dashed straight lines correspond to exponential decay; jagged thin curves are a guide to the eye; only points represent actual data. As summarized in Table~\ref{approxtab}, these results do not represent a meaningful improvement over the classical greedy result (Fig.~\ref{greedyapproxfig}), though in some cases where $P_q \of{N}$ decays exponentially for both approaches, the exponent for TAQC may be better. These results were obtained using the Fujitsu Quantum Simulator, a classical HPC system.}\label{qaoafig}
\end{figure*}

\begin{figure*}
    \includegraphics[width=0.49\textwidth]{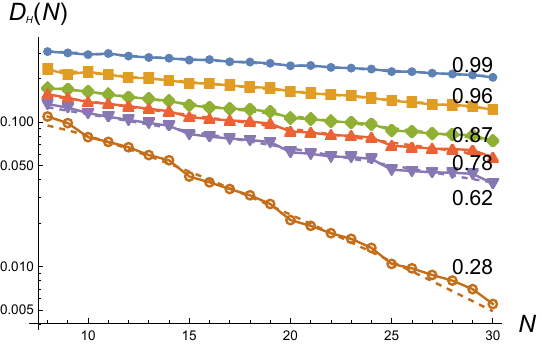}
    \includegraphics[width=0.49\textwidth]{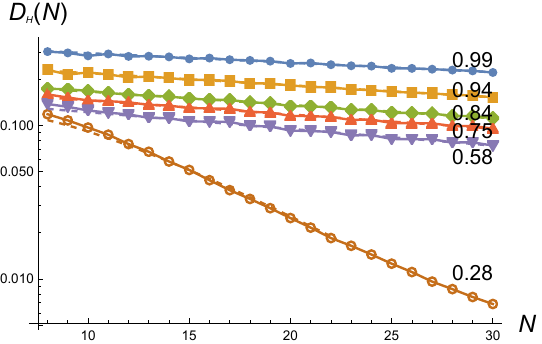}

    \includegraphics[width=0.49\textwidth]{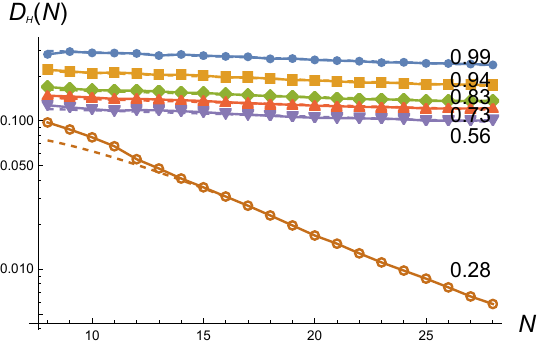}
    \includegraphics[width=0.49\textwidth]{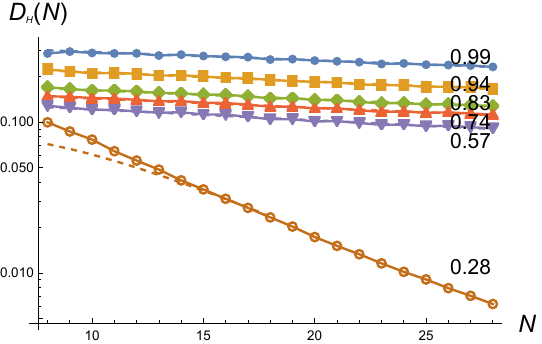}
    \caption{Probabilities of returning states within Hamming distance $D_H = \cuof{2/5,1/3,1/4,1/5,1/8,0}N$ flips from the ground state $G$ for QAOA (top to bottom curves), for $N_C = \cuof{2,4,6,3 \sqrt{N}/2}N$ (clockwise from top left) and application parameters in section~\ref{qaoasec}. As seen in the figures, the per-shot probability of finding the ground state is essentially independent of constraint density and tracks the prediction $N 2^{-0.28 N}$ in Eq.~\ref{Om0AQC}. On the other hand, the probabilities of returning states at various extensive fractional Hamming distances, e.g. $N/4$ or fewer flips, decay much more slowly, owing to the exponentially large number of target states, and the task of finding states comparatively close to $G$ becomes easier as $N_C/N$ increases. That said, these probabilities all decay exponentially with $N$ in all cases, consistent with the worsening approximation ratios observed in Figure~\ref{qaoafig}. In particular, as discussed in section~\ref{qaoasec}, these results allow us to test the prediction in section~\ref{AQChard} that the per-state matrix elements to mix with excited states can collapse as the energy increases. The numbers adjacent to each curve on the plots are the best fit for the decay exponent $b$ in $\avg{\Omega_{0j}^2}_j \propto 2^{-b N}$; as the Hamming distance (and thus average energy, per Eq.~\ref{avgE}) increases, these matrix elements decay much more rapidly, exceeding the exponential growth of the number of target states and confirming our expectation that AQC is a poor approximation algorithm in the worst case.}\label{QAOA_HD_fig}
\end{figure*}

To make firm points of comparison, alongside the simulations of spectrally folded optimization itself we extensively benchmarked high-depth TAQC on these instances. For our high-depth TAQC simulations, we formulated our algorithm to approximate continuous time evolution over a total time $t_F$, with:
\begin{align}\label{QAOAschedule}
&\ket{\psi \of{t+dt}} = e^{-2 \pi i f \of{t} dt H_D   } e^{-2 \pi i g \of{t} dt H_P   } \ket{\psi \of{t}}, \\
&f \of{t} = \sqrt{1-t/t_F}, \; g \of{t} = \sqrt{t/t_F}.\nonumber
\end{align}
In all the presented data we used $t_F=N/32$ and $dt \simeq 0.05$. Individual shots use a random evolution time between $2t_F/3$ and $4t_F/3$; this runtime averaging produces substantially more reliable scaling, particularly when probabilities are small. These parameters were chosen by trial and error for smaller systems; we observed that the probabilities of finding the ground state and other low-energy states increased sublinearly with $t_F$ beyond this point. The relative improvements of the probabilities of returning low-lying states were similarly sublinear. As discussed in section~\ref{nofinetune}, we focused TAQC rather than QAOA to avoid the overhead of angle optimization and to have easily interpretable results. No numerical or iterative optimization methods were used, nor was any per-instance tuning employed, for this algorithm or any other algorithm in this work.

The results of our TAQC simulations, of 1000 random PPSPs for each choice of $N$ running from 8 to 30, are shown in Fig.~\ref{qaoafig}. The probability of finding the ground state, shown in Fig.~\ref{QAOA_HD_fig}, decays exponentially with an exponent very close to that in Eq.~(\ref{Om0AQC}), which we find remarkable given the simplifying assumptions in that derivation, and that it does not use the more sophisticated techniques used to compute tunneling rates between semiclassical minima. Further, this exponent displays only small variations with constraint density and is nearly identical in all four cases. Smaller system studies for other constraint densities all yielded very similar results for $P_{\mathrm{GS}}$, as predicted by Eq.~(\ref{Om0AQC}).

Turning to approximation hardness, being relatively sparse, the $N_C = 2N$ problems are fairly well-approximated by TAQC, with the algorithm returning strings within $q=0.75$ with constant or saturating probability; we attribute this to the presence of many competing minima with energies not far from $E_{\mathrm{GS}}$. In contrast, for $N_C = 4N$ the algorithm's performance for approximation degrades, with clear exponential decay for approximation ratios better than $q = 0.55$. For higher constraint densities approximation becomes even more difficult, decaying exponentially below $q=0.45$ for $N_C = 6N$ and 0.25 for $N_C = 3 N^{3/2} / 2$. We expect decay at sufficiently large $N$ for any constant fraction in that case, but cannot simulate larger system sizes. Crucially, the thresholds $q_a$ we measure are nearly identical to those found by the classical greedy algorithm (figure~\ref{greedyapproxfig}), and no signatures of an exponential quantum advantage in these instances can be seen.

Interestingly, as $N_C/N$ increases, the probabilities of finding states comparatively close to $\ket{G}$ in Hamming distance improve (see Fig.~\ref{QAOA_HD_fig}), but the probabilities of finding states close in energy worsen. We attribute this behavior to the high energy clustering phase conjectured in Secs.~\ref{3xordef} and~\ref{greedysec}. Empirically for our PPSPs there is a high density of local minima with energies $E \geq q_a E_{\mathrm{GS}}$, and if $q_a$ is relatively close to 1 it becomes harder for high-depth TAQC to find local excitations near the planted ground state, as the probability amplitudes will be spread over increasingly many competing minima and their own basins of attraction. Conversely, as $q_a$ decreases with increasing $N_C/N$, the probability of finding local excitations relatively near $\ket{G}$ increases though still decays exponentially, as the low energy minima far from $\ket{G}$ are at proportionally higher energies and thus do not compete directly with few-flip states. That the thresholds $q_a$ for TAQC match those of the greedy classical algorithm further supports the interpretation of an energy threshold above which local minima become common.

Further, these results allow us to directly test the qualitative predition in section~\ref{AQChard} that the matrix elements for the initial paramagnetic state to mix with excited states can collapse as the energy increases, due to the weakening transverse field and destructive interference. As shown in FIG.~\ref{QAOA_HD_fig}, the probabilities of finding states within fractional Hamming distances $\cuof{2/5,1/3,1/4,1/5,1/8}N$ from $G$ all decay exponentially, albeit with much smaller exponents than the probability of finding $G$. We emphasize that these smaller exponents mask the fact that the per-state mixing rate $\avg{\Omega_{0j}^2}_j$ is decaying extremely rapidly, since the total success probability for finding states within $D_H$ flips is given by
\begin{eqnarray}
    P_{D_H} \of{t_F} \propto \frac{\avg{\Omega_{0j}^2}_j t_F}{W} N_{D_H},
\end{eqnarray}
where $N_{D_H}$ is the number of states at this distance and is given by Eq.~\ref{NT}. Since $N_{D_H}$ grows exponentially quickly, for the total $P_{D_H}$ to decay exponentially $\avg{\Omega_{0j}^2}_j$ must decay rapidly, and the resulting fit exponents in FIG.~\ref{QAOA_HD_fig} are stark evidence of this. In section~\ref{AQChard}, we identified weakening fields and destructive interference as good reasons to not be confident in TAQC as a potent approximation algorithm, even though we were not able to perform a rigorous analytical calculation due to the complexity involved. Our numerical results confirm this suspicion. Spectrally filtered optimization was originally formulated to resolve these two issues, and as shown in the next section, leads to dramatically different behavior and approximation performance.

Comparing the classical and established quantum methods, we find that, for these approximation hard instances, TAQC performs very poorly for finding the ground state but less poorly for approximation below the classical hardness threshold $q_a$ (see Table~\ref{approxtab}), with decay exponents that are much closer to the classical result. In some higher constraint density cases our fits produced favorable exponents for approximation with TAQC but our range of $N$ here is smaller than we would prefer to claim any relative quantum advantage \emph{absent rigorous theoretical justification}. Nonetheless, both methods show clear super-polynomial decay per-shot for approximate optimization below the energy range where local minima are dense. With these results in hand, we now turn to folded quantum optimization, which maintains an approximation guarantee regardless of the problem's constraint density.

\section{Numerical tests of spectrally filtered quantum optimization}\label{specfoldsims}

Having numerically confirmed the expectation that our PPSPs are superpolynomially hard for classical and prior quantum approaches, for both exact and approximate optimization, we now present the results of our folded quantum optimization simulations. We simulated both the trial minimum annealing and interpolation (e.g. AQC) variations. In all cases aside from MSFO, we chose runtimes increased linearly in $N$, albeit with larger prefactors than in the TAQC simulations (which used $t_F = N/32$). Following the discussion in section~\ref{nofinetune}, a longer runtime further helps reduce the potential influence of diabatic local heating as the Hamiltonian parameters are varied. Run and ramp times that are too short can lead to artificially poor scaling, arising from the formation of local excitations as the transverse field is turned on or off too quickly. This is fundamentally a different, and much more prosaic, issue, than decaying collective tunneling rates, but can be difficult to distinguish when our only measures are energy and Hamming distance from $\ket{G}$.

\subsection{Spectrally folded adiabatic interpolation performance}

\begin{figure*}
\includegraphics[width=0.49\textwidth]{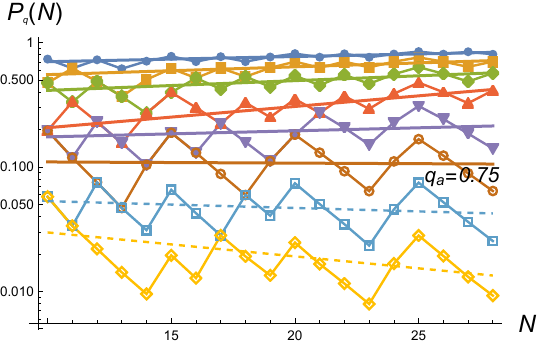}
\includegraphics[width=0.49\textwidth]{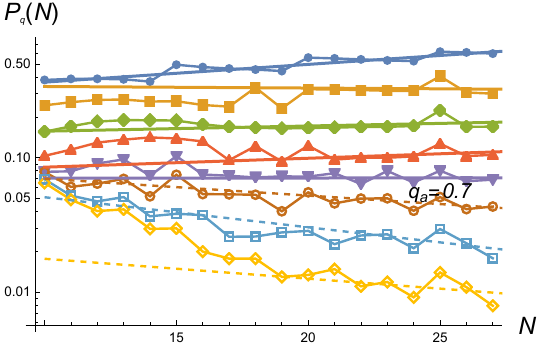}

\includegraphics[width=0.49\textwidth]{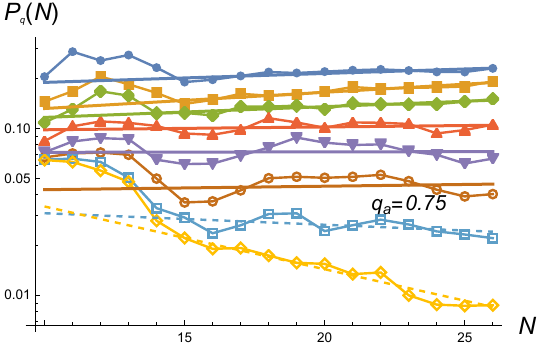}
\includegraphics[width=0.49\textwidth]{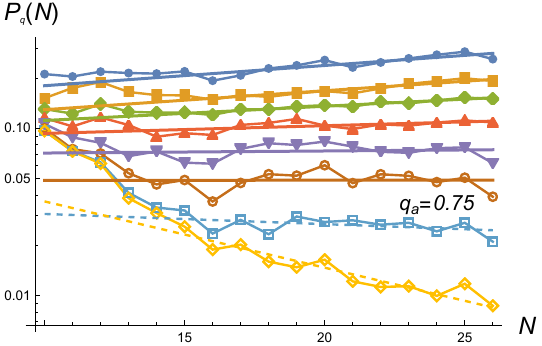}
\caption{Performance of the quadratic AQC formulation of spectral folding, for $N_C = \cuof{2,4,6,3 \sqrt{N}/2}N$ (clockwise from top left), with $N$ running from 10 to 27 (top row) or 26 (bottom row), $A=0.75$, $dt=0.0325$, $t_f = N/24$ and other parameters as stated in text. In each plot the 8 curves plot $P_q \of{N} = P \of{E \leq q E_{\mathrm{GS}}}$ for $q$ running from 0.5 to 0.85 (top to bottom) in steps of 0.05. Thick straight lines correspond to simple exponential fits where $P_q \of{N}$ is not decaying, dashed straight lines correspond to exponential decay, and thin lines between points are included for visual clarity. In all four cases spectrally folded optimization is able to meet its approximation target of $A=0.75$, returning states at or below this energy with constant probablity in a linearly growing number of cost function calls. This is in stark contrast to our classical greedy algorithm (FIG.~\ref{greedyapproxfig}) and QAOA (FIG.~\ref{qaoafig}) results, where the achievable polynomial time approximation ratio steadily worsens with increasing $N_C/N$, and supports the theoretical analysis of section~\ref{theorysec}. These results were obtained using the Fujitsu Quantum Simulator, a classical HPC system.}\label{specfoldAQCfig}
\end{figure*}

We first present simulations of the AQC variation of spectral folding, in figures~\ref{specfoldAQCfig} and~\ref{AQC_HD_fig}. For the AQC variation, we followed the procedure in Eq.~(\ref{QAOAschedule}), with a quadratic $H_{\mathrm{fold}}$ (Eq.~\ref{nlfold}, without the symmetrization at high energies) in place of $H_P$ and $f \of{t} = \of{1-t/t_F}^{1/4} $ instead of $\sqrt{1-t/t_F}$, with an average runtime $t_F = N/24$ and $dt=0.0325$. This schedule modification was found to improve scaling at higher approximation ratios. In all cases individual shots are runtime averaged between $2 t_F/3$ and $4 t_F/3$ as in our TAQC simulations. 

As shown in figure~\ref{specfoldAQCfig}, in all but one case we were able to meet the approximation target $A=0.75$, but not exceed it; for $N_C = 4N$ we were limited to $q_a \simeq 0.7$. Our spectral folding methods are also much better at reliably returning states close to $G$ in Hamming distance, consistent with the approximation guarantee (see Fig.~\ref{AQC_HD_fig}). We likewise tested increasing $A$ to 0.85 in simulations up to $N=24$ with this variation, and found improved prefactors but no improvement in scaling (see Fig.~\ref{specfoldaqcfig-85}). This suggests that we have found the performance ceiling for this approach. Interestingly, the worst case $q_a = 0.7$ observed for this method is not far below the approximation ratio of $\sim 0.8$ predicted in section~\ref{ssec:caveats}, using a much more simplified analysis than was employed for the TMA and MSFO variations.

\subsection{Trial minimum annealing performance}

\begin{figure*}
    \includegraphics[width=0.49\textwidth]{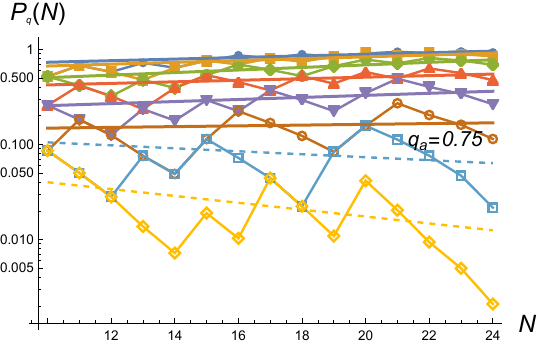}
    \includegraphics[width=0.49\textwidth]{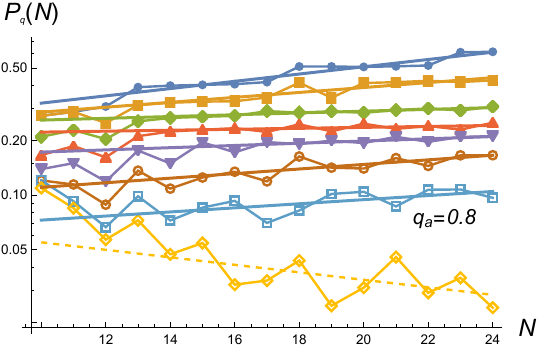}

    \includegraphics[width=0.49\textwidth]{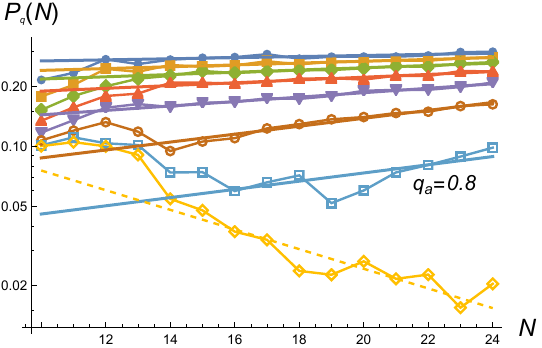}
    \includegraphics[width=0.49\textwidth]{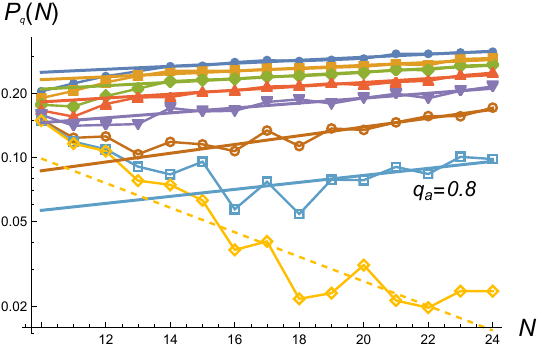}
    \caption{Performance of the trial minimum annealing formulation of spectral folding with a 3-XORSAT lowering Hamiltonian $H_L$, with $A=0.85$, plotted for constraint densities (clockwise from top left) $N_C = \cuof{2N,4N,6N,1.5\times N^{3/2}}$ and approximation ratios between $q=0.5$ and $0.85$ with $N$ running from 10 to 24, for the parameters detailed in the text. All data is derived from averaging over 960 random instances and choices of $t_f$. Compared to the AQC formulation shown in figure~\ref{specfoldAQCfig}, the achievable approximation ratio $q_a$ is often slightly higher, though the total gate count in this formulation is larger by a constant prefactor. In all cases $q_a$ well exceeds the value of 0.6 conservatively predicted for this formulation.}\label{specfoldtmafig}
\end{figure*}

We also tested the TMA formulation of spectral folding, with results plotted in Fig.~\ref{specfoldtmafig}. For these variations we used runtimes $t_F = N/12$ and $dt=0.025$; note that this choice of $t_F$ is a factor of $8/3$ larger than in our TAQC simulations but with the same scaling. In smaller system studies similar qualitative performance was observed for shorter $t_F$ (such as $N/24$ or $N/32$). Following the principles in section~\ref{nofinetune}, these longer runtimes were chosen to ensure we did not need to worry about performance degradation due to heating from the ramps themselves. For this variation we used a 3-XORSAT $H_L$--the formulation for which we could predict performance in Sec.~\ref{theorysec}--with the minimum energy set to $-2N$ via $C \of{t}$, which was linearly ramped down to zero by $t_f$, and simple sinusoidal ramp profiles with $t_r = N/24$; the transverse field strength $\kappa$ during the main evolution was 1.3. 

The careful reader may note that choosing $C \of{t}$ to set the minimum energy to $-2N$ instead of $-N$ is naively suboptimal, as all other things being equal sweeping over a larger energy range increases $W$ in Eq.~(\ref{PTF}), and should reduce the returned probabilities $P_q \of{N}$ by an appropriate prefactor. However in our simulations this choice consistently improved both the prefactors and scaling, e.g. the value of $q_a$, as compared to choosing a minimum energy $-N$ for $H_L$. We suspect this has to do with the band structure considerations described in Appendix~\ref{bandstructure} but due to the complexity of the problem, are unable to make a quantitative prediction.

The performance of the two approaches is qualitatively similar with subtle differences as we vary the returned approximation ratio $q$. At lower approximation ratios the AQC formulation returns higher probabilities, at lower total gate count since there are no ramping steps and no additional gates associated with adding $H_L$. However, at higher approximation ratios the TMA formulation appears to be better able to approach the approximation target $A=0.85$. As discussed in the algorithm definition, folded optimization will definitionally fail to consistently return energies significantly below $A E_{\mathrm{GS}}$, and absent warping we expect it to break down as $A$ gets too close to 1 given that TAQC and similar methods fail to reliably approximate these problems. Choosing the best value for $A$ is thus a subtle issue that depends on the problem class; for extensions of this method to hard CSPs it will necessarily change from one problem class to the next.

\begin{figure*}
    \includegraphics[width=0.49\textwidth]{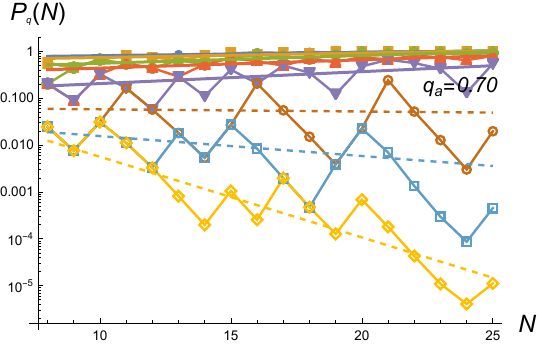}
    \includegraphics[width=0.49\textwidth]{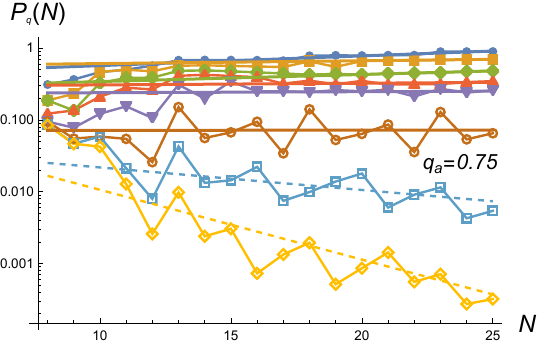}

    \includegraphics[width=0.49\textwidth]{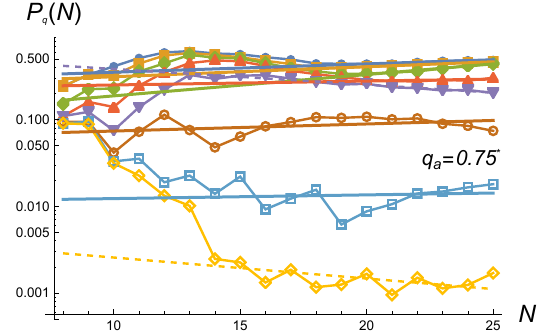}
    \includegraphics[width=0.49\textwidth]{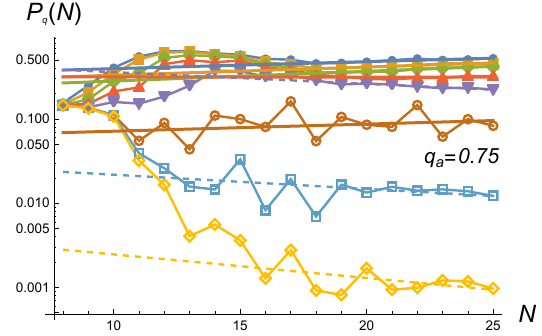}
    \caption{Performance of the trial minimum annealing formulation of spectral folding with a local $Z$ bias $H_L$, with $A=0.75$, for $N$ running from 8 to 25. $q_a$ in each case is comparable to other variations, though greater non-monotonicity in the individual $P_q \of{N}$ curves makes fitting more difficult.}\label{specfoldtmalinfig}
\end{figure*}

Likewise, as mentioned in section~\ref{ssec:caveats}, one can replace the random 3-XORSAT lowering Hamiltonian in TMA for a simpler set of linear $Z$ biases, reducing the gate count per timestep and, potentially, increasing the per-state tunneling rate by implementing a shallower cost-per-flip curve. In comparison to the 3-XORSAT variation discussed in the previous paragraph, to achieve good performance we needed to double the ramp time.  This protocol seemed to be more sensitive to performance degradation from heating during ramps. The algorithm also benefitted from adjusting $C \of{t}$ so that the minimum energy of $H_L$ was $-3N$, as compared to $-2N$ for the 3-XORSAT $H_L$. Relative performance for $A=0.75$ is comparable to the other variations, as illustrated in Fig.~\ref{specfoldtmalinfig}, though the individual $P_q \of{N}$ show more significant non-monotonicity that makes reliable curve fitting challenging. This issue is even more pronounced for $A=0.85$ (data not shown), to the point that we did not quote $q_a$ values for that variation in the Table~\ref{approxtab}.

\subsection{Multi-stage filtered optimization performance}\label{MSFOsims}

\begin{figure*}
    \includegraphics[width=0.75\textwidth]{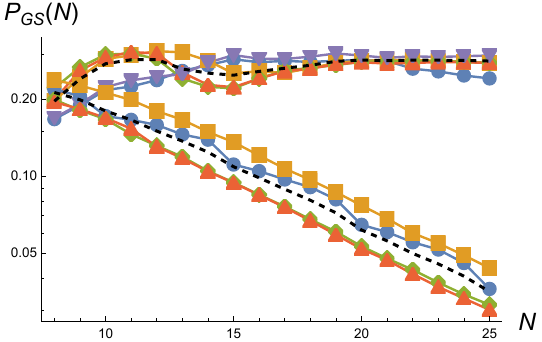}
    \caption{Probability of finding the ground state with multi-stage filtered optimization (top, approximately constant data), as compared to Trotterized adiabatic quantum evolution (bottom, exponentially decaying data), for $N_C = \cuof{2,4,6,3\sqrt{N}/2}N$ (blue, gold, green and red points), demonstrating the rough independence of constraint density--and, critically, the scaling of $P_{GS} \of{N}$ for the runtime specified--predicted through many-body theory in the text. The TAQC data, which all shows clear exponential decay, is taken from FIG.~\ref{QAOA_HD_fig} and rescaled by a multiplicative factor so that the total per-shot quantum evolution time is the same for all data. This runtime is sub-quadratic (scaling approximately as $N^{1.5}$) and given by Eq.~\ref{MSFOruntime}. For $N_C = 2N$ we observe performance degradation at larger $N$, which turns out to be due to Trotter error heating; reducing $dt$ from 0.025 to 0.02 restores constant $P_{GS} \of{N}$ (purple triangles). We attribute the larger sensitivity to Trotter error at low $d_C$ to the increased local variance of the density of constraints, and it may be the case that $dt$ needs exhibit slow polynomial decay with $N$ to maintain asymptotically constant scaling; both issues are discussed in section~\ref{sec:multistage}. Black dashed lines plot the average over constraint densities for the two methods. Each MSFO datapoint is the average of 480 randomly generated instances.}\label{MSFOmainfig}
\end{figure*}

For our final set of simulations, we studied the multi-stage filtered optimization of section~\ref{sec:multistage}, which for $d_C \gg 1$ and $\epsilon \ll 1$ is predicted to find the ground state in sub-quadratic total evolution time. In that section, we made two claims, the first  concerning the stability of the transition point and the second concerning the validity of mean-field methods for predicting the minimum gap. If the second claim in particular asymptotically holds then the time to solution must scale polynomially, roughly as $N^{1.46}$ for the parameters ($x=2$ and $w=0.6$ in Eq.~\ref{composedfilters}) chosen. Our goal in these sections was to test the basic veracity of these claims. In particular, we wanted to verify that for a polynomial quantum evolution time $t_F \of{N}$, predicted from a mean field analysis, the probability of finding the ground state $P_{GS} \of{N}$ would be asymptotically constant with increasing $N$.

To determine $t_F \of{N}$, we needed to take into account a few effects that are not relevant in the large $N$ limit. First, as discussed in appendix~\ref{sec:costflip}, the average cost per flip away from $G$ has meaningful small-$N$ deviations from the asymptotic result in Eq.~\ref{avgE}. To take this into account, we numerically calculated the average $E \of{x}$ for $x$ flips away from $G$ and used that result as the cost function in computing $\Omega_{\mathrm{mf}} \of{N}$. For $x=2$, $w=0.6$ and $N \leq 30$, we obtained $\Omega_{\mathrm{mf}} \of{N} \propto N^{-0.16}$. Second, the slope $W \of{N}$ was not perfectly linear, but instead best fit by $W \of{N} \propto \sqrt{N^2 + w_0^2}$. Third, let the avoided crossing occur at $s_c \of{N}$ (for total Hamiltonian $H = \kappa H_D + s H_{\mathrm{cost}}^{x,w}$). It turns out for $N$ in the range accessible for classical simulation that the scaling of the gap with $s$ is substantially asymmetric around $s_c$ (with a much sharper slope with $s>s_c$), and $s_c \of{N}$ likewise empirically varies as $s_c \of{N} = s_0 + a/N$ for a constant $a$ and $s_0 \simeq 0.3$ is the asymptotic result. Putting all these effects together, we posited a runtime scaling
\begin{eqnarray}\label{MSFOruntime}
t_F \of{N} \propto \frac{ \sqrt{N^2 + w_0^2}}{\Omega_{\mathrm{mf}}\of{N}^2 s_c \of{N}}.
\end{eqnarray}

This function approximately scales as $N^{1.5}$ for the parameters chosen here, and was used as the input to set the runtime of our simulations. We set $dt=0.025$ to minimize heating from Trotter error (though as seen in figure~\ref{MSFOmainfig}, we needed to reduce it further for $d_C = 2$), and let the runtime of the first stage be $t_F \of{N}$ and the second and third stages be $t_F \of{N}/2$ for each. Finally, we chose the prefactor of $t_F \of{N}$ so that $t_F \of{8} = 1/4$, matching the runtime used in the TAQC simulations of section~\ref{qaoasec}. 

The results of our simulations are plotted in FIG.~\ref{MSFOmainfig}, and support our assertion of a polynomially scaling time to find the true ground state for PPSPs in the approximation-hard regime. With suitably small $dt$ we find that $P_{GS} \of{N}$ is asymptotically independent of constraint density and asymptotically constant with $N$, at least up to $N=25$, the largest sizes simulated. Both the constant scaling, and the approximate $d_C$ independence, are implied by our theory. The performance advantage of MSFO is particularly stark when compared to TAQC, as seen in the figure. Per Eq.~\ref{PTF} we assume that the TAQC success probability scales roughly as $P_{GS} \propto \Omega_0 \of{N}^2 t_F \of{N}/W\of{N}$, and therefore to make the comparison fair we took the probabilities computed in section~\ref{qaoasec} and multiplied them by the appropriate factor $\of{8/N}\of{t_F \of{N}/t_F \of{8}}$ to afford them the same total quantum evolution time as MSFO.

We find the result for $d_C = 2$ to be particularly surprising, in some sense, given that the local constraint density variation is proportionally large in that case, and that the clustering energy is 75-80\% of $E_{GS}$ so we do not think transverse field chaos can be a priori ignored once filtering is applied. The only algorithm modification we needed to make to match the performance at higher constraint densities was to reduce $dt$ (and thus increase the circuit depth, so the total evolution time was the same) by $20\%$. We attribute the increased sensitivity to Trotter error to the larger local variations in the cost per flip of the classical $H_P$; recall that we have normalized each problem so that $E_{GS} = -N$, and the variance of the local energy cost scales as $1/\sqrt{d_C}$ as a result. 

We emphasize that we do not guarantee that MSFO will always be able to find the ground state for any appropriately normalized MAX-3-XORSAT instance, given that it cannot overcome TFC. We presume that there are also additional hypergraph properties that ensure exponential scaling for MSFO (and our other filtered optimization methods), for both exact and approximate optimization; we just do not know what they are. If the transition to the ground state becomes first order and is missed, MSFO could likely suffer from the same field weakening and interference problems as TAQC, though this could be addressed by choosing an approximation target $A<1$, something that could be interesting to explore in future work.

\subsection{Discussion of our spectrally folded optimization results}

The reader may note that all of our spectral folding simulations targeted energies in the range $A=0.75$ to $A=0.85$, well above the theoretical prediction of $\sim 0.6$. We also tested $A=0.65$ at smaller scales (data not shown) for both formulations but found no region of the parameter space where it showed meaningful improvements over choosing $A=0.75$ or higher. When increasing $A$ to 0.85 we were not able to meet this target with constant probability, though for some cases we did see improvements in the returned achievable approximation ratio $q_a$ when compared to $A=0.75$. Aside from our MSFO simulations where $A=1$, we did not run tests with $A > 0.85$ due to system size constraints, though comparing performance between $A=0.75$ and $A=0.85$ in table~\ref{approxtab} suggests we have found the performance ceiling for these methods. Given the comparatively small system sizes available to classical simulation our ability to draw meaningful scaling distinctions with very small changes in $A$ is limited. Considering the results of all our numerical simulations, the minimum achievable approximation ratio for our PPSP problem classes is at least 0.7 (often 0.75) for the quadratic AQC variation, and at least 0.75 (often 0.8) for the 3-XORSAT TMA variation.

As summarized in table~\ref{approxtab}, the contrast between the clear super-polynomial decay of higher approximation ratios with classical methods and TAQC (Figs.~\ref{greedypgsfig}-\ref{QAOA_HD_fig}), and the constant probabilities returned by folded quantum optimization (Figs.~\ref{specfoldAQCfig}-\ref{MSFOmainfig}) is stark. In particular, for the established methods, the polynomial time approximation threshold is set by the problem structure, specifically the relative energy of the high energy clustering phase, and as our PPSPs become more extremal this threshold steadily worsens as a fraction of the energy of the planted ground state. We found no evidence for an exponential separation in approximation power between local classical searches and TAQC. We assume, but did not test, that this would hold for linear-depth QAOA as well, though we do not rule out the possibility of meaningful speedups given fine tuning. At lower constraint densities, the clustering energy is not far above the ground state and spectrally folded quantum optimization provides no benefits for approximation, though it can still show scaling advantages for returning states close in Hamming distance to the ground state (see appendices). At higher constraint densities however, spectral folding is able to return states close to $G$ (and often find it directly, given quadratic runtimes and multiple stages), in both energy and Hamming distance, consistently and without degradation as $N_C/N$ increases. The performance of all spectral folding variations tested was broadly similar, with the 3-XORSAT TMA variation returning the highest approximation ratio but at the highest prefactor cost in gate count.
 
This illustrates the fundamentally different structure of collective tunneling in this approach. Where TAQC and other direct methods attempt to find the ground state, asymptotically fail (given exponential scaling), and return approximate states passed after this missed opportunity, spectrally folded optimization deforms the cost function to search for states in an exponentially large hyperspherical shell, avoiding the interference and weakening field issues that are qualitatively responsible for TAQC's lack of obvious quantum advantage. In the interest of fair comparisons the total runtime per-shot of all routines besides MSFO is $O \of{N}$ Hamiltonian calls. With this simple linear scaling we are able to ensure that any exponential scaling of the time to solution not visible from our figures or fits must involve very small exponents, though of course we cannot rule out such behavior on numerics alone. We thus do not see our numerical simulations as a precision scaling benchmark of folded quantum optimization, but rather a test of the basic veracity of our conservative analytical predictions in section~\ref{theorysec}. 

Likewise, if our goal is to find ground states and not just good approximations thereof, it may well be the case that algorithms targeting $A<1$ are actually superior in the approximation-hard regime, even if MSFO with $A=1$ can find the ground state directly in sub-quadratic time! This is because those algorithms are capable of reliably finding states \emph{near} the ground state, e.g. in its basin of attraction, and from there simple classical methods such as our quasi-greedy algorithm or simulated annealing can likely relax to the ground state quickly. If the polynomial runtime scaling to find approximate solutions is favorable (e.g. $O \of{N}$ vs. $O \of{N^{1.46}}$), then relaxing from an approximate solution may be more efficient. The choices explored in this paper likely only scratch the surface of what is possible in filtered optimization. We present all of these simulations--and new quantum algorithms--as constructive evidence for the fast approximability of MAX-3-XORSAT through a novel speedup mechanism. The results of our simulations, for a wide range of parameters, all suggest that our theory is sound.

\subsection{Extensions of these results}

Having thoroughly explored MAX-3-XORSAT in this work, it would be interesting to test spectral filtering methods on other hard CSPs or problems that can be straightforwardly formulated as such. We expect that the derivation of the achievable approximation ratio through the overlap of dressed states could generalize with some modifications to many other problem classes. We focused on MAX-3-XORSAT due to the simplicity of its structure, classical approximation hardness, and the fact that its exponential difficulty scaling is obvious at small $N$ for more standard classical and quantum approaches.

It also strikes us as noteworthy that while spectral folding can be implemented in traditional classical heuristics, there is no benefit to doing so. As we discussed at length in Sec.~\ref{introsec}, classical algorithms generally start from high energy states and attempt to cool towards the ground state through local updates. In the case of linear spectral folding, the folding procedure makes no difference whatsoever to the returned energy until the fold energy has been reached, and in the worst case reaching that energy is an exponentially difficult task for classical computers unless P=NP, due to entropic barriers as discussed in Sec.~\ref{introsec} and confirmed in our simulations. Changes to the high energy spectrum from quadratic folding are similarly not expected to make approximation easier. So while spectral folding is not in and of itself a quantum operation, we expect that it is only valuable in quantum algorithms and we consider it an irreducibly quantum method as a result. Further, because of its nonlocal nature, even in the locally gapped dressed problem phase we expect volume-scaling entanglement in low-lying states of the folded quantum spin glass, and significant classical simulation difficulty. 

One interesting potential exception to this argument is quantum Monte Carlo. Being a stoquastic problem, a folded quantum spin glass can in principle be efficiently simulated using QMC~\cite{isakov2016understanding,andriyash2017can,jiang2017scaling,jiang2017path,king2019scaling}, which for a uniform field has in some cases proven to be an effective quantum-inspired classical solver~\cite{heimronnow2015,albash2018demonstration}. Incorporating spectral folding into these algorithms is possible, though the loss of locality makes evaluating each update much more expensive. As QMC is, fundamentally, an energy-based classical local update rule combined with many replicas, we do not expect it to overcome the entropic barrier in MAX-3-XORSAT the way that true quantum evolution can, though of course have not tested this here. In this vein, we note also the results of Hastings, and subsequently Gily\'{e}n and Vazirani, that a super-polynomial separation exists between stoquastic AQC (the class containing some of the algorithms in this work) and classical computation \cite{hastings2021power,gilyen2021sub}; the stoquasticity of our algorithms thus does not rule out super-polynomial speedups. All that said, there may be some narrow cases where incorporating spectral filtering in a QMC calculation could prove useful as a classical solver; this would be an interesting avenue for future research. 

If it turns out that many or all of these instances are efficiently approximable classically through QMC with a filtered spectrum (or any other classical method, for that matter), that would be a very significant discovery in its own right. Nothing in this work rules that out, and we think the issue deserves further inquiry. 

\section{Conclusions and Outlook}

Using the NP-hard MAX-3-XORSAT problem class, we explored the question of classical and quantum approximation hardness from a practical, mechanism-focused point of view. Guided by theoretical intuition, we proposed a class of instances called planted partial solution problems (PPSPs) which we showed are empirically hard for both exact and approximate optimization for classical searches and established quantum methods such as adiabatic quantum computing (AQC) and, we presume (but did not directly test), the quantum approximate optimization algorithm (QAOA). Through extrapolation and qualitative analysis of a rigorous calculation of the typical-case minimum gap at the first order paramagnet to spin glass transition bottlenecking MAX-3-XORSAT, we were able to identify two effects that significantly impede the ability of high-depth Trotterized AQC (TAQC) to approximate the hardest instances, namely weakening transverse field and destructive interference. We then proposed a novel algorithmic advance, called \emph{spectrally filtered quantum optimization}, that provides multiple routes to circumvent these issues. The filtering transformation itself is conceptually simple; the analytical analysis of the full algorithm's performance is perhaps less so. And rather intriguingly, it works by applying classical modifications to the classical cost function being optimized, which only provide benefits for quantum optimization.

Using a resummed extensive-order perturbation theory, we were able to predict a constant fraction approximation guarantee for our difficult random hypergraph PPSPs, and consequently, an exponential quantum speedup in the classically hard regime. To further support our claims, we performed a series of numerical simulations of high-depth TAQC, classical optimization routines, and spectrally filtered optimization methods, out to the largest sizes that we could feasibly reach while still being able to gather good statistics. These numerical results support our claims and we did not discover any meaningful discrepancies or red flags in them; every major prediction here has been numerically checked to the extent we found reasonably possible.

The implications of an effective fast approximation guarantee through spectrally filtered optimization--or any quantum algorithm, for that matter--are profound, and even restricting ourselves to MAX-3-XORSAT it is important to ask what types of hypergraph might cause it to fail. To be clear, ``failure" in this case means that, for a class of hypergraphs, some property invalidates the analytical predictions we made and restricts polynomial-time approximation to values of $q_m$ near zero. Such a property, whether present in large random instances or specific to a tiny subclass thereof, would have to break both folded optimization through collective tunneling, and the softening of the first order phase transition in the multistage composition of folding and warping. 

Focusing on the trial minimum annealing formulation, if a subsequent, more sophisticated analysis shows, for example, that due to some subtle effect missed in our resummed extensive order perturbation theory, the variations we propose are asymptotically limited to $q_m=0.4$ for random hypergraphs instead of our prediction of $\geq 0.6$, we would not consider that a general failure of the algorithm. Given the exponential density of target states, for folded quantum optimization to fail the per-state tunneling rate needs to be much worse than what our calculations and simulations return; even doubling the per-state tunneling decay exponent in Eq.~(\ref{PTAScondition}) yields $q_m \simeq 0.18$, a much worse approximation ratio than we predict and observe in simulations, but still a constant fraction better than random guessing. And this estimate is, again, for a formulation of our trial minimum annealing algorithm that did not use all the tools available to us, such as spectral warping, that would further improve performance. And finally, we emphasize again that neither of our two filtered optimization algorithms used all the techniques available to improve approximation performance--targeting excited states by folding in the case of multistage filtered optimization, and spectral warping in trial minimum annealing.

We are not so hubristic as to claim it is impossible that there are hidden effects that substantially worsen scaling at large $N$ for random hypergraphs, which are not captured by our theories and are invisible in our numerics. But we see no evidence for it, have no idea what such an issue could be, and formulated our theory such that the simplifying approximations we made were more likely to underestimate the achievable approximation ratio $q_a$ than overestimate it. Excellent empirical performance in simulation, and the fact that two completely different calculations analyzing completely different filtered optimization algorithm formulations yield similar speedups, both bolster the strength of our claims. We see it as much more likely that one can structure PPSP hypergraphs in some non-random way as to violate the core assumptions of our calculations and become inapproximable. The conditions on such hypergraphs are, however, fairly strict; besides needing to more than double the exponential decay rate of per-state tunneling in folded optimization (as compared to random hypergraphs), whatever property is responsible for the slowdown must be resilient both to the addition of a random, uncorrelated problem with similar constraint density and ground state energy, and to spectral deformations such as nonlinear folding and spectral warping. Further, such graphs must be relatively dense, as sparse problems are easy to approximate by solving random sub-problems, and their construction rules must not inadvertently render them amenable to classical optimization. 

Identifying hypergraph properties that cause this entire class of methods to break down could lead to valuable new discoveries about macroscopic quantum tunneling physics and problem hardness, and further algorithm innovations in the steps needed to mitigate their effects. As we stated in the introduction, the underlying reasons for classical approximation hardness are generic and intuitive if often only applicable to extremal problem instances, but their quantum equivalents are not, and much more opaque. The correctness of our predictions here would imply that quantum approximation hardness may not be generic at all, and instead specific to as-yet undiscovered sets of problem properties. Let's go exploring.

\section{Acknowledgments}

We would like to thank Tameem Albash, David Huse, Matthew Jones, Chris Laumann, Gianni Mossi, Ojas Parekh, Eleanor Rieffel, Luca Trevisan, Antonello Scardicchio and Davide Venturelli for valuable discussions of the issues in this work. We would like to thank Alex Dalzell and Caleb Rotello for detailed feedback on the manuscript. We would also like to thank Takuto Komatsuki and Joey Liu for support with HPC calculations. This work was supported by the DARPA Reversible Quantum Machine Learning and Simulation program under contract HR00112190068, as well as by National Science Foundation grants PHY-1653820, PHY-2210566, DGE-2125899, and by the U.S. Department of Energy, Office of Science, National Quantum Information Science Research Centers, Superconducting Quantum Materials and Systems Center (SQMS) under contract number DE-AC02-07CH11359. The SQMS Center supported EK's advisory role in this project, as well as time improving and fine tuning the algorithms (including the formulation of the MSFO algorithm) and writing the paper. Many of the numerical simulations in this work were performed with a generous grant of HPC access from the Fujitsu Corporation. Part of this research was performed while the one of the authors (BAB) was visiting the Institute for Pure and Applied Mathematics (IPAM), which is supported by the National Science Foundation (Grant No. DMS-1925919).
The Flatiron Institute is a division of the Simons Foundation.
\appendix

\section{Appendix: small-$N$ deviations from the asymptotic cost per flip curve}\label{appendix:costflip}

\begin{figure}
\includegraphics[width=\columnwidth]{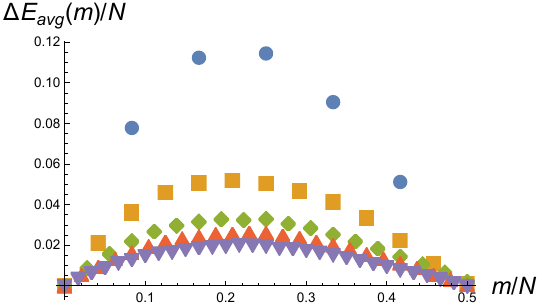}
\caption{Fractional deviations from the average energy Eq.~\ref{avgE} for random MAX-3-XORSAT instances, as a function of $m$ flips away from the planted ground state $G$, numerically estimated for $N= \cuof{12,24,36,48,60}$ (blue circles through purple triangles and monotonically decreasing).}\label{dEfig}
\end{figure}

In this section, we consider small-$N$ deviations from the asymptotic mean cost per flip scaling, Eq.~\ref{avgE}. As discussed in that subsection, the statement $E_{\mathrm{avg}} \of{x} = E \of{G} \of{1- \frac{2x}{N}}^p$ is only asymptotically exact, owing to the approximation made in taking the step size to zero in Eq.~\ref{perflip}. Empirically, as plotted in FIG.~\ref{dEfig}, the real average energy at finite $N$ tends to be larger (e.g. less negative) than the asymptotic scaling predicts, though as fraction of the total energy these deviations decrease with system size, with an overall envelope that scales roughly as $1/N$. 

While this effect has no impact on the asymptotic scaling of our methods, it can potentially pose challenges for small $N$ simulation. As the average energy costs (e.g. $E_{\mathrm{avg}} \of{m} - E \of{G}$) enter in products into the denominators of collective tunneling rate calculations, even small fractional increases can produce significant reductions in the associated minimum gaps. They can similarly alter how the system responds to the spectral filtering transformations. However, this is at least partially balanced in scaling by the fact that the scale of the cost per flip increases is decreasing with $N$, so for the algorithms considered in this work the resulting deviations from the predicted asymptotic scaling are typically modest.

We caution however that at larger $p$ these deviations become more severe, even if they too vanish as $N \to \infty$. This issue is one of the main reasons we focused exclusively on $p=3$ MAX-3-XORSAT in this work, as the methods and techniques we introduced can readily generalize to higher degree constraints. We feel that the generalization from mean-field like analytical calculations to real disordered problems is highly nontrivial, and thought it important to be able to test the basic veracity of our predictions using detailed numerical simulations. For $p=5$ we were not confident in being able to access the asymptotic scaling of the system in full wavefunction simulations with $N \leq 25$, so we did not explore MAX-5-XORSAT in this work.

\section{Tunneling between two $p$-spin wells}\label{appendix:pspin}.

\begin{figure}
\includegraphics[width=\columnwidth]{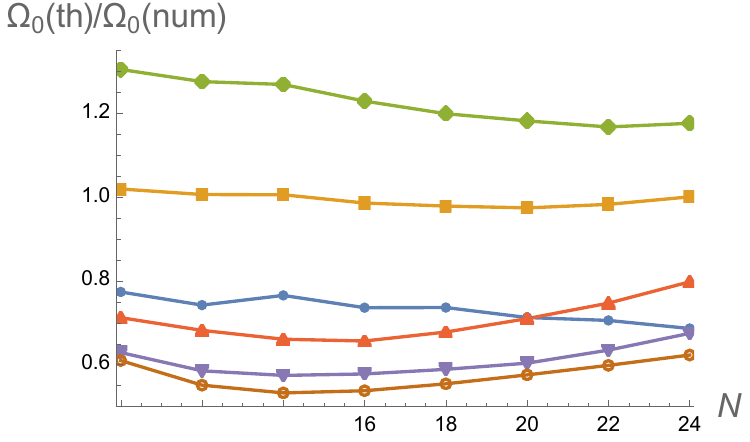}
\caption{Comparison of Eq.~\ref{Om0full} to the tunnel splitting found from exact diagonalization of the Hamiltonian in Eq.~\ref{Hpsym}, for $p= \cuof{2,3,4,5,7,9}$ (blue, gold, green, purple, red, brown) and $\kappa$ stated in the text. There are no free parameters in this; exponents found by extrapolating Eq.~\ref{Om0full} out to $N=80$ agree with the numerical results to within few percent. To make sensible predictions for $p \geq 4$ an additional modification was needed to regularize the transverse field corrections to state energies-- see text for details. Only $p=3$ is directly relevant to this work; simulation of other $p$ values is to demonstrate the accuracy and flexibility of our theory.}\label{pspinfig}
\end{figure}

To benchmark our prediction in Eq.~\ref{Om0full}, we performed a series of numerical simulations, shown in FIG.~\ref{pspinfig}. For these calculations, we computed the degeneracy splitting $2 \Omega_0$ for two p-spin wells centered at classical states $a$ and $b$ separated $N/2$ flips apart, with a total Hamiltonian given by
\begin{align}\label{Hpsym}
&H = 
- \frac{1}{N^{p-1}} \sqof{ \of{\sum_j Z_j a_j}^p + \of{\sum_j Z_j b_j}^p       } 
\nonumber \\
&- \kappa \sum_j X_j,
\end{align}
The $\of{p,\kappa}$ values we used were $\cuof{ \of{2,1.25},\of{3,1.25},\of{4,1.2},\of{5,1.1},\of{7,1.06},\of{9,1.03} }$, and the resulting exponents found from numerical fitting of $\Omega_0 \of{N} = a \sqrt{N} 2^{-b N}$, as compared to Eq.~\ref{Om0full}, are (listed as $\of{p,b_{num},b_{th}}$) $\{ \of{2,0.175,0.180},\of{3,0.235,0.239},\of{4,0.275,0.287}$, $\of{5,0.482,0.512},\of{7,0.591,0.623}, \of{9,0.676,0.722} \}$, demonstrating excellent overall agreement, even with the prefactor, as shown in FIG~\ref{pspinfig}.

While fully sufficient for our purposes here (where $p=3$), we want to highlight two issues with the formulation in Eq.~\ref{umn} that suggest a more refined treatment should be developed to tackle $p>3$ and/or asymmetric minima. First, $\partial_m E_{m,n}$ is guaranteed to vanish at some point, since the energy curve at the peak of the barrier is smooth, and since it enters the calculation in the denominator, it predicts a diverging scale for energy corrections. In symmetric cases, this divergence is canceled by the factor of $N/2 - 2n$ in the numerator and the result is finite, but if the energy curve is not symmetrical between the two minima those two factors won't generally coincide (as the barrier peak won't occur at $M/2$ flips for asymmetric minima) and the divergence will not be canceled.

Second, even we consider the symmetric case and the divergence \emph{is} canceled, this formulation of perturbation theory can still lead to unphysical results. For $p\geq 4$, while the energy corrections remain finite throughout, for $\kappa$ relatively close to $\kappa_c$, the renormalized energy barrier peaks at some $m= m_c < M/2$, and decreases from there. This is unphysical, and we can mitigate it with the ad hoc prescription that $u_{m,0}$ simply stops increasing beyond $m=m_c$ and maintains that value until $m=M/2$. The sum in Eq.~\ref{defxi} is similarly modified so that the insertion of an additional spin flip cannot lower the energy. Using this prescription produces quantitatively good results for $p$ running from 4 to 9, as shown in FIG.~\ref{pspinfig}. 

One can also work through this analysis for the splitting $p=2$ mean field all-to-all ferromagnet, where the transition out of the ferromagnet phase occurs when $\kappa=2$ for this normalization. In that case all $N$ spins must flip and there are no secondary spin corrections; further, due to a symmetry cancellation the self energy corrections (e.g. Eq.~\ref{umn}) are energy-independent and thus do not renormalize the tunneling barrier. Repeating the same steps for this somewhat simpler calculation yields
\begin{equation}
\label{om0p2ata}
\Omega_0 \of{N\kappa} \simeq \sqrt{\frac{N}{2 \pi} } \frac{\kappa^N}{w \of{N, \kappa}^2} \of{ \frac{e}{4} }^N. 
\end{equation}
We of course still need to evaluate the normalization factor $w \of{N,\kappa}$. Explicitly, it is given by
\begin{eqnarray}\label{wMp2ata}
w \of{N,\kappa} = \sqrt{1 + \sum_{k=1}^{N/2} \binom{N}{k} \kappa^{2k} \of{k! \prod_{j=1}^k \epsilon_k^{-1} }^2 }
\end{eqnarray}
This function is extremely well fit by a simple exponential in $N$, with a coefficient that depends on $\kappa$. Empirically, it can be well approximated as, 
\begin{eqnarray}
w \of{N,\kappa}^2 \propto \of{1 + a \kappa^b}^N,
\end{eqnarray}
where fitting a range of $\kappa$ values from $0.4$ to $1.6$ gives $a=0.066$ and $b=2.25$, in fairly good agreement with the result $\of{1+\kappa^2/16}^M$ one can derive from simple second order perturbation theory. 

Of note is that Eq.~\ref{wMp2ata} is both of similar quantitative accuracy as the results plotted in FIG.~\ref{pspinfig}, and it predicts that the decay exponent smoothly approaches zero as $\kappa \to 2$, at which point the minimum gap decays inverse polynomially. Similar behavior--a crossover to polynomial scaling when theory predicts the decay exponent must vanish--was observed both analytically and numerically in the 1d transverse field Ising chain in~\cite{grattan2023exponential}. We see this as lending further indirect support to our polynomial time approximation hardness prediction in Eq.~\ref{PTAScondition}--though the situation is fairly different all these calculations identify a crossover to polynomial scaling by the point at which a predicted decay exponent vanishes in careful many-body theory.

\section{Band structure considerations for linear spectral folding}\label{bandstructure}

In the simulations in section~\ref{specfoldsims}, we noticed that the quadratic AQC formulation of spectrally folded quantum optimization performed similarly to the rough prediction in section~\ref{theorysec}, returning a worst case approximation ratio of $q_a = 0.7$ compared to the approximate theoretical estimate 0.68. In contrast, the linear trial minimum annealing versions more notably outperformed the analyitical prediction (returning $q_a = 0.75$ as compared to 0.6), and in numerical experiments the best choice for the lowering parameter $C \of{t}$ in Eq.~\ref{TMAhamdef} was 2-3 times that one would naively expect. We suspect that this has to do with the band structure of the dressed eigenstates in $\mathcal{T}$ (the band of classical states at or very near the fold energy) when the transverse field $\kappa$ is nonzero, and in this subsection argue why this might be the case. We present these arguments as suggestions and not a rigorous proof, and think more work on this issue could be valuable for shedding light on the detailed structure of these optimization algorithms.

Let $\cuof{\ket{T_j}}$ be the set of all $N_{\mathcal{T}}$ bitstring states in $\mathcal{T}$; states in $\MT$ all have $E \simeq A E_{\mathrm{GS}}$ before any corrections from the lowering Hamiltonian. Further let $\ket{\psi_k}$ be a dressed eigenstate whose spectral weight is concentrated in $\MT$. Thus
\begin{eqnarray}
\ket{\psi_k} \simeq \sum_{j \in \MT}^{N_{\MT}} c_{jk} \ket{T_j}, \; \; \avg{\abs{c_{jk}^2}} \simeq \frac{1}{N_\MT}.
\end{eqnarray}
Finally, let $\Omega_{Lj}$ be the $M$-spin tunneling matrix element into a bitstring state $\ket{T_j}$ from $\ket{L_D}$, where $\avg{\Omega_{Lj}}_j = \Omega_0$ as calculated in Eq.~\ref{Om0full}. 

We choose our gauge and basis so that all transverse field terms are negative and $H$ is real, which is straightforward since $H$ is stoquastic. In this gauge we can assume all the $\Omega_{Lj}$ matrix elements are negative, though their magnitudes can of course vary substantially over $\MT$ (but we expect are fairly well-correlated locally, when considering states in $\MT$ only a few flips apart). Because the states in $\MT$ are all near-zero energy, the local transverse field-induced ``hopping" matrix elements connecting them are either direct (if a given pair of states in $\MT$ are one flip apart), or the result of short ranged, few-flip tunneling through local excited states. In either case, the resulting matrix element is negative, and the band of $\ket{\psi_k}$ states are (approximately) the eigenstates of a hopping-like model on a sparse, disordered graph of $N_\MT$ sites, where most hopping matrix elements are negative and there is local potential disorder. With all these quantities defined, we now want to estimate the tunneling matrix element $\tilde{\Omega}_{Lk}$ from $\ket{L_D}$ into $\ket{\psi_k}$.

We first consider states $\ket{\psi_k}$ near the band center, e.g. where the energy shifts from the transverse field are solely due to the second order local processes captured in section~\ref{theorysec} and the ``hopping energy" is nearly zero. For such states, we can assume the signs of the amplitudes $c_{jk}$ are randomized. Then by the law of large numbers:
\begin{eqnarray}\label{bandcentertunnel}
\avg{\tilde{\Omega}_{Lk}^2}_k &\simeq& \avg{ \of{ \sum_{j \in \MT}^{N_\MT} c_{jk} \Omega_{Lj} }^2 }_k \\
&\propto& N_\MT \avg{ c_{jk}^2 \Omega_{Lj}^2}_j \propto \avg{\Omega_{Lj}^2}. \nonumber
\end{eqnarray}
Thus, the average squared matrix element to tunnel into a state in the band center has the same scaling as one obtains from a more naive calculation that ignores the band structure entirely. Since the majority of states in the band are near the center in this respect, this calculation shows that ignoring the local band structure in $\MT$ is a decent approximation for obtaining the average collective tunneling rate that enters into equations like~\ref{PTAScondition}.

But what happens when we consider states that are more extremal, e.g. near the top or bottom of the band? Near the top of the band, we can assume significant local alternation in the signs/phases of the $c_{jk}$ coefficients, and thus the assumption of randomization is still fairly good and Eq.~\ref{bandcentertunnel} likely accurately captures the scaling. For states near the bottom of the band, however, the situation changes. First, these states will be lower in energy by an $O \of{N}$ factor, which means $C \of{t}$ must be further reduced for $\ket{L_D}$ to cross them, and the individual tunneling matrix elements $\Omega_{Lj}$ into such states may scale differently as a result. We label these matrix elements $\Omega_{Lj}'$ and do not attempt to predict what, if any, changes to scaling may arise. 

Second, to minimize the energy the signs/phases of the $c_{jk}$ of nearby (in Hamming distance) configurations will be synchronized given that the matrix elements that couple them are real and (mostly) negative (again, in this gauge choice; see \cite{bravyi2006complexity}). And this synchronization can dramatically enhance tunneling rates. Let us guess, for example, that in the band ground state $\ket{\psi_0}$, \emph{most} $c_{j0}$ are positive. Then:
\begin{eqnarray}
\abs{\tilde{\Omega}_{L0}^2} \simeq \avg{ \of{ \sum_{j \in \MT}^{N_\MT} c_{j0} \Omega_{Lj}' }^2 } \propto N_\MT \avg{\of{\Omega_{Lj}'}^2}.
\end{eqnarray}
Since $N_\MT$ is exponentially large in $N$ (Eq.~\ref{NT}), the tunneling matrix element into the band ground state can be exponentially larger than the average matrix element for tunneling into the band center, though if $\Omega_{Lj}'$ decays more quickly with $N$ that may reduce or erase this effect. We expect that collective tunneling into low-lying states in the band could be similarly enhanced even if there is more variation in the signs of the $c_{jk}$ terms, as the matrix elements $\Omega_{Lj}$ are locally correlated in magnitude. And though there are exponentially fewer states near the band bottom compared to the band center, if the tunneling matrix elements are exponentially larger the weighted average over all states in the band can be substantially increased--we saw this effect already in averaging over the relative distance to $\ket{L}$ of different states in $\MT$ (Eq.~\ref{Om0avg}). It is thus quite plausible that band structure effects could further increase collective tunneling rates and be responsible for the relative overperformance of TMA spectral folding in our simulations, as compared to the analytical prediction.

All that said, we did not include this analysis in the main text because we do not presently have the mathematical tools to quantitatively predict the band structure and nature of the dressed eigenstates, nor can we predict the potential scaling changes in the collective tunneling matrix elements $\Omega_{Lj}'$ to states near the bottom of the band. So we present these results to justify our claim that ignoring band structure is likely to underestimate the achievable approximation ratio, and to suggest interesting directions for further research.

\begin{figure*}
    \includegraphics[width=0.49\textwidth]{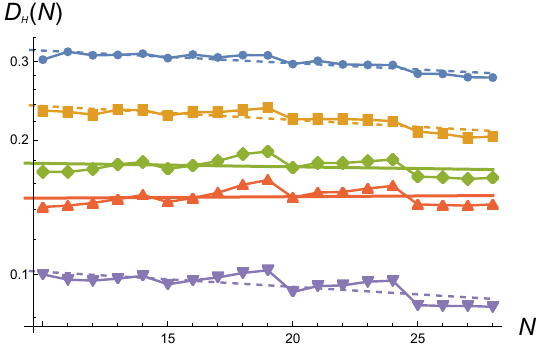}
    \includegraphics[width=0.49\textwidth]{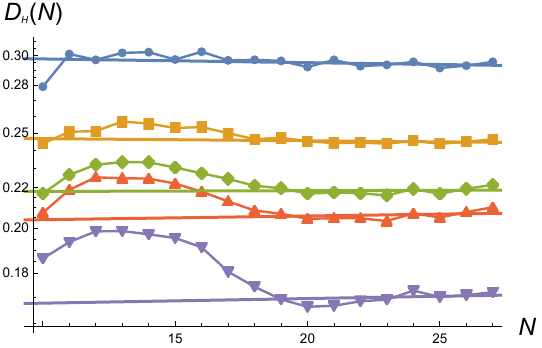}

    \includegraphics[width=0.49\textwidth]{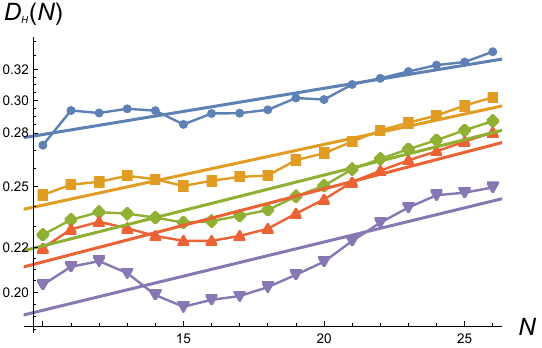}
    \includegraphics[width=0.49\textwidth]{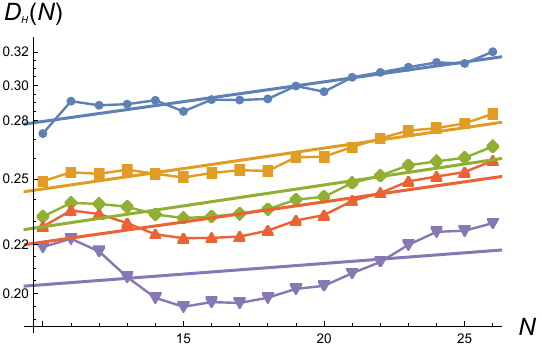}
    \caption{Probabilities of returning states within Hamming distance $D_H = \cuof{2/5,1/3,1/4,1/5,1/8}N$ flips from the ground state $G$ for the AQC formulation of spectral folding in figure~\ref{specfoldAQCfig} (top to bottom curves), for $N_C = \cuof{2,4,6,3 \sqrt{N}/2}N$ (clockwise from top left) and application parameters in text. For $N_C=2N$ some probabilities decay slowly with system size, due to competition with other minima (though non-monotonicity makes the fitting somewhat ambiguous here); for all other cases they are constant or increase toward some large $N$ saturation value, indicating that this variation is finding states near the global minimum with constant probability. The probability of finding $G$ is not plotted, as for spectrally folded optimization it's essentially zero.}\label{AQC_HD_fig}
\end{figure*}

\begin{figure*}
    \includegraphics[width=0.49\textwidth]{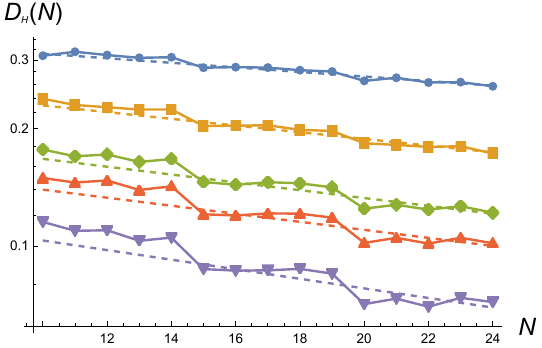}
    \includegraphics[width=0.49\textwidth]{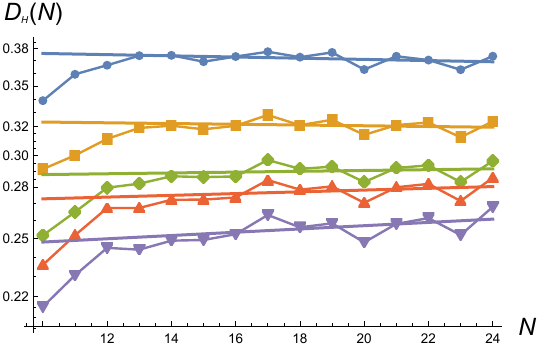}

    \includegraphics[width=0.49\textwidth]{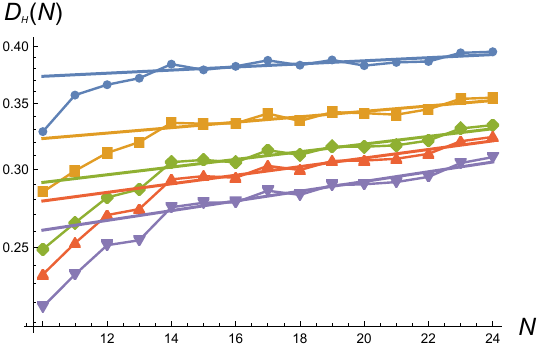}
    \includegraphics[width=0.49\textwidth]{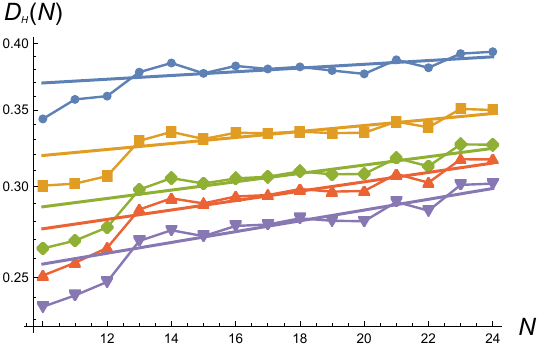}
    \caption{Probabilities of returning states within Hamming distance $D_H = \cuof{2/5,1/3,1/4,1/5,1/8}N$ flips from the ground state $G$ for the TMA formulation of spectral folding in figure~\ref{specfoldtmafig} (top to bottom curves), for $N_C = \cuof{2,4,6,3 \sqrt{N}/2}N$ (clockwise from top left) and application parameters in text. For $N_C=2N$ the probabilities decay slowly with system size, due to competition with other minima; for all other cases they are constant or increase toward some large $N$ saturation value, indicating that this variation is finding states near the global minimum with constant probability. The probability of finding $G$ is not plotted, as for spectrally folded optimization it's essentially zero.}\label{TMA_HD_fig}
\end{figure*}

\begin{figure*}
    \includegraphics[width=0.49\textwidth]{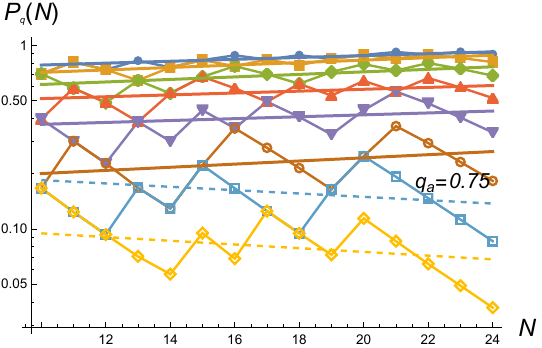}
    \includegraphics[width=0.49\textwidth]{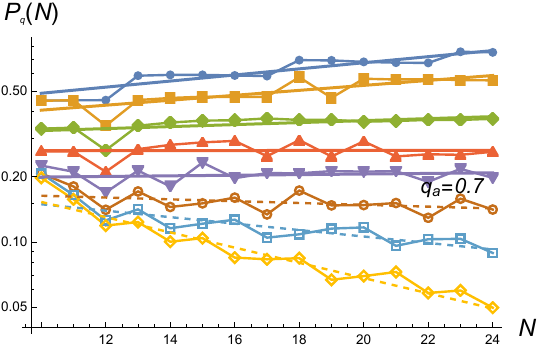}

    \includegraphics[width=0.49\textwidth]{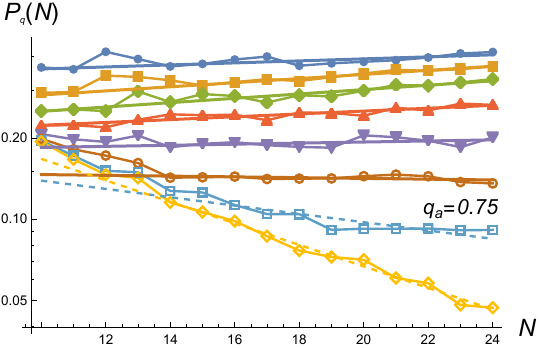}
    \includegraphics[width=0.49\textwidth]{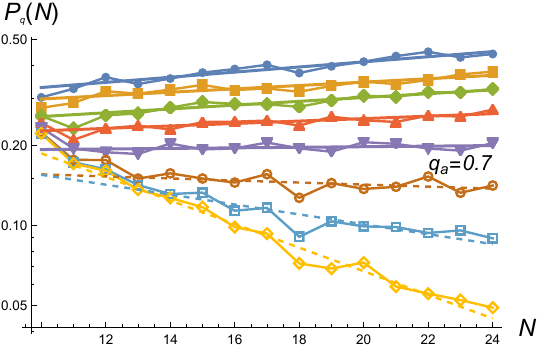}
    \caption{Performance quadratic AQC spectral folding, for $N_C = \cuof{2,4,6,3 \sqrt{N}/2}N$ (clockwise from top left) with increased approximation target $A=0.85$ and all other parameters equal. In comparison to the $A=0.75$ results in figure~\ref{specfoldAQCfig}, increasing $A$ improves prefactors but does not improve scaling and in the case of $N_C/N = 6$, modestly reduces $q_a$. This suggests we have reached the performance ceiling for this variation on the tested PPSP classes.}\label{specfoldaqcfig-85}
\end{figure*}

\bibliography{fullbib}

\end{document}